# Metal Halide Perovskites for Violet and Ultraviolet Light Emission

*Sebastian Fernández,[1,†] Manchen Hu,[1,†] Tyler K. Colenbrander,[1,†] Divine Mbachu,[1] Daniel N. Congreve[*,1]*

1: Department of Electrical Engineering, Stanford University, Stanford, CA, 94305
†: These authors contributed equally
*Email: congreve@stanford.edu

**Abstract**

Emissive metal halide perovskites (MHPs) have emerged as excellent candidates for next-generation optoelectronics due to their sharp color purity, inexpensive processing, and bandgap tunability. However, the development of violet and ultraviolet light-emitting MHPs has lagged behind due to challenges related to material and device stability, charge carrier transport, tunability into the ultraviolet spectrum, toxicity, and scalability. Here, we review the progress of both violet and ultraviolet MHP nanomaterials and light-emitting diodes, including materials synthesis and device fabrication across various crystal structures and dimensions (e.g., bulk thin films, 2D thin films, nanoplatelets, colloidal nanocrystals, and more) as well as lead-free platforms (e.g., rare-earth metal halide perovskites). By highlighting several pathways to continue the development of violet and ultraviolet light-emitting MHPs while also proposing tactics to overcome their outstanding challenges, we demonstrate the potential of state-of-the-art violet and ultraviolet MHP materials and devices for important applications in public health, 3D printing, nanofabrication, and more.

**Key Words:** Metal halide perovskites, Violet emission, Ultraviolet light, Light-emitting diodes, Lead-free, Colloidal nanocrystals, 2D perovskites, Rare-earth, Optoelectronics

## Introduction

Metal halide perovskites (MHPs) have emerged as a promising semiconductor material for a variety of applications such as solar cells,[1–4] photodetectors,[5] photocatalysis,[6] lasers,[7,8] and light-emitting diodes (LEDs)[2,9–13] due to their favorable electrical, optical, and magnetic properties accessible through inexpensive processing methods. In particular, perovskite LEDs (PeLEDs) have strong potential for next-generation LED technologies due to their sharp color purity, high luminance, bandgap tunability, defect tolerance, and potential for low-cost fabrication.[2,9,10,12–15] Following the rapid development of perovskite solar cells, PeLED efficiencies have likewise experienced significant improvements over the past decade, with external quantum efficiencies (EQEs) now exceeding 32% for red and green PeLEDs[16–18] and 26% for blue PeLEDs,[19] compared to 0.1% for some of the first green room-temperature PeLEDs in 2014.[20,21]

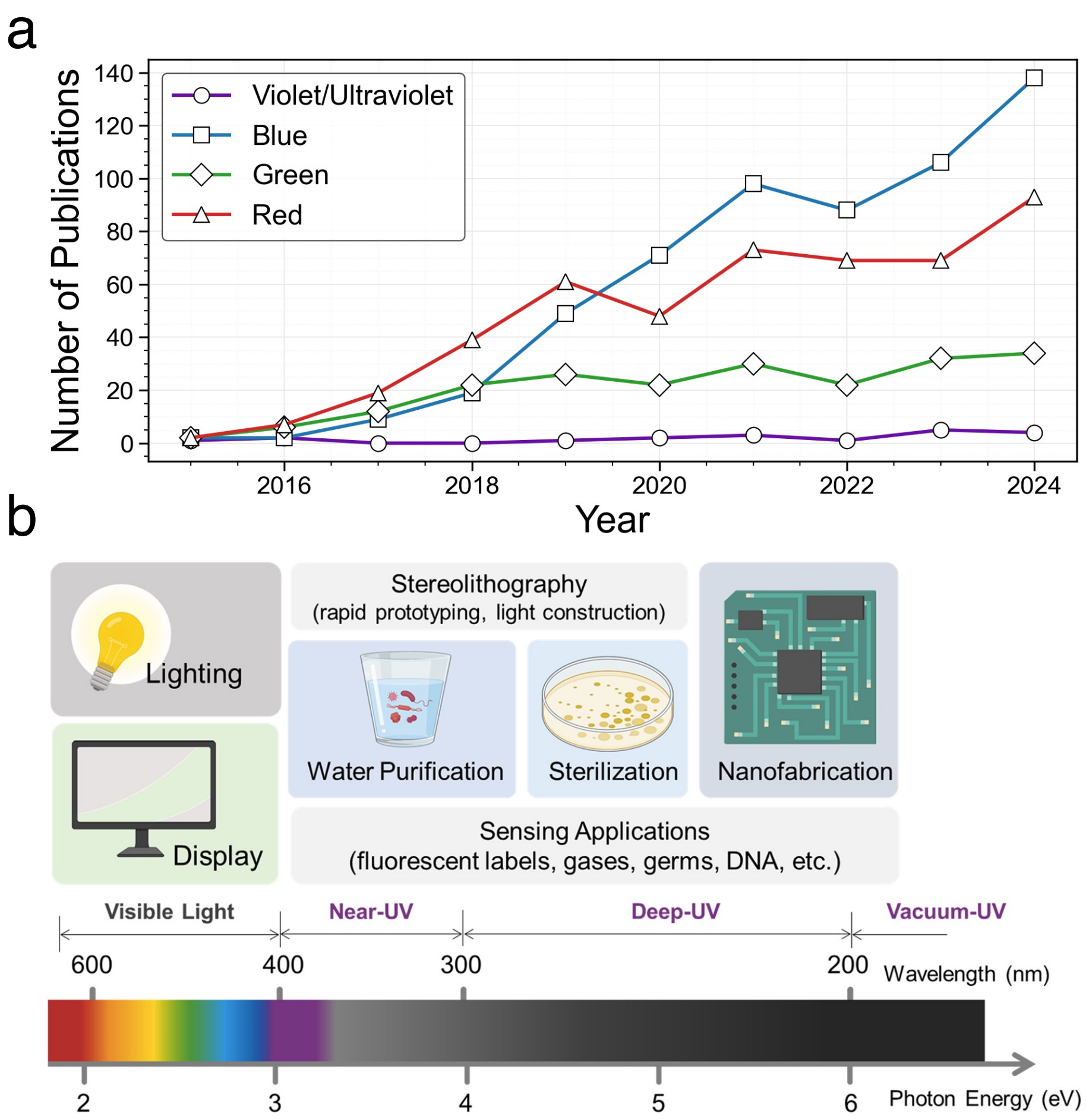

**Figure 1**. Progress of perovskite light-emitting diodes (PeLEDs). **(a)** The number of PeLED publications each year sorted by violet/ultraviolet, blue, green, and red PeLEDs. To systematically quantify the research disparity across the emission spectrum of PeLEDs, a bibliometric analysis was performed using the Web of Science Core Collection (https://www.webofscience.com/wos/woscc/smart-search) on February 3, 2025. We utilized a rigorous Boolean search strategy that intersected device definitions with specific spectral indicators, employing the 'Topic' field for material identification and the 'Title' field for strict chromatic filtering. The dataset was generated using the base string Topic: (perovskite*) AND Topic: ("light-emitting diodes"*) refined by AND Title: (color*), where the color variable was iteratively substituted with blue*, green*, red*, or violet* to capture relevant variations. Note: only publications with a listed color are represented and thus this does not fully represent the overall field of PeLED publications. **(b)** Summary of the applications of light in both the visible and ultraviolet energy regimes. Adapted with permission from ref. [22]. Copyright 2023 Elsevier Inc.

Despite the success of red, green, and blue PeLEDs, studies on violet (defined here as between 400 nm and 435 nm) and ultraviolet (UV) (<400 nm) emissive MHPs have lagged behind (**Figure 1a**) due to challenges related to material and device stability, charge carrier transport, tunability into the UV spectrum, toxicity, and scalability. However, light emission in the violet and UV (V/UV) range is crucial for many applications beyond lighting and display technologies, motivating the need for further research to improve these emitters. Solving these current challenges could unlock low-cost, scalable V/UV PeLEDs for applications such as public health (sterilization[23–26] and water purification[26]), 3D printing,[27] micro/nanofabrication,[28] optical communications,[29] bio-sensing,[26] and more[30–32] (**Figure 1b**).

Traditionally, V/UV light emission for these applications has been achieved by fluorescent mercury lamps, which are inefficient, bulky, and slow to turn on and off.[33] To replace sources like mercury lamps, V/UV LEDs have become increasingly common as they offer a low-power, compact solution with quick turn on and off speeds and the ability to tune the emission spectrum.[33] Currently, commercial V/UV LEDs are based on III-V material systems such as InGaN/GaN/AlGaN materials grown with metal organic chemical vapor deposition (MOCVD).[34] However, MOCVD is a costly and complex process that requires high temperatures and involves toxic and pyrophoric precursor gases like trimethylaluminum ($Al(CH_3)_3$) and trimethylgallium ($Ga(CH_3)_3$), which significantly contribute to safety concerns and manufacturing complexity.

Furthermore, the requirement of lattice-matched substrates and precise thermal budgets elevates the cost and scalability challenges of producing high-performance V/UV emitters.

Compared to fluorescent mercury lamps and III-V LEDs, the MHP material system discussed here can be fabricated using simpler methods via solution-based processing or thermal evaporation, paving the way for low-cost, non-toxic, scalable V/UV PeLEDs. In this review, we first highlight several application spaces (e.g., sensing, nanofabrication, and public health) for V/UV emitters and explore the fundamentals of the MHP structure that dictate light emission. We then discuss the strategies employed to achieve V/UV emission in MHPs through compositional engineering of the halide (X-site), B-site, and A-site as well as dimensional control. We survey the current progress on a variety of material systems that demonstrate V/UV photoluminescence (PL) across different crystal structures and dimensions (e.g., three-dimensional (3D), two-dimensional (2D), one-dimensional (1D), or zero-dimensional (0D) crystal structures in thin films, nanoplatelets, colloidal nanocrystals, and more) as well as lead-free platforms (e.g., rare-earth metal halide perovskites). We then discuss LED device architectures and the different charge injection strategies utilized to achieve V/UV electroluminescence (EL). By reviewing the current state-of-the-art performance of V/UV PeLEDs, we highlight the promising work conducted so far and the challenges and limitations that motivate areas of further research. We propose tactics to overcome some of the outstanding challenges and suggest several pathways to continue the development of V/UV light-emitting MHPs. Lastly, we present an outlook on the potential impacts of improved V/UV light-emitting MHPs and their corresponding PeLEDs.

## Applications for Violet and Ultraviolet Light Emission

V/UV materials and LEDs have important applications in numerous technology spaces including sensing, nanofabrication, public health, and more. We highlight some of these promising applications in **Figure 2.** From **Figure 2a**, Shur *et al.* illustrate a UV LED fluorimeter to measure the fluorescence from particles dispersed in water or air.[33] The use of UV LEDs within fluorimeters can lead to sharper contrast by inducing a stronger fluorescence response from the sample.[35] Moreover, UV LEDs can be integrated into nanofabrication processes, including photolithography. In **Figure 2b**, Feng *et al.* develop AlGaN UV micro-LEDs to enable maskless photolithography.[36] Specifically, they construct micro-displays made of UV micro-LEDs and transfer the image on a photoresist-coated wafer, showcasing the possibilities for UV micro-LEDs to reduce time and expenses incurred by the semiconductor industry. Lastly, UV light can deactivate microbes (e.g., bacteria and viruses) and thus improve public health, broadly. From **Figure 2c**, Bhattarai *et al.* show the main mechanisms in which UV light can inflict damage on microbes, including thymine formation, protein cross-linking, oxidative damage, protein structure disruption, and enzyme inactivation.[37] The UV-C regime (200-280 nm) is particularly attractive since these emission wavelengths are known to inactivate harmful viruses.[33,37] For instance, 270-280 nm light can greatly reduce *Escherichia coli* (E. Coli) populations with doses on the order of a few mJ $mL^{-1}$.[33] However, emissive MHPs are understudied at this important spectrum regime and new materials are needed to achieve such short emission wavelengths. In the following sections, we suggest pathways to tune the optical properties of V/UV MHPs, and corresponding PeLEDs, to enter this uncharted territory.

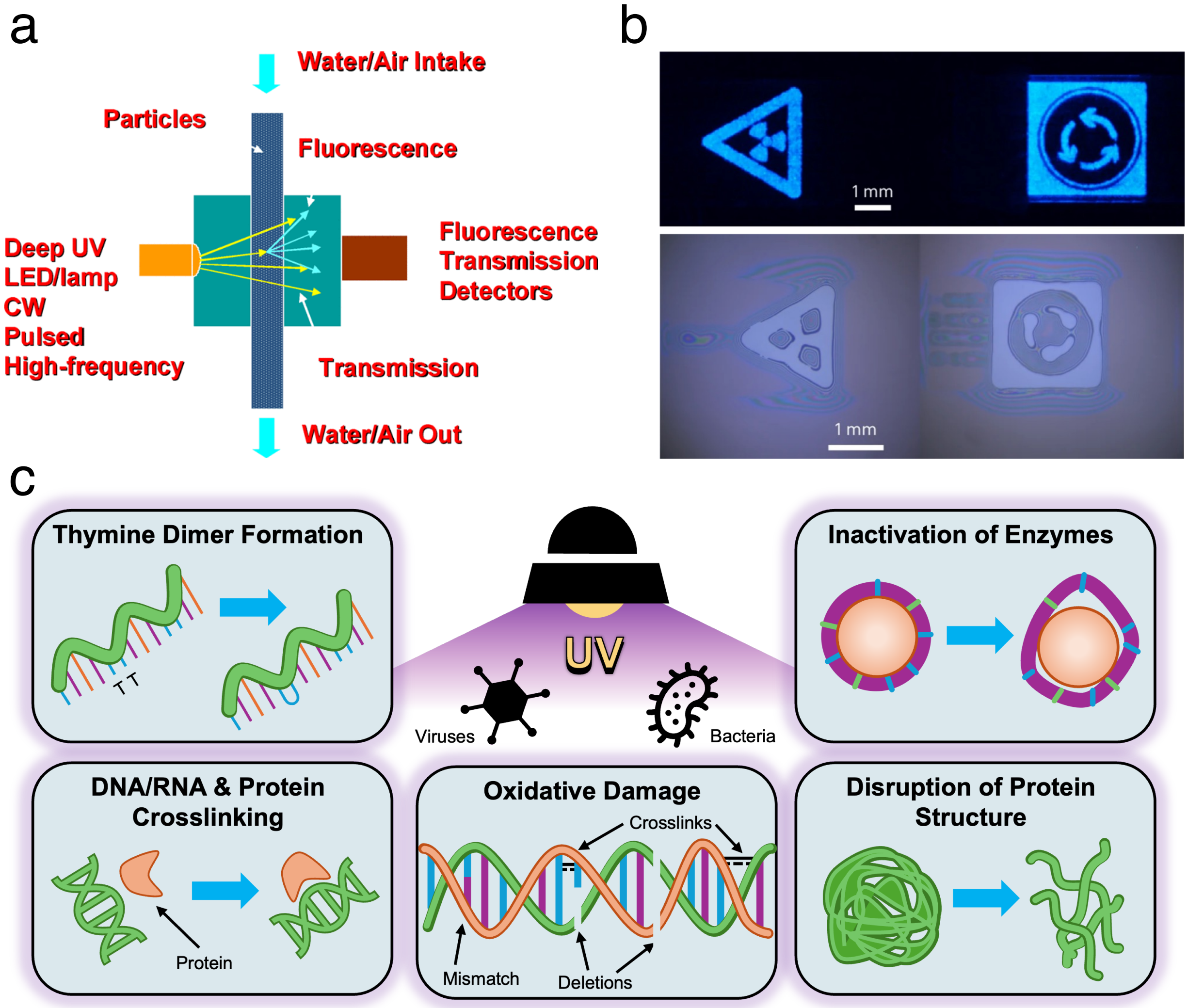

**Figure 2.** Applications of V/UV materials and LEDs. **(a)** Schematic of a UV LED fluorimeter. Adapted with permission from ref. [33]. Copyright 2010 IEEE. **(b)** AlGaN UV micro-LED displays (top) used to pattern photoresist-coated silicon wafers and corresponding maskless photolithography images (bottom). Adapted with permission from ref. [36]. Copyright 2024 Springer Nature. **(c)** Five UV light-induced microbial deactivation mechanisms. Adapted with permission from ref. [37]. Copyright 2024 The Authors.

## Fundamentals of Metal Halide Perovskites for Light Emission

The standard 3D perovskite structure is based on the formula $ABX_3$, where A and B are cations of different sizes and X is an anion. For traditional metal halide perovskites (MHPs), the A-site is typically an organic or inorganic cation (e.g., methylammonium ($MA^+$), formamidinium ($FA^+$), $Cs^+$), the B-site is a divalent metal (e.g., $Pb^{2+}$, $Sn^{2+}$), and the X-site is a halide anion (e.g., $I^-$, $Br^-$, $Cl^-$). However, due to the vast parameter space of available materials that influence both composition and lattice structure, there are an increasing number of alternative structures that are typically referred to as perovskites or perovskite derivatives.

As shown in **Figure 3a**, the standard 3D $ABX_3$ perovskite structure can be modified to form unique materials by changing the relative amounts of halides at the X-site or by partially or fully replacing cations at the A- or B-sites. The choices in chemical composition for these sites, along with material synthesis and processing parameters, can alter the crystal structure and geometric structure of the material, affecting its electronic and optical properties. Beyond the traditional inorganic or organic cations at the A-site, the crystal structure dimensionality can be controlled by employing large, bulky organic cations like phenethylammonium ($C_6H_5CH_2CH_2NH_3^+$, $PEA^+$) to form layered 2D ($PEA_2BX_4$) or quasi-2D ($PEA_2(ABX_3)_{n-1}BX_4$) crystal structures.[22,38] At the B-site, partial doping of lead halide perovskites with ions like $Ni^{2+}$ and $Cd^{2+}$ has been shown to improve the luminescent properties of these materials.[39,40] Alternatively, fully replacing the divalent metal (e.g., $Pb^{2+}$) with a trivalent metal such as $Sb^{3+}$ or $Bi^{3+}$ can form materials with an $A_3B_2X_9$ structure.[41,42] Other trivalent metals such as $Ce^{3+}$ can form structures with a different stoichiometry, $A_3BX_6$.[43] The B-site can also be occupied by a monovalent metal such as $Cu^+$, which has been demonstrated in different crystal structures such as $A_2BX_3$ or $A_3B_2X_5$.[44,45] As evident from the variety of possible crystal structures that arise from changing the A- or B-site cations, the structure formed not only depends on the charge and size of the new cations, but also on the properties of the other ions in the structure that influence the bonding between the atoms and the resulting crystal structure. The choices in chemical composition for these ionic sites, along with material synthesis and processing parameters, can alter the crystal structure and geometric structure of the material, affecting its electronic and optical properties. These choices enable access to a wide array of perovskite, double-perovskite and perovskite inspired structures, all of which will be broadly included in this review.

For example, these derivatives of the standard 3D $ABX_3$ structure tend to have a reduced-dimensional crystal structure like those shown in **Figure 3a** (e.g., 2D layers in $PEA_2PbBr_4$, 1D chains in $Rb_2CuBr_3$, or 0D octahedra in $Cs_3CeBr_6$).[22,43,46,47] There are two main types of 2D perovskite crystal structures: Ruddlesden-Popper structures with offset layers separated by two monoammonium ligands and Dion-Jacobson structures with compact, aligned layers separated by one diammonium ligand.[48] In both types, the alternating layers of organic ligands and 2D perovskite octahedra create a multiple quantum well structure with a high exciton binding energy.[49] There are also two main types of 1D crystal structures, face-sharing chains and edge-sharing chains, where the perovskite structures are confined in two dimensions and form isolated 1D nanowires.[50] When confined in three dimensions, 0D isolated perovskite octahedra are formed. In these reduced-dimensional structures, the strong confinement of excitons leads to higher exciton binding energies and more effective radiative recombination.[43,51]

Along with the dimensionality of the crystal structure, the dimensionality of the geometric structure (i.e., nanoscale morphology) can be controlled by multiple factors including the composition or synthesis method. As shown in **Figure 3a**, the geometric structure can be composed of a thin film (3D), nanoplatelets (2D), nanowires (1D), or quantum dots (0D), which may or may not match the dimensionality of the crystal structure. Thus, it is important to distinguish between crystal structure and geometric structure dimensionality when discussing dimensionality effects. For example, a bulk $CsPbCl_3$ thin film could have a 3D crystal structure and 3D geometric structure. Alternatively, many of the studies reviewed here synthesize CsPbCl3 nanocrystals (NCs) that are geometrically confined in three dimensions and exhibit quantum confinement effects resulting from a 0D geometric structure when their side lengths become small enough, typically on the order of 10 nm.[52] As the NCs become smaller, the strength of quantum confinement effects increase, and the term quantum dot (QD) is used to describe a subset of NCs with sizes on the order of the excitonic Bohr radius, depending on the material system.[53] These quantum confinement effects often blue-shift the emission to higher energies, making dimensional control one of the key levers toward achieving V/UV emission.[54,55] Thus, in addition to 3D $ABX_3$ perovskites, reduced-dimensional metal halide crystal and geometric structures will also be considered for this review.

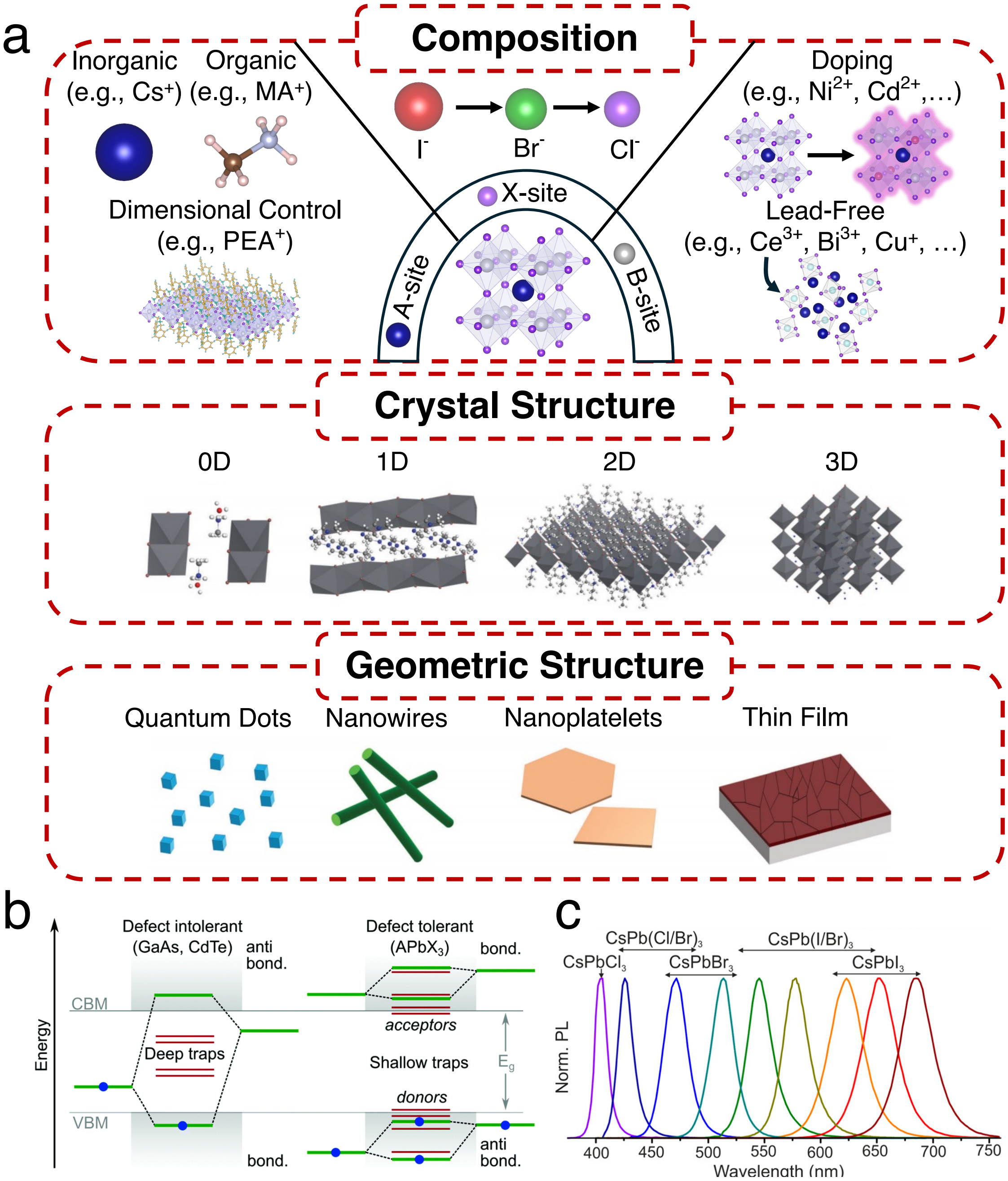

**Figure 3.** Tunable composition and structure, defect tolerance, and tunable emission of MHPs. (a) MHPs can be engineered by tuning the chemical composition at the A-site, B-site, or X-site. These changes, along with other factors like the synthesis method, can result in different crystal structures across 0D, 1D, 2D, and 3D dimensionalities as well as geometric structures across quantum dots, nanowires, nanoplatelets, and thin film morphologies. Structures in the "Composition" section

were made with VESTA.[56] Adapted with permission from ref. [47]. Copyright 2018 Royal Society of Chemistry. **(b)** Location of trap states with respect to the band edges for III-V (e.g., GaAs)/II-VI (e.g., CdTe) semiconductors (left) and MHPs (right). The deep trap states within the bandgap in III-V and II-VI semiconductors can severely impair optoelectronic device performance. Conversely, MHPs have shallow trap states, making them a more defect-tolerant material. Adapted with permission from ref. [57]. Copyright 2019 Royal Society of Chemistry. **(c)** Tunable PL emission from $CsPbX_3$ (X = Cl, Br, I) nanocrystals. Adapted with permission from ref. [52]. Copyright 2015 American Chemical Society.

Along with tunable structural properties, defect tolerance within MHPs increases their viability for optoelectronic devices like solar cells or LEDs. Unlike III-V or II-VI semiconductor materials (e.g., GaAs and CdTe) which have deep trap states, MHPs have shallow traps near the band edges (**Figure 3b**). The formation of deep trap states within the bandgap can impair device performance due to increased non-radiative recombination and immobilization of charge carriers, but since the defect states of MHPs are typically near or within the conduction and valence bands, the bandgap for MHPs is relatively unobstructed and free of traps.[57] Thus, MHPs are not a defect- or trap-free material, but the energetic location of the defect states makes MHPs much more defect tolerant as compared to adjacent semiconductor classes.

Another promising characteristic of MHPs for light emission is their tunable bandgap and peak emission wavelength that can be altered by modifying the composition of the halide anion (X-site). Specifically for lead halide perovskites, it is well known that increasing halide electronegativity ($Cl^- > Br^- > I^-$) lowers the energy of the valence band maximum but does not significantly alter the conduction band minimum, resulting in a higher bandgap.[58,59] Therefore, by sweeping the halide from iodide to bromide to chloride in $CsPbX_3$ NCs, the emission can be tuned across the visible spectrum (700 nm to 410 nm) as shown in **Figure 3c**.[52,60–62]

The optical properties of MHPs can be further modified by engineering at the A-site or B-site in the $ABX_3$ perovskite structure.[13] Changing the A-site cation can expand, contract, or distort the lattice structure, leading to changes in bandgap.[63] Typically, increasing the size of the A-site cation ($Cs^+ < MA^+ < FA^+$) leads to smaller bandgaps, such as in lead iodide ($APbI_3$) systems.[64] However, this effect is typically less pronounced as compared to the X-site halide effect because the electronic structure of halide perovskites is mainly influenced by the B-site metal and X-site halide.[58,59] Changes at the B-site can alter the valence and conduction bands, as well as the octahedral factor (i.e., the ratio between the ionic radii of the B-site and X-site ions), which can influence the bandgap. For example, when comparing lead ($Pb^{2+}$) and tin ($Sn^{2+}$) perovskites, lead has a higher electronegativity than tin, which impacts the orbital energy levels and causes the valence band to shift downwards, leading to a higher bandgap in the lead-based compounds.[63,65] In another example, doping $Cd^{2+}$ into the B-site in $CsPbCl_3$ NCs was shown to increase the bandgap, likely due to lattice contraction from the smaller radius $Cd^{2+}$ compared to $Pb^{2+}$.[40]

Overall, along with their varied structural compositions, beneficial defect tolerance, and tunable optical properties, emissive MHPs show promise for a plethora of applications due to their high color purity, high photoluminescence quantum yield (PLQY), direct bandgap, and potential for low-cost fabrication. For more details on the general characteristics of MHPs and further explanation of the fundamentals and mechanisms governing light emission in MHPs, the authors direct the reader to these other reviews.[2,9–13,66] In this review, we will next introduce the specific strategies utilized to achieve V/UV emission by engineering the composition, crystal structure, and geometric structure of metal halide perovskite and metal halide perovskite derivative materials.

## Strategies for Achieving Violet and Ultraviolet Emission

Enabled by the tunable nature of MHPs, various strategies targeting their composition and structure have been implemented to achieve wide bandgap violet (defined here as having a luminescence emission peak between 400 nm and 435 nm) or UV (<400 nm) emission. Starting with the common $CsPbX_3$ material system, this review explores strategies to create material systems for V/UV emission by altering the composition of the halide (X-site), B-site, or A-site, which can induce changes in the crystal or geometric structure. We then further explore the dimensional control of the geometric structure through synthetic and processing parameters. The materials discussed are categorized by their crystal structure. **Table 1** and **Figure 4** summarize the $ABX_3$ perovskite materials with a 3D crystal structure, such as $CsPbX_3$ NCs. While most references discussed refer to these materials broadly as NCs, they are also sometimes referred to as QDs because the NCs are confined in three dimensions and small enough to demonstrate quantum confinement effects, depending on the exciton Bohr radius of the material system.[52,53] Thus, within our dimensionality framework, these colloidal NCs exhibit a 3D crystal structure and a confined 0D geometric structure. There is also growing interest in exploring lead-free or reduced-dimensional systems that break from the standard $ABX_3$ structure. **Table 2** and **Figure 5** summarize these perovskite or perovskite derivative materials (e.g., 3D double perovskite or reduced-dimensional crystal structures) which represent a variety of crystal and geometric structures.

**Table 1.** Optical properties of V/UV-emitting lead halide perovskite materials with the standard 3D $ABX_3$ crystal structure. Entries are left blank when values are not reported in a reference.

| Perovskite | PL Peak | FWHM | PLQY | Abs. Peak | Reference |
|---|---|---|---|---|---|
| $CsPbCl_3$ | 405 nm* | | | | [52] |
| $CsPbCl_3$ | 404 nm | | | | [61] |
| $CsPbCl_3$ | 390 nm | | 1 % | | [62] |
| $CsPbCl_3$ | 420 nm* | | | | [60] |
| $CsPbCl_3$ | 405 nm | 12 nm | 10 % | | [67] |
| $CsPbCl_3$ | 410 nm | | 3.8 % | | [68] |
| $CsPbCl_3$ (Ce-doped) | 410 nm, 430 nm | | 24.3 % | | [68] |
| $CsPbCl_3$ | 408 nm | 11 nm | 65 % | | [69] |
| $CsPbCl_3$ (Cd-doped) | 406 nm | 10-12 nm | 98 % | 400 nm | [70] |
| $CsPbCl_3$ | 402 nm | 14 nm | 50 % | | [71] |
| $CsPbCl_3$ (Ni-doped) | 406 nm* | | 96.5 % | | [39] |
| $CsPbCl_3$ | 404 nm | 11 nm | 60 % | 394 nm | [72] |
| $CsPbCl_3$ (La/F-doped) | 410 nm | | | 397 nm | [73] |
| $CsPbCl_3$ | 406.1 nm | 10 nm | 77.1 % | | [74] |
| $CsPbCl_3$ (Cu-doped) | 403 nm | 14 nm | 60 % | | [75] |
| $CsPbCl_3$ | 405 nm* | | 97.2 % | 402 nm* | [76] |
| $CsPbCl_3$ | 405 nm* | | 50 % | | [77] |
| $CsPbCl_3$ (Cu-doped) | 403 nm | 12 nm | 12.3 % | | [78] |
| $CsPbCl_3$ (Mg-doped) | 405 nm* | | 79 % | | [79] |
| $CsPbCl_3$ | 410 nm | 13 nm | | | [80] |
| $CsPbCl_3$ (Zn-doped) | 409 nm | 10-15 nm | 88.7 % | 407 nm | [81] |
| $CsPbCl_3$ | 404 nm | 12 nm | 71 % | 402 nm | [82] |
| $CsPbCl_3$ | 404 nm | 9 nm | 80 % | | [83] |
| $CsPbCl_3$ (Mg-doped) | 402 nm | | 75.8 % | 385 nm | [84] |
| $CsPbCl_3$ (Cd-doped) | 381 nm | | 60.5 % | | [40] |
| $CsPbCl_2Br$ | 431 nm | 12 nm | 87 % | | [85] |
| $CsPbCl_3$ | 406 nm | 10 nm | 70 % | | [85] |
| $CsPbCl_3$ (Sr-doped) | 410 nm* | 12 nm | 82.4 % | | [86] |
| $CsPb(Cl/Br)_3$ | 405 nm | 24 nm | | | [87] |
| $CsPbCl_3$ | 405 nm | 10.6 nm | 87 % | 402 nm | [88] |
| $CsPbCl_3$ | 408 nm | 11 nm | 42.5 % | 393 nm | [89] |
| $CsPbCl_3$ (Ce-doped) | 411 nm | | 65 % | | [90] |
| $CsPbCl_3$ | 409 nm | | | | [91] |
| $MAPbCl_3$ | 428 nm | | | 400 nm | [92] |
| $MAPbCl_3$ | 404 nm | 15 nm | 5 % | | [69] |
| $MAPbCl_3$ | 397 nm | 11 nm | 3.3 % | | [85] |
| $MAPbCl_3$ | 405 nm | 9 nm | | 400 nm | [93] |
| $FAPbCl_3$ | 400 nm | 13 nm | 1.2 % | | [85] |
| $FAPbCl_3$ | 407 nm | 16 nm | 2 % | | [69] |

*value estimated from figure.

*Halide Composition.* Bandgap-tunable cesium lead halide ($CsPbX_3$) colloidal nanocrystals (NCs) have demonstrated emission across the visible spectrum (410 to 700 nm), with violet emission achieved by employing chloride at the X-site to form $CsPbCl_3$ NCs. A summary of representative studies to date can be found in **Table 1**, which details the optical properties, including photoluminescence (PL) and absorption when reported, of lead halide perovskite NCs with emission in the V/UV range. Early studies on $CsPbX_3$ NCs across the visible spectrum reported narrow PL emission linewidths (10 to 40 nm) and high photoluminescence quantum yields (PLQYs) as high as 95%,[52,60–62] but performance of $CsPbCl_3$ NCs with wide bandgap emission in the V/UV range lagged behind that of red/green/blue $CsPbI_3$ or $CsPbBr_3$ emitters. Akkerman *et al.* reported that $CsPbBr_3$ NCs with a slight addition of chloride through an anion exchange reaction achieved a PLQY of 95% with emission around 495 nm, but further replacement of bromide anions with chloride yielded a PLQY as low as 1% with emission at 390 nm.[62] Since then, considerable efforts have been made to improve the PLQY of violet-emitting $CsPbCl_3$ NCs, and various techniques such as doping of metals (e.g., Ni, Cd) into the crystal lattice structure have been utilized to demonstrate near-unity PLQYs in $CsPbCl_3$ NCs.[39,70,76]

As shown in **Figure 4a**, Zhang *et al.* measured $CsPbCl_3$ NCs (undoped and synthesized using a $PhPOCl_2$ precursor) with a sharp excitonic absorption peak around 402 nm and narrow PL emission around 404 nm (FWHM of 12 nm).[82] This relatively small Stokes shift (~15 meV) suggests that emitted photons come from exciton recombination. The direct radiative emission of photons is further supported by other studies on $CsPbCl_3$ NCs that demonstrate that photoluminescence excitation (PLE) measurements align closely with the absorption spectra.[77] Note that changes in the synthesis or composition, such as through doping at the B-site discussed in the next section, can alter the optical properties of the $CsPbCl_3$ NCs, with reported photoluminescence emission peaks as low as 381 nm ($Cd^{2+}$-doped $CsPbCl_3$ NCs)[40] to as high as 430 nm ($Ce^{3+}$-doped $CsPbCl_3$ NCs)[68] in different studies.

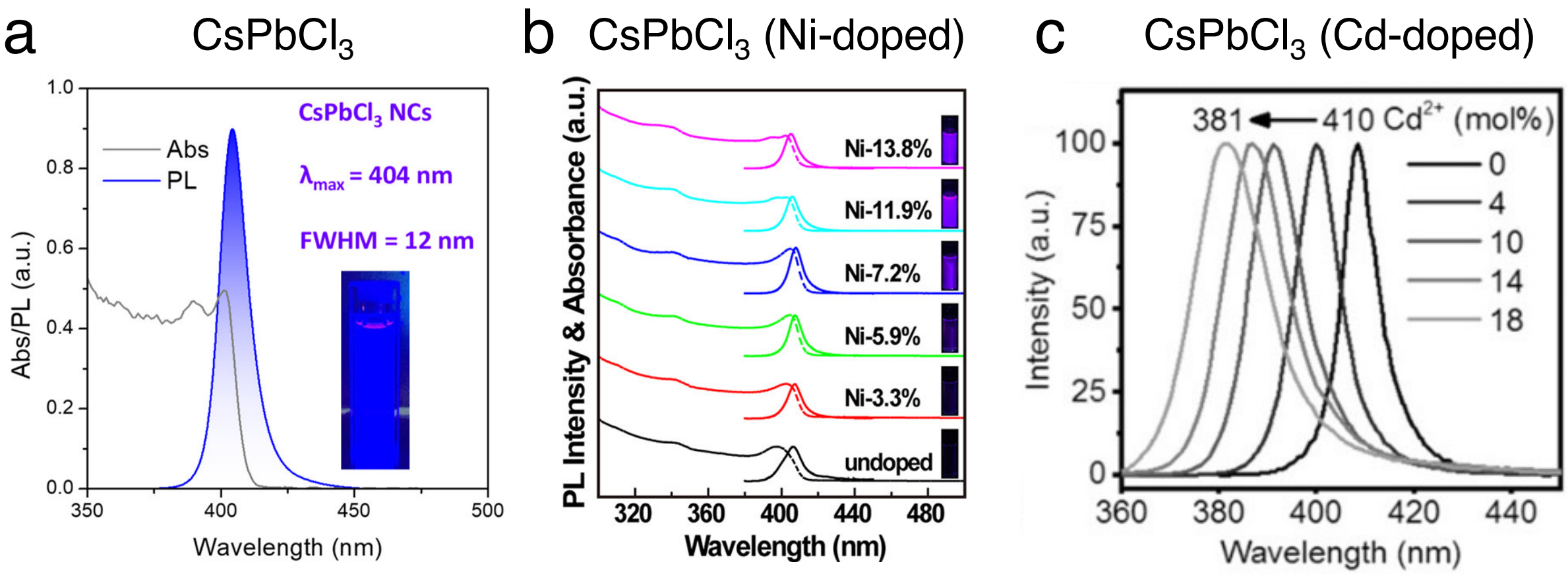

**Figure 4.** Optical properties of lead-based V/UV light-emitting MHPs. **(a)** Absorption and PL spectra of $CsPbCl_3$ (350 nm excitation wavelength). Inset shows a photograph of $CsPbCl_3$ NCs under UV illumination. Adapted with permission from ref. [82]. Copyright 2021 American Chemical Society. **(b)** Absorption and PL spectra of undoped and Ni-doped $CsPbCl_3$ NCs. Insets show the corresponding photographs of NC solutions under UV illumination. Adapted with permission from ref. [39]. Copyright 2018 American Chemical Society. **(c)** PL spectra of $CsPbCl_3$ and $CsPbCl_3$:$Cd^{2+}$ NCs with different doping content of $Cd^{2+}$. Adapted with permission from ref. [40]. Copyright 2021 John Wiley and Sons.

Similar to the effects observed for $CsPbX_3$, employing smaller halides also blue-shifts the PL emission in some of the other metal halides summarized in **Table 2**. In 2D phenethylammonium lead halide ($PEA_2Pb(Br/Cl)_4$) thin films, we showed that the PL emission can be blue-shifted from 407 nm ($PEA_2PbBr_4$) to 394 nm ($PEA_2PbCl_1Br_3$) to 384 nm ($PEA_2PbCl_2Br_2$) by increasing the amount of chloride relative to bromide.[94] Increasing the relative amount of chloride at the X-site causes a blue-shift in PL across a variety of other metal halide systems as well, such as

$MAPbX_3$,[69,85] $FAPbX_3$,[69,85] $MA_3Bi_2X_9$,[55,95,96] $FA_3Bi_2X_9$,[97] $Cs_3CeX_6$,[98,99] and $(DFPD)_4CeX_7$,[100] which are discussed in more detail in the following sections.

Unlike most of the materials discussed, replacing bromide with chloride can lead to a red-shift in PL in some lead-free material systems in **Table 2**. For example, copper halide systems such as $Rb_2CuCl_3$ and $K_2CuBr_3$ demonstrate shorter wavelength PL for the bromide compounds compared to chloride compounds. These copper halides have 1D crystal structures. For example, $K_2CuBr_3$ is made up of 1D chains made up of corner sharing $[CuBr_3]^{2-}$ tetrahedra separated by $K^+$ cations,[46] and the same type of structure is seen for the Rb-based copper halides as well.[44] Creason *et al.* reported $Rb_2CuBr_3$ with PL centered at 385 nm and $Rb_2CuCl_3$ with PL centered at 395 nm,[44] and then in a later study, Creason *et al.* reported $K_2CuBr_3$ with PL centered at 388 nm and $K_2CuCl_3$ with PL centered at 392 nm.[46] Similar studies from other groups on $Rb_2CuX_3$ and $K_2CuX_3$ systems show the same trends.[101–105] The demonstrated red-shift in emission when replacing bromide with chloride, contrary to the trends for most other MHP systems, can be attributed to a combination of changes in the bandgap associated with structural deformation and a strong excitonic effect arising from self-trapped exciton (STE) states in these materials.[45]

**Table 2.** Optical properties of lead-free or reduced-dimensional V/UV-emitting metal halide perovskite or perovskite derivative materials that do not follow the standard 3D $ABX_3$ structure. Entries are left blank when values are not reported in a reference.

| Perovskite | PL Peak | FWHM | PLQY | Abs. Peak | Crystal Structure | Ref. |
|---|---|---|---|---|---|---|
| $Cs_4CuIn_2Cl_{12}$ | 381 nm | 73 nm | 1.7 % | 273 nm | 2D | 106 |
| $Rb_2CuBr_3$ | 385 nm | 54 nm | | | 1D | 44 |
| $Rb_2CuCl_3$ | 395 nm | 52 nm | 100 % | | 1D | 44 |
| $Rb_2CuBr_3$ | 385 nm | | 98.6 % | 300 nm | 1D | 101 |
| $Rb_2CuCl_3$ | 397 nm | 50.2 nm | 99.4 % | 276 nm | 1D | 102 |
| $K_2CuBr_3$ | 391 nm | 61.4 nm | 86.98 % | | 1D | 103 |
| $K_2CuCl_3$ | 386 nm | 52 nm | 98.8 % | | 1D | 104 |
| $K_2CuBr_3$ | 388 nm | 54 nm | 55 % | | 1D | 46 |
| $K_2CuCl_3$ | 392 nm | 54 nm | 96.58 % | | 1D | 46 |
| $K_2CuBr_3$ | 383 nm | 49 nm | 79.2 % | | 1D | 105 |
| $Cs_3Bi_2Br_9$ | 410 nm | 48 nm | 19.4 % | 396 nm | 2D | 42 |
| $Cs_3Bi_2Cl_9$ | 393 nm | 59 nm | 26.4 % | | 1D | 42 |
| $FA_3Bi_2Br_9$ | 437 nm | 65 nm | 52 % | 404 nm | 2D | 97 |
| $FA_3Bi_2Cl_9$ | 399 nm | 63 nm | 39 % | | | 97 |
| $MA_3Bi_2Br_9$ | 423 nm | 62 nm | 12 % | 376 nm | 2D | 55 |
| $MA_3Bi_2Cl_9$ | 360 nm | 50 nm | 15 % | | 1D | 55 |
| $MA_3Bi_2Br_9$ | 422 nm | 62 nm | 13.5 % | 376 nm | 2D | 95 |
| $MA_3Bi_2(Cl_{0.33}Br_{0.67})_9$ | 422 nm | 41 nm | 54.1 % | 388 nm | | 95 |
| $MA_3Bi_2(Cl_{0.5}Br_{0.5})_9$ | 412 nm | | 19.1 % | 366 nm | | 95 |
| $MA_3Bi_2(Cl_{0.67}Br_{0.33})_9$ | 399 nm | | 22.4 % | 353 nm | | 95 |
| $MA_3Bi_2Cl_9$ | 370 nm | | 24.7 % | 333 nm | 1D | 95 |
| $MA_3Bi_2Br_9$ | 400 nm | | 50.1 % | 376 nm | 2D | 96 |
| $MA_3Bi_2Br_6Cl_3$ | 379 nm | | 35.4 % | 355 nm | | 96 |
| $Cs_3Sb_2Br_9$ | 410 nm | 41 nm | 46 % | | 2D | 41 |
| $Cs_3Sb_2Br_9$ | 409 nm | | 51.2 % | 368 nm | 2D | 107 |
| $Cs_3CeBr_6$ | 400 nm | | | | | 98 |
| $Cs_3CeBr_7$ | 410 nm | | | | | 98 |
| $Cs_3CeCl_6$ | 385 nm | | | | | 98 |
| $Cs_3CeCl_7$ | 375 nm | | | | | 98 |
| $Cs_3CeBr_6$ | 391 nm, 421 nm | | 91.2 % | | 0D | 43 |
| $Cs_3CeI_6$ | 430 nm, 470 nm | | 71.4 % | | 0D | 99 |
| $Cs_3CeBr_2I_4$ | 410 nm, 450 nm* | | | | 0D | 99 |
| $Cs_3CeI_6$ | 424 nm, 470 nm* | | 74.6 % | | 0D | 108 |
| $Cs_3CeCl_6$ | 380 nm, 409 nm | | 58.83 % | | 0D | 109 |
| $Cs_3CeBr_6$ | 392 nm, 421 nm | | | 357 nm | 0D | 110 |
| $Cs_3CeCl_6 \cdot 3H_2O$ | 334 nm, 356 nm | | 93.54 % | 305 nm | 1D | 111 |
| $Cs_3CeCl_6$ | 380 nm, 410 nm | | 91.82 % | 365 nm | 0D | 111 |
| $Cs_3CeCl_6 \cdot 3H_2O$ | 373 nm, 406 nm | | ~100 % | | 1D | 112 |
| $Rb_3CeI_6$ | 427 nm, 468 nm | | 51 % | | | 113 |
| $(DFPD)_4CeBr_7$ | 400 nm, 430 nm | 30-40 nm | 98 % | | 0D | 100 |
| $(DFPD)_4CeCl_7$ | 375 nm, 400 nm | 30-40 nm | ~100 % | | 0D | 100 |
| $(DFPD)CeCl_4 \cdot 2MeOH$ | 353 nm, 368 nm | 30-40 nm | 95 % | | 1D | 100 |
| $Cs_3MnBr_5$ (Ce-doped) | 387 nm, 419 nm | 22.5 nm, 30.5 nm | | | | 114 |
| $Cs_2ZnCl_4$ (Ce-doped) | 370 nm | | ~100 % | 225 nm | 0D | 115 |
| $Cs_2ZnBr_4$ (Ce-doped) | 342 nm, 367 nm | | 97 % | | 0D | 116 |
| $Cs_2NaCeCl_6$ | 370 nm, 410 nm | | 6 % | 340 nm | 3D | 117 |
| $Cs_2NaPr_{0.99}Ce_{0.01}Cl_6$ | 265 nm, 276 nm, 302 nm, 370 nm, 410 nm | | | | 3D | 117 |
| $Cs_2NaPrCl_6$ | 265 nm, 276 nm, 302 nm | | 14 % | 245 nm | 3D | 117 |
| $PEA_2PbBr_4$ | 405 nm* | | | | 2D | 118 |
| $PEA_2PbCl_4$ | 350 nm* | | | | 2D | 118 |
| $PEA_2PbBr_4$ | 407 nm | | 26 % | 404 nm | 2D | 54 |
| $PEA_2PbBr_4$ | 410 nm | | 11 % | 400 nm | 2D | 119 |
| $PEA_2PbBr_4$ | 408 nm | 11.6 nm | 15-25 % | | 2D | 51 |
| $PEA_2PbBr_4$ | 437 nm | | | 433 nm | 2D | 120 |
| $PEA_2PbBr_4$ | 406 nm | 8 nm | 8 % | 402 nm | 2D | 22 |
| $PEA_2PbBr_4$ | 407 nm | | | | 2D | 94 |
| $PEA_2PbCl_1Br_3$ | 394 nm | | | | 2D | 94 |
| $PEA_2PbCl_2Br_2$ | 384 nm | | | | 2D | 94 |
| $PEA_2PbCl_1Br_3$ | 393 nm | 14 nm | 1 % | 382 nm | 2D | 121 |
| $PEA_2PbCl_1Br_3$ | 401 nm | 16.82 nm | | 385 nm | 2D | 122 |
| $PEA_2EuCl_4$ | 401 nm | 29 nm | 7.73 % | 350 nm | 2D | 123 |
| $MAPbBr_3$ | 436 nm | < 25 nm | | | 2D | 124 |
| $BA_2PbBr_4$ | 406 nm | | 26 % | | 2D | 125 |
| $BA_2PbBr_4$ | 406 nm | 15.9 nm | | 400.5 nm | 2D | 126 |
| $EA_2CdBr_4$ | 388 nm | 32 nm | | | 2D | 127 |
| (5-MBI)$PbBr_3$ | 430 nm | | 34.7 % | 297 nm | 1D | 128 |
| [BAPrEDA]$PbCl_6 \cdot (H_2O)_2$ | 392 nm | 73 nm | 21.3 % | 300 nm | 0D | 129 |
| $MA_2CdBr_4$ | 415 nm | 94 nm | | | 0D | 130 |
| $CdCl_2$-4HP | 416 nm | | 63.55 % | | 2D | 131 |
| $Rb_3InCl_6$ (Cu-doped) | 398 nm | 54 nm | 95 % | | 0D | 132 |
| T-$PbBr_2$ | 423 nm | 20 nm | 16 % | 408 nm | 2D | 133 |
| B-$PbBr_2$ | 413 nm | 19 nm | 53 % | 398 nm | 2D | 133 |
| $Cs_2AgIn_{0.9}Bi_{0.1}Cl_6$ | 390 nm | | 36.6 % | 367 nm | 3D | 134 |
| $KMgF_3$ | 145 nm, 165 nm | | | | | 135 |
| $BaLiF_3$ | 160 nm, 180 nm, 225 nm* | | | | | 135 |

*value estimated from figure.

*B-Site Engineering.* Lead is commonly employed at the B-site cation for metal halide perovskite optoelectronic devices,[2,9–13] but there is increasing interest to partially or fully replace the lead in a metal halide perovskite due to lead toxicity concerns and limited performance of lead systems in the V/UV range. While lead systems demonstrate high performance in the red/green/blue emission ranges,[16–19] moving to wider bandgap emission requires further defect passivation, dimensional control/confinement, or higher energy electronic state transitions that can be accessed by utilizing other metals.

One successful strategy to increase the PLQY within the metal halide perovskite semiconductor class, broadly, has been metal doping at the B-site using metals such as $Ni^{2+}$, $Cd^{2+}$, $Cu^{+}/Cu^{2+}$, $Zn^{2+}$, $Mg^{2+}$, $Sr^{2+}$, or $Mn^{2+}$.[39,40,70,75,78,79,81,86,136,137] **Table 1** shows a variety of different studies in $CsPbCl_3$ NCs employing different doping strategies to achieve high PLQY V/UV emission. As shown in **Figure 4b**, Yong *et al.* utilized $Ni^{2+}$ doping via a modified hot-injection method to replace $Pb^{2+}$ with $Ni^{2+}$ at the B-site, which improved the short-range order of the lattice and eliminated halide vacancy defects, improving the PLQY from 2.4% to 96.5%.[39] Mondal *et al.* demonstrated a room temperature post-synthetic treatment of weakly emitting $CsPbCl_3$ NCs with $CdCl_2$, which improved the PLQY from 3% to 96% (and to 98% when starting with initial samples with a higher PLQY of 32%).[70] The post-synthetic treatment led to facile exchange of $Pb^{2+}$ by $Cd^{2+}$ at the B-site without changing the size and structure of the lattice. The authors showed that this post-synthetic treatment suppressed nonradiative carrier trapping centers and introduced shallow energy levels facilitating radiative recombination, leading to the enhanced PLQY without affecting the PL peak position (406 nm) and spectral width. In a different study where $Cd^{2+}$ was incorporated into the NCs during synthesis as opposed to a post-synthetic treatment, Zhang *et al.* synthesized Cd-doped $CsPbCl_3$ NCs that demonstrated a blue-shift in emission with increasing $Cd^{2+}$ content as shown in **Figure 4c**.[40] The authors state that this blue-shift in emission is likely due to an increase in bandgap caused by lattice contraction from the smaller radius of $Cd^{2+}$ compared to $Pb^{2+}$.

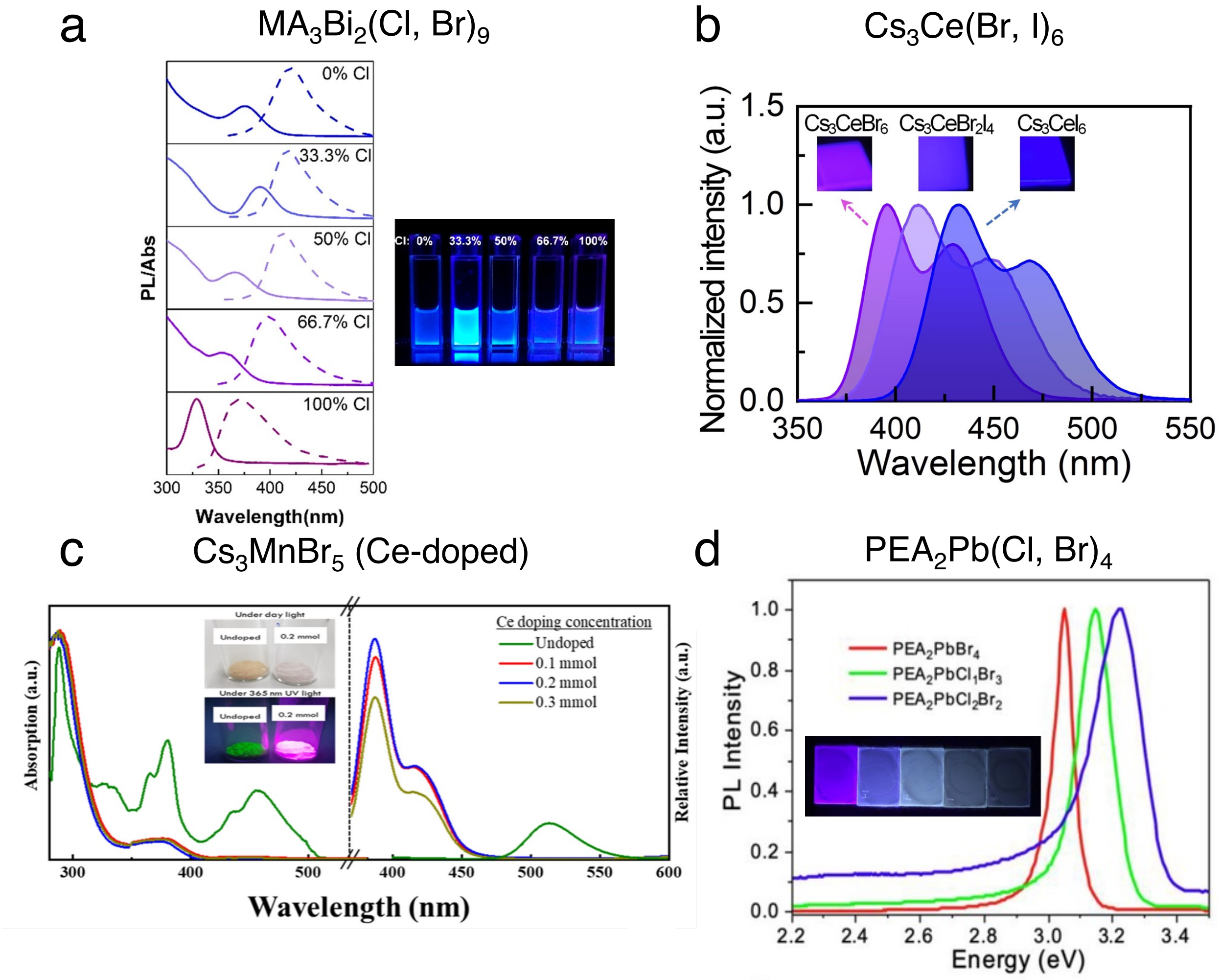

**Figure 5.** Optical properties of V/UV light-emitting bismuth halide perovskites, cerium halide perovskites, manganese halide perovskites, and 2D lead halide perovskites. **(a)** Absorption and PL spectra (left) of $MA_3Bi_2(Cl, Br)_9$ perovskite QDs with different levels of chloride content. Photograph (right) of corresponding QD solutions under UV illumination. Adapted with permission from ref. [95]. Copyright 2018 American Chemical Society. **(b)** PL spectra of $Cs_3CeBr_6$, $Cs_3CeBr_2I_4$, and $Cs_3CeI_6$. Insets show the corresponding photographs of cerium halide perovskite films under UV illumination. Adapted with permission from ref. [99]. Copyright 2022 The Authors. **(c)** Absorption (left) and PL spectra (right) of undoped and Ce-doped $Cs_3MnBr_5$ samples. Insets show the corresponding photograph of pristine and 0.2 mmol Ce-doped $Cs_3MnBr_5$ samples under daylight and under UV illumination. Adapted with permission from ref. [114]. Copyright 2023 Springer Nature. **(d)** PL spectra of $PEA_2PbBr_4$, $PEA_2PbCl_1Br_3$ and $PEA_2PbCl_2Br_2$. Inset shows the corresponding photographs of $PEA_2PbX_4$ (X = Br or Cl) thin films under UV illumination, from left to right: $PEA_2PbBr_4$, $PEA_2PbCl_1Br_3$, $PEA_2PbCl_2Br_2$, $PEA_2PbCl_3Br_1$, and $PEA_2PbCl_4$. Adapted with permission from ref. [94]. Copyright 2023 SPIE, the international society for optics and photonics.

Along with partially replacing $Pb^{2+}$ through doping the B-site in $CsPbCl_3$ NCs, making entirely lead-free systems with other metals such as copper ($Cu^{+}$, $Cu^{2+}$), bismuth ($Bi^{3+}$), or antimony ($Sb^{3+}$) has proven successful for pushing emission deeper into the UV range as shown in **Table 2**. Liu *et al.* reported lead-free $Cs_4CuIn_2Cl_{12}$ double perovskite NCs with 381 nm emission and a PLQY of 1.7%.[106] Copper has also been used in the $K_2CuX_3$ and $Rb_2CuX_3$ systems discussed in the previous section.[44,46] $Bi^{3+}$ is an appealing candidate to replace $Pb^{2+}$ because it is less toxic, isoelectronic to $Pb^{2+}$, and more stable than other commonly explored alternatives like $Sn^{2+}$. Leng *et al.* reported one of the first $MA_3Bi_2X_9$ perovskite QDs, achieving 360 nm emission with a 15% PLQY for $MA_3Bi_2Cl_9$ and 423 nm emission with a 12% PLQY for $MA_3Bi_2Br_9$.[55] In another study, Leng *et al.* further improved upon these QDs by utilizing a Cl-passivation process that greatly improved the PLQY, achieving emission centered at 370 nm with a 24.7% PLQY for $MA_3Bi_2Cl_9$ and centered at 422 nm with a 54.1% PLQY for Cl-passivated $MA_3Bi_2Br_9$ (33% Cl).[95] Note that even after the addition of 33% Cl content to the $MA_3Bi_2Br_9$ QD, the PL peak position remain unchanged, although the absorption peak red-shifted from 376 to 388 nm as shown in **Figure 5a**. The authors attribute the unchanged PL peak to reduced surface defects after Cl introduction, combined with the effect of size increase. However, after a further increase of Cl content past 33%, the authors observed the expected trends of a blue-shift of the PL (and the absorption peak). Wu *et al.* produced $MA_3Bi_2Br_9$ QDs with an even lower PL emission wavelength of 400 nm via a cryo-bonding ligand-assisted reprecipitation (Cb-LARP) method, achieving a PLQY of 50.1%.[96] The authors then demonstrated tunability of the halide to further push emission into the UV range with $MA_3Bi_2Br_6Cl_3$ QDs emitting at 379nm with a PLQY of 35.4%. As discussed further in the next section on A-site engineering, the A-site cation can also be altered to create $FA_3Bi_2X_9$ or $Cs_3Bi_2X_9$ perovskite QDs.[42,97] Similar to $Bi^{3+}$, antimony ($Sb^{3+}$) forms a $Cs_3Sb_2Br_9$ perovskite structure and is another alternative to $Pb^{2+}$ at the B-site. Zhang *et al.* demonstrated violet-emitting $Cs_3Sb_2Br_9$ perovskite QDs with PL centered at 410 nm and a PLQY of 46%,[41] and Ma *et al.* similarly demonstrated PL centered at 409 nm and a PLQY of 51.2%.[107] While $Cs_3Sb_2Br_9$ QDs have not been heavily studied, the study by Ma *et al.* is one of the few lead-free demonstrations of violet EL (408 nm) in a PeLED, which will be discussed further in the section titled, "*State-of-the-Art Violet and Ultraviolet PeLED Performance*".[107]

The other lead-free material system that has been used to demonstrate V/UV EL in PeLEDs incorporates the rare-earth element cerium ($Ce^{3+}$), which has bright violet emission from the Ce-5d to Ce-4f transition.[98] Wang *et al.* produced thermally-evaporated $Cs_3CeBr_6$ thin films in which $[CeBr_6]^{3-}$ octahedra are isolated from each other by $Cs^+$ cations in a 0D crystal structure.[43] Due to this 0D structure and the localized f-orbitals in $Ce^{3+}$, there is small band dispersion, which increases the exciton binding energy and enables high PLQYs. The films exhibited doublet PL emission with peaks at 391 nm and 421 nm and a PLQY of 91.2% and were then used in the first demonstration of V/UV EL from a $Cs_3CeBr_6$-based PeLED. Similar to lead halide systems, the $Cs_3CeX_6$ material system has demonstrated shifts in emission with halide tunability. **Figure 5b** shows the PL spectra of $Cs_3CeBr_6$ with doublet emission at 391 nm and 421 nm, $Cs_3CeBr_2I_4$ with doublet emission around 410 nm and 450 nm, and $Cs_3CeI_6$ with doublet emission around 430 nm and 470 nm.[99] The emission blue-shifts with increasing bromide content, and studies on cerium chloride systems have demonstrated emission even further into the UV range.[100,109,111,112]

Cerium-based perovskites synthesized using solution processing also enable V/UV emission. Aiming to create a simple solution-based method that is cost-effective and environmentally friendly, Dutta *et al.* used a water-based synthesis to produce spin-coated $Cs_3CeBr_6$ thin films, reporting doublet PL emission at 392 nm and 421 nm and also showing EL with a corresponding PeLED.[110] Wang *et al.* developed an anti-solvent vapor-assisted crystallization approach for preparing $(DFPD)_4CeCl_7$ and $(DFPD)CeCl_4{\cdot}2MeOH$ single crystals.[100] The $(DFPD)_4CeCl_7$ exhibited doublet PL emission peaks at 375 nm and 400 nm with a PLQY approaching 100% and $(DFPD)CeCl_4{\cdot}2MeOH$ exhibited doublet PL emission peaks at 353 nm and 368 nm with a PLQY of about 95%. They also reported $(DFPD)_4CeBr_7$ with PL emission at 400 nm and 430 nm and a PLQY of 98%, demonstrating the tunability of emission through the halide site.

Cerium has also been used as a dopant to achieve violet emission, both in the $CsPbCl_3$ NCs discussed previously[68,90] and in other lead-free material systems. For example, Dutta *et al.* reported a water-assisted synthesis of $Ce^{3+}$-doped $Cs_3MnBr_5$ powder that exhibited doublet PL at 387 nm and 419 nm.[114] As shown in **Figure 5c**, the undoped $Cs_3MnBr_5$ has no emission in the V/UV range, but when the $Mn^{2+}$ at the B-site is doped with $Ce^{3+}$, the characteristic doublet emission from $Ce^{3+}$ appears. Cerium doping was also utilized in studies by Wen *et al.* and Zhu *et al.* that incorporated $Ce^{3+}$ into $Cs_2ZnCl_4$ and $Cs_2ZnBr_4$ hosts, respectively, to achieve UV emission centered around 370 nm with near-unity PLQY.[115,116]

Along with cerium, the electronic transitions in the rare-earth element praseodymium ($Pr^{3+}$) are conducive to wide bandgap emission, enabling even shorter wavelength ultraviolet emission. In a study on lanthanide double perovskite NCs such as $Cs_2NaPrCl_6$ and $Cs_2NaCeCl_6$, Saghy *et al.* showed emission in the UV-C range from $Cs_2NaPrCl_6$ NCs with peaks at 265 nm, 276 nm, and 302 nm, much lower wavelength emission as compared to $Cs_2NaCeCl_6$ NCs with peaks at 370 nm and 410 nm.[117] These emission peaks in the $Cs_2NaPrCl_6$ NCs correspond to different electronic transitions in praseodymium, and because emission in the UV-C range is rare for perovskite emitters, this highlights the potential for future studies on praseodymium-based metal halide emitters.

*A-Site Engineering.* While $CsPbCl_3$ NCs have been the most widely researched lead halide perovskites for blue-violet emission, the A-site $Cs^+$ can be replaced by another monovalent cation such as $FA^+$ or $MA^+$, changing the properties of the material. Recall that increasing the size of the A-site cation ($Cs^+ < MA^+ < FA^+$) typically leads to smaller bandgaps.[64] Imran *et al.* demonstrated results that follow this trend for both iodide and bromide systems: $CsPbI_3$, $MAPbI_3$, and $FAPbI_3$ NCs exhibited emission peaks at 691 nm, 730 nm, and 764 nm, and $CsPbBr_3$, $MAPbBr_3$, and $FAPbBr_3$ NCs exhibited emission peaks at 512 nm, 527 nm, and 531 nm.[69] Das and Samanta also report similar trends for $APbI_3$ and $APbBr_3$ NCs.[85] Interestingly, the trend breaks for $APbCl_3$ NCs in both studies (see **Table 1**), as they show $CsPbCl_3$, $MAPbCl_3$, and $FAPbCl_3$ NCs with emission peaks at 408 nm, 404 nm, and 407 nm, respectively, (Imran *et al.*)[69] and 406 nm, 397 nm, and 400 nm, respectively (Das and Samanta).[85] The reasons for this difference are not explicitly explored in these studies, but the results demonstrate that simply changing cation size does not always induce the expected bandgap trends, as there are a variety of competing factors influencing the structural and optical properties of a material. It should also be noted that the PLQY varies drastically between different cation choices, particularly for the lead chloride NCs. For example, Imran *et al.* reported a PLQY of 65% for $CsPbCl_3$ NCs but PLQYs of only 5% and 2% for $MAPbCl_3$ and $FAPbCl_3$, respectively,[69] and similar numbers are reported by Das and Samanta (70% versus 3.3% and 1.2%).[85]

Different A-site cations have been utilized in other lead-free, violet-emitting MHP systems as well as shown in **Table 2**. For the copper halide systems discussed in the previous sections, rubidium ($Rb^+$) and potassium ($K^+$) were used in place of cesium ($Cs^+$) to achieve violet emission. Cs-based copper halides like $Cs_3Cu_2X_5$ do not emit in the violet range, with the lowest achievable wavelength of emission centered around 445 nm for $Cs_3Cu_2I_5$ (recall that larger halides lead to lower emission wavelengths, unlike most other MHPs).[45] But when $Cs^+$ is replaced by $Rb^+$ or $K^+$, the materials with the smaller A-site cations ($K^+ < Rb^+ < Cs^+$) form different perovskite structures that emit in the violet range, with emission wavelengths reported as short as 383 nm for $K_2CuBr_3$.[105]

For Bi-based QDs, Leng *et al.* reported all-inorganic $Cs_3Bi_2Cl_9$ and $Cs_3Bi_2Br_9$ QDs emitting at 393 nm and 410 nm, respectively.[42] In a separate study, Leng *et al.* also demonstrated $MA_3Bi_2Cl_9$ and $MA_3Bi_2Br_9$ QDs emitting at 360 nm and 423 nm, respectively.[55] Shen *et al.* showed $FA_3Bi_2Cl_9$ and $FA_3Bi_2Br_9$ QDs emitting at 399 nm and 437 nm, respectively.[97] Unlike the clear trend between halide size ($Cl^- < Br^- < I^-$) and emission peak, there is no clear trend between A-site cation size ($Cs^+ < MA^+ < FA^+$) and emission peak, suggesting that additional factors influence the bandgap in these Bi-based QD systems. Similar to the lead halide systems, tuning the A-site, B-site, or X-site can all change the bandgap, but the trends may differ depending on the complementary atoms at the other sites and the dimensionality of the material system.

*Dimensional Control.* Utilizing quantum confinement through reduced-dimensional geometric structures is another strategy to achieve wide bandgap violet emission. Utilizing a bulky organic cation such as $PEA^+$ at the A-site creates a 2D layered Ruddlesden-Popper perovskite structure (e.g., $PEA_2PbX_4$). Within our dimensionality framework, these materials have a 2D crystal structure because they consist of perovskite octahedra layers sandwiched between PEA layers. They are also 2D in their geometric structure because they form nanoplatelets, but the morphology can change slightly depending on the how the material is synthesized.[22,122] The 2D crystal structure forms repeating quantum wells, tightly confining excitons with high binding energies and leading to narrow, violet PL centered at 407 nm that was used to create one of the first violet PeLEDs (EL centered at 410 nm) by Liang *et al.* in 2016.[54] Multiple other studies on PEA-based lead halide systems listed in **Table 2** have demonstrated the promise of using reduced-dimensional systems to push emission into the UV range,[22,51,94,118–123] with PL demonstrated as short as 384 nm for $PEA_2PbCl_2Br_2$ as shown in **Figure 5d**.[94] Lead-free systems are also attractive for achieving deeper UV emission, and we demonstrated 401 nm emission for 2D $PEA_2EuCl_4$ nanoplatelets.[123] Furthermore, other A-sites and B-sites have been used in studies on 2D MHPs and other reduced-dimensional structures with emission into the UV range such as $MAPbBr_3$, $BA_2PbBr_4$, $EA_2CdBr_4$, and more shown towards the end of **Table 2**.

Beyond nanoplatelets, 0D geometric structures like QDs with sizes on the order of the excitonic Bohr radius are common for V/UV emission because the additional quantization and confinement allows for wider bandgaps and higher energy emission. For example, Leng *et al.* showed that a $MA_3Bi_2Br_9$ single crystal (3D geometric structure) exhibited broad PL emission with a peak at 550 nm and FWHM of around 100 nm. Then, when synthesizing $MA_3Bi_2Br_9$ QDs (0D), they observed that the PL emission peak shifted to 423 nm with a FWHM of 62 nm, demonstrating a significant blue-shift in emission and a sharper peak compared to the bulk 3D structure.[55]

Overall, the full list of materials in **Table 1** and **Table 2** demonstrate the current state-of-the-art in V/UV-emitting perovskite and perovskite derivative materials. Starting with the common $CsPbX_3$ material system, violet emission can be achieved by employing chloride at the X-site to produce $CsPbCl_3$ NCs. Then, compositional engineering can either alter the $CsPbCl_3$ structure through doping or by fully replacing the A-, B-, or X-site atoms to form new crystal structures, which are typically lower-dimensional. Importantly, to move into the UV range, we can utilize a wide range of lead-free systems employing B-site metals such as $Cu^{+}$, $Bi^{3+}$, $Sb^{3+}$, $Ce^{3+}$, $Mn^{2+}$, $Zn^{2+}$, $Pr^{3+}$, $Eu^{2+}$, $Cd^{2+}$, and $In^{3+}$. Lastly, dimensional control of the geometric structure through synthetic and processing parameters can form structures like nanoplatelets or QDs, in which quantum confinement effects can enable higher energy emission. Additional studies on these materials and exploratory studies on new materials can improve the current state-of-the-art and push emission further into the UV range.

## Device Architecture and Engineering

While the optical properties of MHPs are highly tunable, achieving efficient device operation relies heavily on the ability to inject, transport, and confine charges effectively. Thus, PeLEDs typically employ a structure in which the perovskite emissive layer is sandwiched between two wide-bandgap transport layers: a hole transport layer (HTL) and an electron transport layer (ETL).[121,138–140] These transport layers facilitate charge injection and confine carriers within the emissive region.[141,142] They are broadly categorized as organic or inorganic, with inorganic materials generally preferred due to their higher chemical and thermal stability within V/UV devices.[143–145] This is particularly critical for V/UV PeLEDs, which operate at comparatively higher voltages, because resulting joule heating accelerates structural degradation and induces EQE roll-off, making thermally stable transport layers essential for mitigating device deterioration.[146,147] Common inorganic and organic transport layers used in PeLEDs, including their respective energy levels, are shown in **Figure 6a**. For efficient PeLED operation, the transport layer energy levels must align with the energy levels of adjacent layers to both inject charge carriers into the emissive perovskite layer while also confining them within the perovskite to induce radiative recombination. Specifically, the ETL should possess a deep valence band to prevent hole leakage and nonradiative recombination, while the HTL should have a shallow conduction band to block electron backflow from the ETL.[91] Consequently, the operating voltage and efficiency of PeLEDs depend on the energy-level offsets, carrier mobilities, layer thicknesses, and interface quality of each layer.[19,91,148–150] the x-axis of the bottom left inset was changed to "Wavelength (nm)" from "Voltage (V)". Adapted with permission from ref. [43]. Copyright 2021 American Chemical Society. Energy band diagrams for $PEA_2PbCl_1Br_3$ UV PeLEDs using **(e)** a single ETL (i.e., TPBi device) device structure or **(f)** a dual ETL (i.e., 1,3,5-tri(*m*-pyridin-3-ylphenyl)benzene (TmPyPB)/TPBi device) device structure. **(g)** EL spectra of the TPBi and TmPyPB/TPBi devices show pure 399 nm UV emission from the dual ETL device structure. Adapted with permission from ref. [121]. Copyright 2024 The Authors.

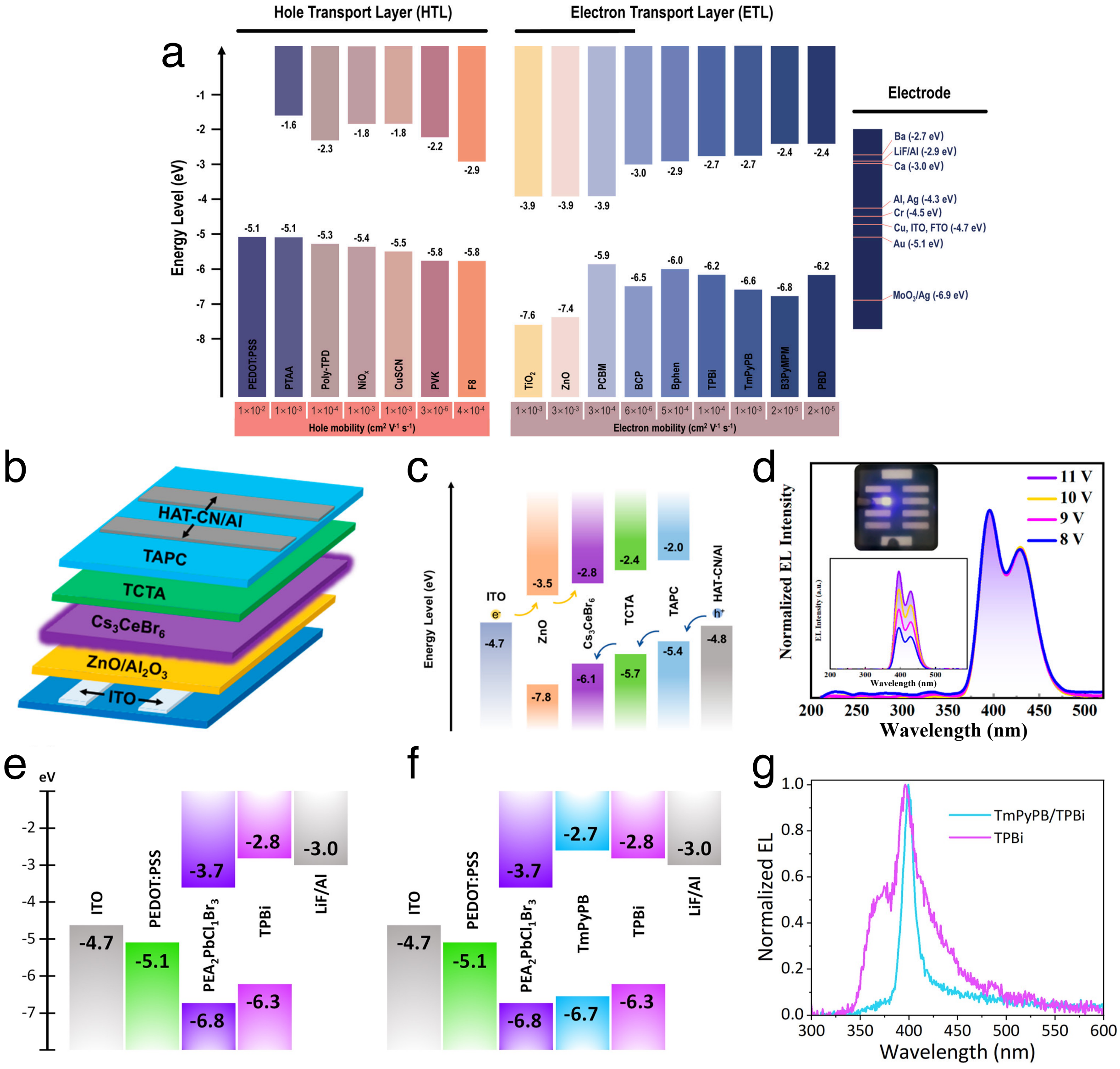

**Figure 6.** Direct charge injection-based device architectures for V/UV PeLEDs. **(a)** Summary of energy level and electron/hole mobilities of widely-used ETLs and HTLs for PeLEDs. Adapted with permission from ref. [151]. Copyright 2021 IOP Publishing Ltd. **(b)** Device structure schematic and **(c)** corresponding energy band diagram of $Cs_3CeBr_6$ V/UV PeLEDs. **(d)** Normalized EL spectra of $Cs_3CeBr_6$ PeLEDs at different applied voltages. Note: the x-axis was changed to "Wavelength (nm)" from "Voltage (V)". Insets show the (top left) photograph of a working $Cs_3CeBr_6$ PeLED and (bottom left) EL spectra of the PeLED operated at different voltages. Note:

However, one persistent challenge that hampers the performance of V/UV PeLEDs is the inefficient injection and transport of charge carriers from the electrodes into the perovskite emissive layer. This issue commonly arises due to (i) energy-level misalignment between the transport and emissive layers, and (ii) imbalance in the rate of electron and hole injection. Improper band alignment at either interface can cause carrier accumulation and exciton quenching, significantly lowering EQE.[122] Moreover, when one carrier type is injected more efficiently than the other, charge imbalance occurs, leading to accumulation of excess carriers within the transport layers. This imbalance also promotes nonradiative pathways such as interfacial exciton quenching and Auger recombination, thereby suppressing EQE.[152]

A practical route to mitigate charge imbalance in PeLEDs is to adjust the thickness of transport layers to achieve balanced injection rates at the perovskite interfaces. Liang *et al.* demonstrated this in a violet PeLED by optimizing the thickness of their ETL (i.e., 2,2',2"-(1,3,5-benzinetriyl)-tris(1-phenyl-1H-benzimidazole); TPBi) and consequently observed a more symmetric carrier flux and improved device EQE.[54] However, for optimal device performance, the ETL/HTL thickness must strike a balance – thin enough to facilitate transport, yet thick enough to effectively block opposite carriers.[91] Another way to ensure balance of charge injections is to tune the mobility of carriers in these devices. Wang *et al.* introduced carbon dots (CDs) as additives within the HTL of their device to reduce the carrier mobility. Owing to their strong affinity for perovskites, the CDs improved film morphology while reducing hole injection, which then balanced the injection rate of carriers in both transport layers, as evident by the hole- and electron-only current-density plots, achieving a maximum EQE of 21.07% for a sky-blue PeLED.[153] Additionally, Ma *et al.* adopted a complementary strategy by incorporating poly(ethylenimine) (PEI) into the ETL of their violet PeLED to lower the cathode work function, facilitating improved electron injection and observed a significant increase in EQE, up to 0.206%.[107] Meanwhile, Ni *et al.* showed that optimizing the HTL (i.e., Poly[N,N' -bis(4-butylphenyl)-N,N' -bis(phenyl)-benzidine]; Poly-TPD) concentration in their device altered the film morphology and surface roughness of the perovskite layer, with the smoothest film (RMS ≈ 5.7 nm) exhibiting the highest EQE of 2.41% at an emission wavelength of 401 nm.[122] Interfacial modification also proves to be effective in improving the balance of carriers in PeLEDs. As seen in **Figures 6b and 6c**, Wang *et al.* introduced a 1 nm $Al_2O_3$ interlayer between the electron transport layer and the perovskite emissive layer which helped decrease electron injection, while the use of Dipyrazino[2,3-f:2',3'-h]quinoxaline-2,3,6,7,10,11-hexacarbonitrile (i.e., HAT-CN; as an electron-blocking/hole-injecting layer) and 4,4',4-Tris(carbazol-9-yl)triphenylamine (i.e., TCTA; as a hole-transport interlayer) further improved hole injection and confinement, thereby suppressing carrier quenching and yielding an improved V/UV PeLED with a doublet emission at 391 nm/421 nm (**Figure 6d)**, EQE of 3.5%, and max luminance of 470 cd $m^{-2}$.[43]

Carrier mobility within the perovskite emissive layer also plays a pivotal role in determining efficiency. Dutta *et al.* found that in $Cs_3CeBr_6$ V/UV PeLEDs, mismatched carrier mobilities caused exciton recombination at the transport layer/perovskite interface. The authors resolved this by inserting a thin Poly(9-vinylcarbazole) (i.e., PVK) layer acting as both an electron-blocking and hole-injection layer, coupled with 2,7-Bis(diphenylphosphoryl)-9,9'-spirobifluorene (i.e., SPPO13) as the electron-injection layer and a 5 nm LiF interlayer to minimize quenching at the cathode side.[110] In our previous work, we demonstrated that adding an additional ETL (**Figures 6e and 6f)** improved charge injection by fine-tuning the energy-level alignment between the perovskite and electrode, leading to a singular electroluminescent peak at 399 nm **(Figure 6g**) and maximum EQE of 0.16% at 2.9 mA $cm^{-2}$.[121]

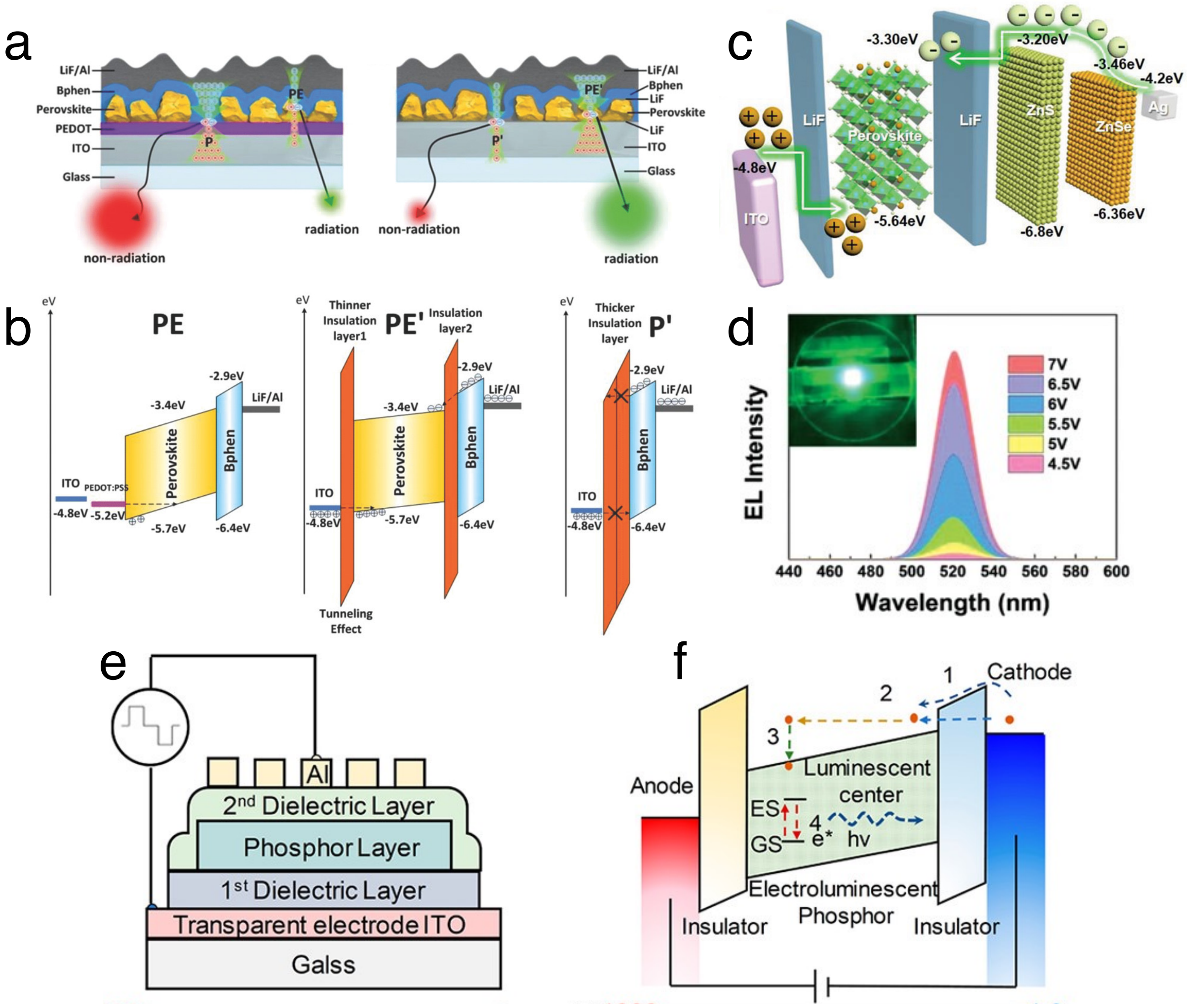

**Figure 7.** Alternative insulator-based LED device architectures. **(a)** The working mechanisms of (left) device A (ITO/PEDOT:PSS/perovskite/Bphen/LiF/Al) and (right) device C (ITO/LiF/perovskite/LiF/Bphen/LiF/Al; insulator-perovskite-insulator (IPI) device structure). **(b)** Energy band diagram analyses for device A (left) and device C (middle and right). (left) PE region (current density through the crystal grains) in device A has a thicker energy barrier for hole injection. (middle) PE′ region (current density through the crystal grains) in device C has a thinner energy barrier for hole injection, where (right) P′ region (current density through the gaps) in device C has a thick LiF layer that prevents tunneling of charge carriers and effectively prevents leakage current. Adapted with permission from ref. [139]. Copyright 2018 John Wiley and Sons. **(c)**

Device structure schematic of a fully inorganic IPI PeLED. **(d)** EL spectra of the fully inorganic IPI PeLED operated at different voltages. Inset shows the photograph of a working fully inorganic IPI PeLED. Adapted with permission from ref. [154]. Copyright 2020 John Wiley and Sons. **(e)** Device structure schematic for thin-film electroluminescence (TFEL) devices. **(f)** EL mechanism of TFEL devices: 1. tunnel injection, 2. carrier acceleration, 3. luminescence center excitation, 4. radiative recombination and light emission. Adapted with permission from ref. [155]. Copyright 2022 American Chemical Society.

Another effective approach to enhance carrier injection and suppress interfacial recombination is embedding insulating interlayers at the perovskite interfaces.[139] Shi *et al.* demonstrated this strategy by incorporating LiF on both sides of their perovskite layer and analyzed this structure with two other structures**,** Device A (ITO/Poly(3,4-ethylenedioxythiophene)-poly(styrenesulfonate) (i.e., PEDOT:PSS)/perovskite/4,7-Diphenyl-1,10-phenanthroline (i.e., Bphen)/LiF/Al), Device B (ITO/PEDOT:PSS/PVK/perovskite/Bphen/LiF/Al), and Device C (ITO/LiF/perovskite/LiF/Bphen/LiF/Al), all emitting in the green spectral regime. Device C exhibited markedly higher EQEs—5.53%, 2.36%, and 2.99% for $FAPbBr_3$, $MAPbBr_3$, and $CsPbBr_3$, respectively—compared to Devices A and B, which showed significantly lower values. This improvement stems from reduced leakage currents, as evidenced by distinct percolation pathways (P, PE, P′, and PE′) in Devices A and C (**Figure 7a**). In Device A, hole injection was hindered due to a large interfacial barrier, whereas in Device C, tunneling through the insulating layers enabled improved injection while simultaneously blocking opposite-carrier leakage (**Figure 7b**). Time-resolved PL further confirmed that these insulating layers mitigate interfacial quenching,[139] however, the thicknesses of these insulating layers should be such that the carriers tunnel through but do not leak out of the perovskite layer into the transport layers. Luo *et al.* showed this by analyzing different thicknesses of LiF as the insulating layers on both sides of a deep-blue $CsEuBr_3$ PeLED. They observed that the tunneling current was highly dependent on LiF thickness. Optimal performance was achieved using a 10 nm LiF layer on the hole-injection side and a 15 nm layer on the electron-injection side, resulting in a maximum EQE of approximately 6.5%.[156] In addition to enhancing charge balance, the Insulator-Perovskite-Insulator (IPI) structure can enhance device operational stability by limiting the migration of ions from the perovskite layers into the transport layers. As shown in **Figure 7c,** the perovskite layer was sandwiched between thin LiF layers in a fully inorganic device, resulting in green EL independent of voltage bias (**Figure 7d**), consequently achieving a maximum brightness of 156,155 cd $m^{-2}$, an EQE of 11.05%, and half-lifetime of about 255 hours at an initial luminance of 120 cd $m^{-2}$.[154] Thus, given the difficulty of finding suitable ETL/HTL pairs that can yield efficient V/UV PeLEDs, IPI holds the potential to circumvent this obstacle by taking advantage of tunneling instead of charge injection.

Incorporating thin-film electroluminescent (TFEL) phosphor centers between insulating layers (**Figure 7e**) is another device-engineering strategy that could help fabricate V/UV PeLEDs, as TFEL phosphors are mainly synthesized by integrating lanthanide ions into inorganic host materials, such as perovskites and other inorganic lanthanide compounds (ILC), which circumvents the solubility problems of lanthanide-based MHPs in commonly used solvents. This structure has been studied extensively within the LED community.[157,158] **Figure 7f** demonstrates that, when a sufficiently high electric field is applied, electrons tunnel through into the emissive layer and are subsequently accelerated, leading to impact excitation of lanthanide ions within the phosphor centers. These excited lanthanide ions then relax radiatively, resulting in light emission.[155,159] Van Haecke *et al.* demonstrated this mechanism using a red-emitting $Ca_{0.5}Sr_{0.5}S$:Eu TFEL phosphor, reporting a luminance of 80 cd $m^{-2}$ at 40 V above the threshold voltage with an optimal Eu doping concentration of 0.1 mol%. They observed that at higher $Eu^{2+}$ concentrations, Eu–Eu interactions became stronger, increasing the density of shallow trap states and quenching both the PL and EL of their devices.[160] Although TFEL devices can achieve reasonably high luminance, their high turn-on voltages, large charge-injection barriers, and comparatively low efficiencies have largely limited their use in modern optoelectronic applications.

# State-of-the-Art Violet and Ultraviolet PeLED Performance

To overview the status of V/UV PeLEDs, we categorize the previous reports as reduced-dimensional (RD) MHPs, 3D MHPs, or rare-earth metal halide perovskites. Focusing first on RD MHPs, we examine the reports related to layered perovskites (e.g., reports where alternating organic (e.g., bulky ligand) and inorganic layers (e.g., metal halide octahedra) are present). By synthesizing phenethylammonium lead bromide ($PEA_2PbBr_4$) nanoplatelets coupled with a solvent vapor annealing process, Liang *et al.* demonstrated 410 nm violet PeLEDs with an EQE of 0.04%.[54] Similarly, we altered the solvent annealing procedure with water additives and incorporated dual electron transport layers to fabricate violet 408 nm $PEA_2PbBr_4$ and UV 399 nm $PEA_2PbCl_1Br_3$ (**Figures 8a and 8b)** PeLEDs with champion EQEs of 0.41% and 0.16%, respectively.[22,121] By varying the concentration of Poly-TPD hole transport layers, Ni *et al.* demonstrated tunable electroluminescence between 394 and 406 nm via $PEA_2PbBr_3Cl$ PeLEDs and achieved EQEs as high as 2.41%.[122] Using an electric-field-deposition technique, Deng *et al.* showcased 408.8 nm $PEA_2PbBr_4$ PeLEDs with an EQE of 0.31%.[51] Moving away from PEA-based PeLEDs, Sun demonstrated centimeter scale 2D $BA_2PbBr_4$ (BA = butylammonium) PeLEDs with an electroluminescence peak of 406 nm and EQE of 0.083%.[126] In order to quantum confine $MAPbBr_3$ (MA = methylammonium) MHPs, Kumar *et al.* incorporated oleic acid and octylamine into the synthesis procedure which then led to 2D $MAPbBr_3$-based PeLEDs with electroluminescence peaks as short as 432 nm with a corresponding EQE of 0.004%.[124] Moving away from lead, layered lead-free $Cs_3Sb_2Br_9$ quantum dots fabricated by Ma *et al.* yielded 408 nm electroluminescence with an EQE of 0.206% **(Figures 8c and 8d)**.[107] Lastly, the fabrication of a non-layered 1D (5-MBI)$PbBr_3$ (5-MBI = 5-methylbenzimidazole) PeLEDs by Cheng *et al.* led to 430 nm electroluminescence with and EQE of 0.009%.[128] Thus, RD perovskite LEDs have successfully enabled the development of V/UV PeLEDs, but more work is needed to improve device performance, including maximum EQEs and tunability into the UV spectrum.

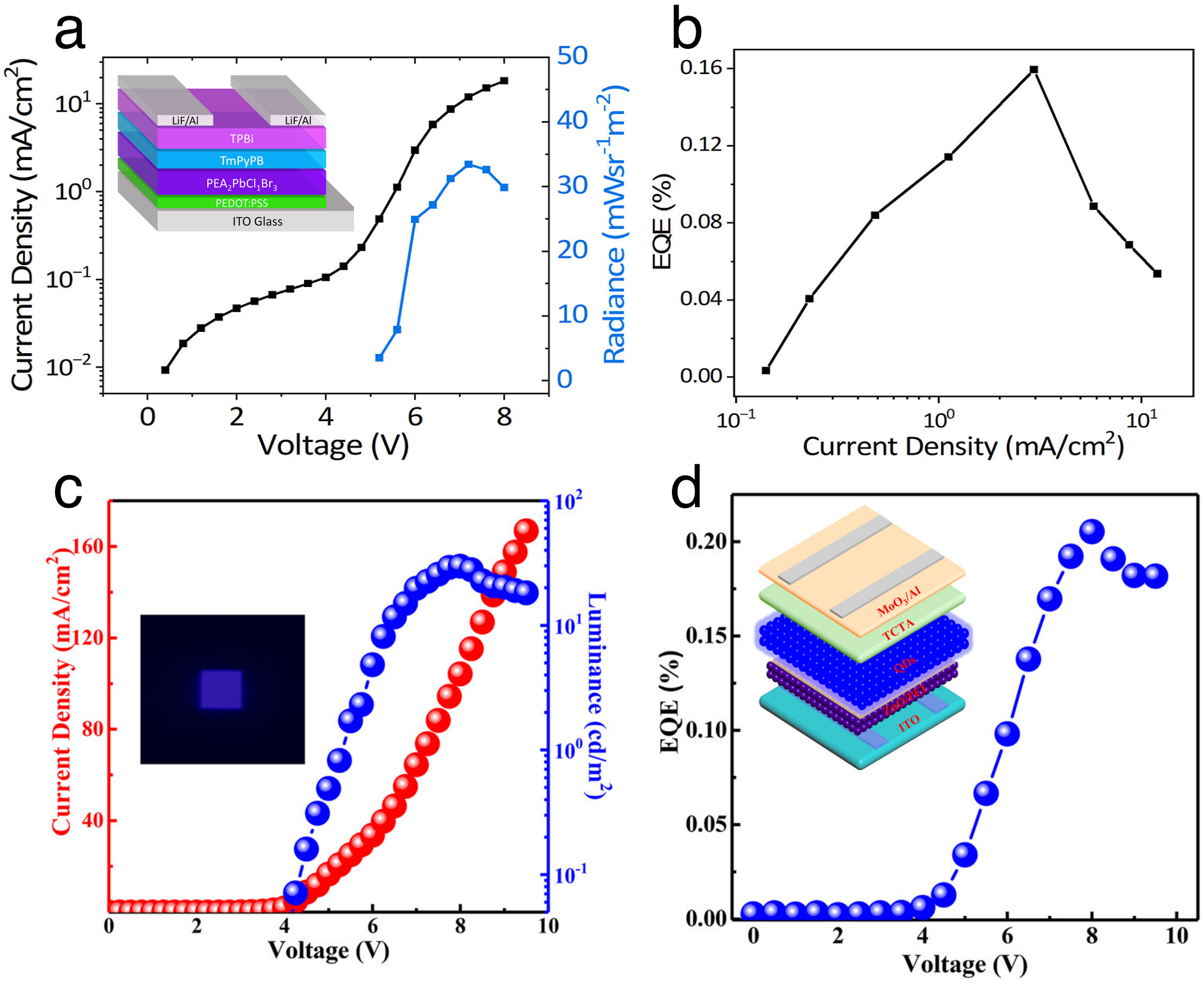

**Figure 8.** Reduced-dimensional V/UV PeLEDs. **(a)** Current density-voltage-radiance characteristics of the champion $PEA_2PbCl_1Br_3$ UV PeLED. Inset shows the device structure of the dual ETL $PEA_2PbCl_1Br_3$ UV PeLED. **(b)** EQE-current density characteristics of the champion $PEA_2PbCl_1Br_3$ UV PeLED. Adapted with permission from ref. [121]. Copyright 2024 The Authors. **(c)** Current density-voltage-luminance characteristics of the champion $Cs_3Sb_2Br_9$ violet PeLED. Inset shows a photograph of the PeLED operated at 7.0 V. **(d)** EQE-voltage characteristics of the champion $Cs_3Sb_2Br_9$ violet PeLED. Inset shows the device structure of the $Cs_3Sb_2Br_9$ violet PeLED. Adapted with permission from ref. [107]. Copyright 2019 American Chemical Society.

Next, we summarize the developments of 3D MHP V/UV PeLEDs. Sadhanala *et al.* employed halide engineering within the $MAPb(Br_xCl_{3-x})$ system which resulted in a 425 nm $MAPbCl_3$ PeLED but low corresponding EQE that could not be measured.[92] Along the same lines, 408.8 nm $MAPbCl_3$ PeLEDs were fabricated using both spin-coating and thermal evaporation of perovskite precursors by Sun *et al.* with an EQE of 0.036%.[93] Moving onto inorganic 3D MHPs, Bai *et al.* thermally evaporated $CsPbCl_3$ PeLEDs with electroluminescence at 410 nm.[80] Similarly, Zhang *et al.* thermally evaporated $CsPbCl_3$ double-layer films (i.e., one unannealed $CsPbCl_3$ film and one annealed $CsPbCl_3$ film) which led to 412 nm PeLEDs.[161] By incorporating an atypical $Mg_{0.71}Zn_{0.29}O$ electron transport layer with thermally-evaporated $CsPbCl_3$ emitters, Zhang *et al.* demonstrated 409 nm PeLEDs.[91] Alternatively, Zhang *et al.* synthesized $CsPbCl_3$ nanocrystals to be integrated into violet 405 nm PeLEDs with an EQE of 0.18%.[82] By introducing $Mg^{2+}$ dopants into $CsPbCl_3$ nanocrystals, Hu *et al.* achieved 402 nm PeLEDs with an EQE of 0.1%.[84] Analogously, Wang *et al.* doped $Ce^{3+}$ ions into $CsPbCl_{3-x}Br_x$ nanocrystals and demonstrated 414 nm PeLEDs with an EQE of 0.84% (**Figures 9a and 9b**).[90] While both 3D and RD MHPs have demonstrated the capability to move towards and into the UV spectrum, the majority of these PeLEDs are dominated by lead halides, which raise significant toxicity concerns. In order to make this technology more commercially viable, alternative MHP material classes should be considered.

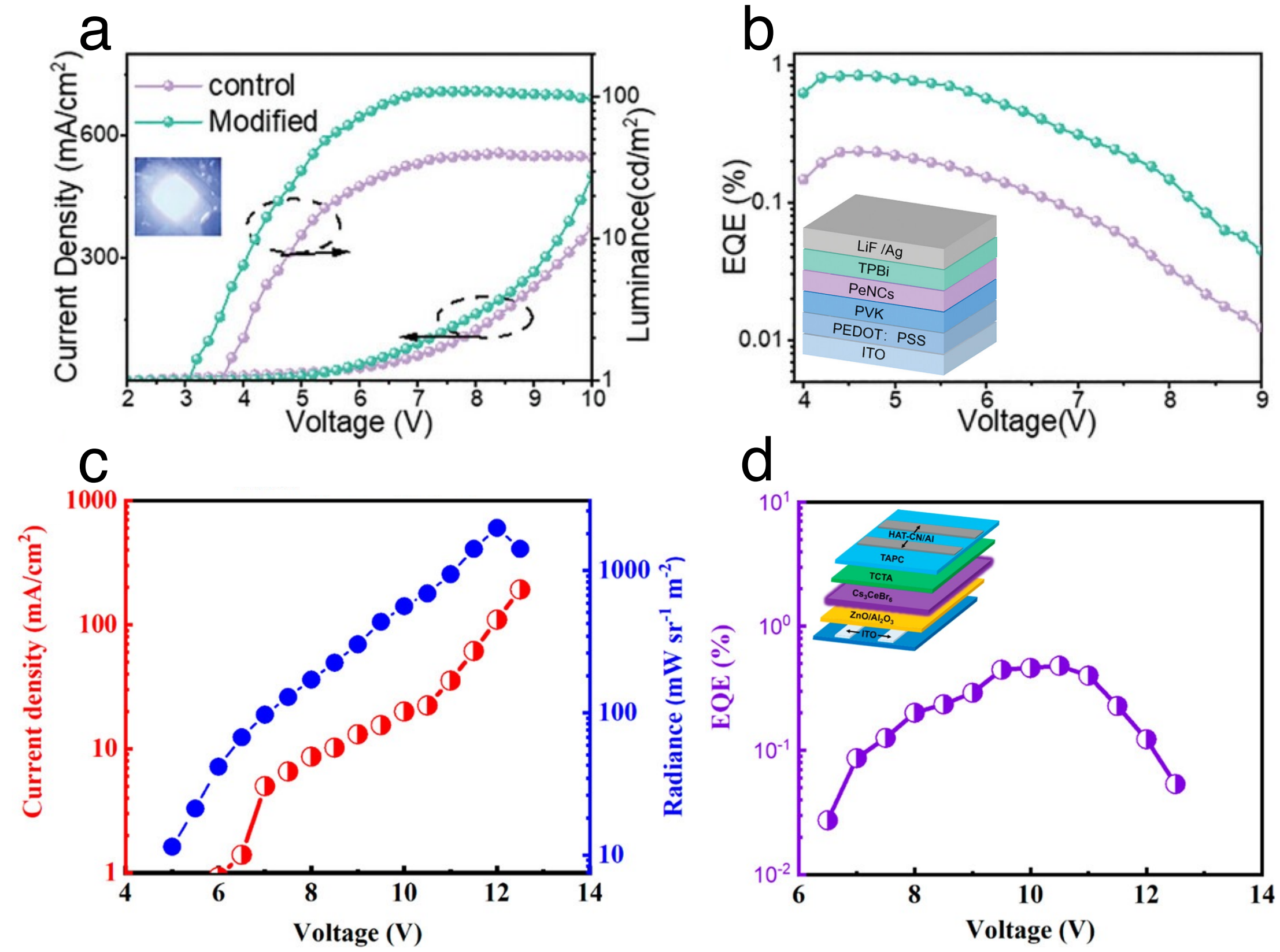

**Figure 9.** 3D and rare-earth metal halide V/UV PeLEDs. **(a)** Current density-voltage-luminance characteristics of both the control and modified (i.e., champion) $Ce^{3+}$-doped $CsPbCl_3$ violet PeLEDs. Inset shows a photograph of a representative PeLED under operation. **(b)** EQE-voltage characteristics of both the control and modified (i.e., champion) $Ce^{3+}$-doped $CsPbCl_3$ violet PeLEDs. Inset shows the device structure of the $Ce^{3+}$-doped $CsPbCl_3$ violet PeLEDs. Adapted with permission from ref. [90]. Copyright 2024 John Wiley and Sons. **(c)** Current density-voltage-radiance characteristics of the champion $Cs_3CeBr_6$ V/UV PeLED. **(d)** EQE-voltage characteristics of the champion $Cs_3CeBr_6$ V/UV PeLED. Inset shows the device structure of the $Cs_3CeBr_6$ V/UV PeLED. Adapted with permission from ref. [43]. Copyright 2021 American Chemical Society.

Rare-earth metal halide PeLEDs are particularly attractive given their potential to address toxicity concerns from lead halide PeLEDs. Specifically, $Ce^{3+}$ has strongly localized f-orbitals which paired with the allowed 5d → 4f electronic transitions make $Ce^{3+}$-based halide perovskites a strong candidate for high-energy light emission.[117] Additionally, the 5d → 4f excited-state lifetime for $Ce^{3+}$ is short (17 ns in $CeBr_3$),[162] which could suggest enhanced luminous efficiency and yield reduced LED degradation.[163] Cerium is also known to be abundant and nontoxic, as the Earth's Ce crust abundance is 0.006 wt %, comparable to that of Pb (0.010 wt %).[164] Regarding Ce halide PeLEDs, Yang *et al.* fabricated thermally-evaporated $Cs_3CeI_6$ PeLEDs with a doublet electroluminescence peak at approximately 430 nm and 470 nm and EQE of 7.9%.[108] Moving towards bromide-based perovskites, Guo *et al.* demonstrated thermally-evaporated $Cs_3CeBr_xI_{6-x}$ PeLEDs with tunable doublet electroluminescence peaks as short as 411 nm with a corresponding EQE of 0.93% for $Cs_3CeBr_2I_4$.[99] Fully eliminating iodide, Wang *et al.* fabricated thermally-evaporated $Cs_3CeBr_6$ PeLEDs with a doublet electroluminescence peak at 391 nm and 421 nm with an EQE of 0.46% (**Figures 9c and 9d)**.[43] By replacing cesium with rubidium, Du *et al.* fabricated thermally-evaporated $Rb_3CeI_6$ PeLEDs with a doublet electroluminescence peak at 427 nm and 468 nm with an EQE of 0.67%, showcasing the effects of A-site engineering within cerium halide PeLEDs.[113] Moving away from thermal evaporation, Dutta *et al.* spin coated $Cs_3CeBr_6$ thin films which resulted in PeLEDs with a doublet electroluminescence peak at 392 nm and 421 nm and champion EQE of 0.44%.[110] Finally, Sun *et al.* spin coated $Cs_3CeCl_6{\cdot}3H_2O$ thin films which resulted in UV PeLEDs with a singular electroluminescence peak at 380 nm and champion EQE of 0.13%.[112] Thus, rare-earth metal halide PeLEDs show great potential to enable nontoxic and efficient UV PeLEDs. Looking forward, additional strategies, such as halide composition modification towards Cl-based rare-earth metal halide PeLEDs and opportunities for dimensionality control should be explored to push deeper into the UV spectrum. **Table 3** summarizes the V/UV RD, 3D, and rare-earth metal halide PeLEDs discussed here while **Figure 10** illustrates the EQE performance of V/UV RD, 3D, and rare-earth metal halide PeLEDs across EL peak wavelength.

**Table 3.** Summary of V/UV metal halide perovskite LEDs.

| Perovskite | EL Peak | EQE | Type | Fabrication | Reference |
|---|---|---|---|---|---|
| $PEA_2PbBr_4$ | 410 nm | 0.04 % | RD | Spin-coating | [54] |
| $PEA_2PbBr_4$ | 408 nm | 0.41 % | RD | Spin-coating | [22] |
| $PEA_2PbBr_3Cl$ | 399 nm | 0.16 % | RD | Spin-coating | [121] |
| $PEA_2PbBr_3Cl$ | 406 nm | 1.63 % | RD | Spin-coating | [122] |
| $PEA_2PbBr_3Cl$ | 401 nm | 2.41 % | RD | Spin-coating | [122] |
| $PEA_2PbBr_3Cl$ | 394 nm | 1.96 % | RD | Spin-coating | [122] |
| $PEA_2PbBr_3Cl$ | 394 nm | 0.90 % | RD | Spin-coating | [122] |
| $PEA_2PbBr_4$ | 408.8 nm | 0.31 % | RD | Electric-field-deposition | [51] |
| $BA_2PbBr_4$ | 406 nm | 0.083 % | RD | Spin-coating | [126] |
| $MAPbBr_3$ | 432 nm | 0.004 % | RD | Spin-coating | [124] |
| $Cs_3Sb_2Br_9$ | 408 nm | 0.206 % | RD | Spin-coating | [107] |
| (5-MBI)$PbBr_3$ | 430 nm | 0.009% | RD | Spin-coating | [128] |
| $MAPbCl_3$ | 427 nm | N/A | 3D | Spin-coating | [92] |
| $MAPbCl_3$ | 408.8 nm | 0.036 % | 3D | Spin-coating/ Thermal Evaporation | 93 |
| $CsPbCl_3$ | 410 nm | N/A | 3D | Thermal Evaporation | 80 |
| $CsPbCl_3$ | 412 nm | N/A | 3D | Thermal Evaporation | 161 |
| $CsPbCl_3$ | 409 nm | N/A | 3D | Thermal Evaporation | 91 |
| $CsPbCl_3$ | 405 nm | 0.18 % | 3D | Spin-coating | 82 |
| $Mg^{2+}$ : $CsPbCl_3$ | 402 nm | 0.1 % | 3D | Spin-coating | 84 |
| $Ce^{3+}$ : $CsPbCl_3$ | 414 nm | 0.84 % | 3D | Spin-coating | 90 |
| $Cs_3CeI_6$ | ~430 & ~470 nm | 7.9 % | Rare-earth | Thermal Evaporation | 108 |
| $Cs_3CeI_6$ | 430 & 470 nm | 3.5 % | Rare-earth | Thermal Evaporation | 99 |
| $Cs_3CeBr_2I_4$ | 411 nm & unreported | 0.93 % | Rare-earth | Thermal Evaporation | 99 |
| $Cs_3CeBr_6$ | 391 & 421 nm | 0.46 % | Rare-earth | Thermal Evaporation | 43 |
| $Rb_3CeI_6$ | 427 & 468 nm | 0.67 % | Rare-earth | Thermal Evaporation | 113 |
| $Cs_3CeBr_6$ | 392 & 421 nm | 0.44 % | Rare-earth | Spin-coating | 110 |
| $Cs_3CeCl_6 \cdot 3H_2O$ | 380 nm | 0.13 % | Rare-earth | Spin-coating | 112 |

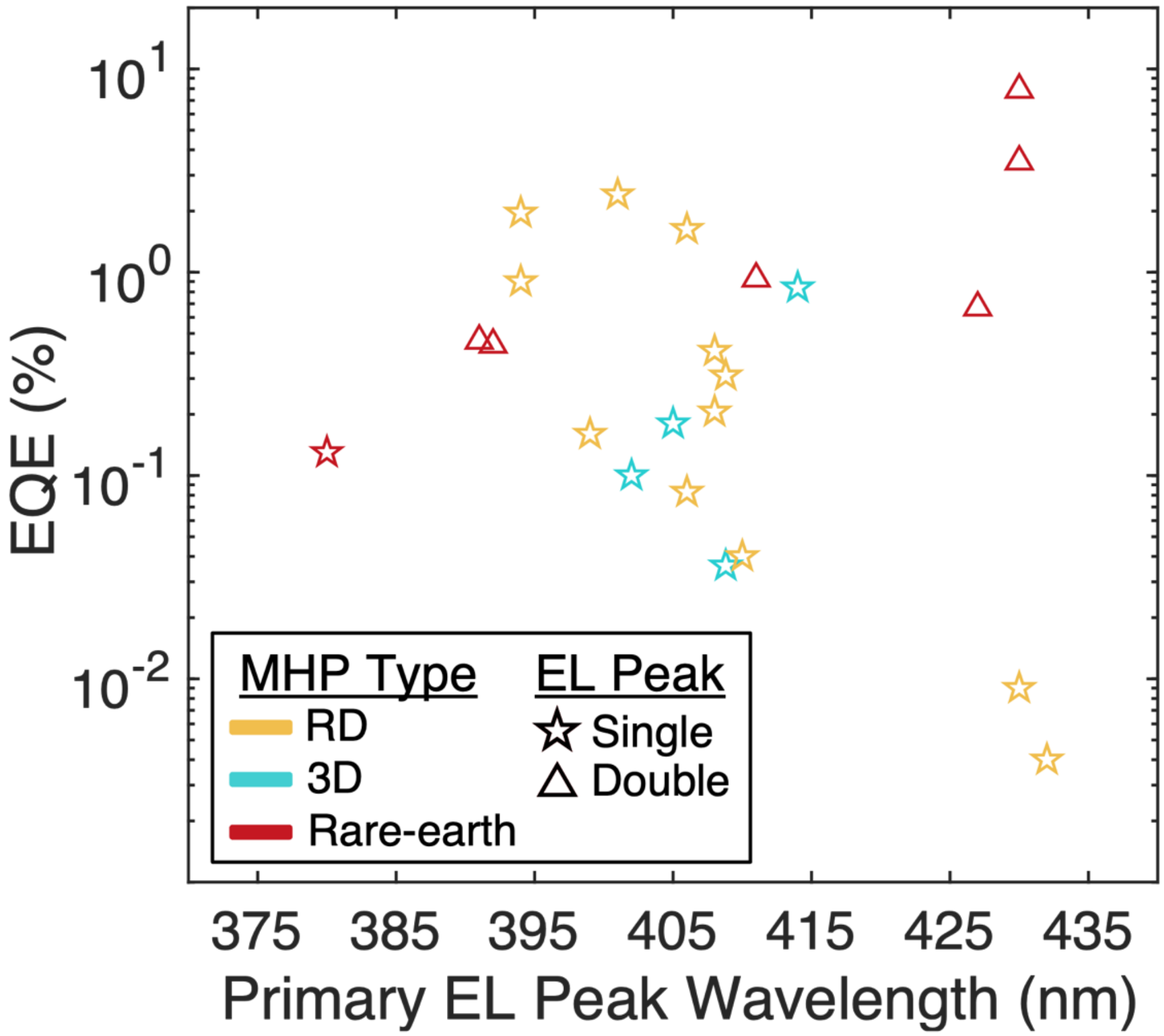

**Figure 10**. Summary of V/UV PeLED EQE performance across EL peak wavelength. Data points are sorted by MHP type (i.e., reduced-dimensional (RD), 3D, and rare-earth metal halide perovskites) as well as single or double EL peaks. All PeLEDs shown are plotted according to their shortest (i.e., primary) EL peak since most of the rare-earth metal halide PeLEDs result in a double EL emission peak. Illustrated from **Table 3**.

## Challenges and Limitations of Violet and Ultraviolet PeLEDs

*Efficiency Roll-Off.* Most PeLEDs exhibit a phenomenon known as efficiency roll-off where the EQE decreases as the current density increases past a certain threshold. V/UV PeLEDs are not immune to this effect, as evidenced from **Figures 8 and 9**. While the causes of this effect can vary, it has been attributed to Auger recombination,[165,166] unbalanced charge carrier injection,[167,168] and Joule heating.[146,169] Given the comparatively lower PLQYs and wider bandgaps of the emissive V/UV perovskite layer as well as the limited optimized transport layers, efficiency roll-off must be carefully addressed as it pertains to V/UV PeLEDs.

In order to address detrimental Auger effects as well as joule heating, the use of phosphonic acid (PA) modifiers for hole charge carrier injection (e.g., [2-(9*H*-carbazol-9-yl)ethyl]phosphonic acid (2PACz)) provides a viable alternative to traditional hole transport layers (HTLs). For instance, Shen *et al*. showed that 2PACz and [2-(3,6-di-*tert*-butyl-9*H*-carbazol-9-yl)ethyl]phosphonic acid (*t*-Bu-2PACz) can limit both harmful Auger and joule heating effects in green perovskite quantum dot LEDs.[166] By performing current density-dependent PL measurements on both PA-based and control (i.e., PEDOT:PSS/PVK as the HTL) PeLEDs, these authors show that Auger effects and/or joule heating are greatly suppressed for PA-based PeLEDs. As a result, the PA-based PeLEDs reach maximum luminances as high as 373,000 cd $m^{-2}$ while prolonging the onset of efficiency roll-off at a 1000-fold higher current density as compared to control PeLEDs. Additionally, PA modifiers can yield low voltage operation in PeLEDs, which could greatly enhance the operation of V/UV PeLEDs considering their inherently higher voltage operation. Wang *et al*. showed that integrating [2-(3,6-dibromo-9*H*-carbazol-9-yl)ethyl]phosphonic acid (Br-2PACz) into green 3D PeLEDs yields turn-on voltages as low as 1.9 V.[170] The authors argue that Br-2PACz facilitates hole charge carrier injection as well as passivating interfacial defects between the perovskite and ITO layers. Lastly, the incorporation of PA modifiers in V/UV PeLEDs can address unbalanced charge carrier injection, given that they induce a lower valence band maximum energy level[166,170–172] (e.g., between -5.5[170] and -6.0[171] eV for Br-2PACz) as compared to traditional HTLs (e.g., PEDOT:PSS has an energy level of -5.1 eV, which is poorly aligned with the UV perovskite layer's valence band maximum) and thus these materials hold the potential to enhance hole charge carrier injection. Overall, PA modifiers deserve further investigation within V/UV PeLEDs given their potential to address Auger effects, joule heating, and unbalanced charge carrier injection.

For joule heating effects, several techniques have been shown to enhance thermal management during PeLED operation. For example, Li *et al.* showed that coupling red light-emitting perovskite nanocrystals treated with diphenylphosphoryl azide (DPPA) and employing thermally conductive sapphire substrates greatly decreased joule heating effects and efficiency roll-off.[173] Specifically, DPPA increases charge injection into the nanocrystals by modifying the nanocrystals' coordination with insulating carbon-chain ligands while replacing glass substrates with sapphire ($K = 46\ W\ m^{-1}\ K^{-1}$) yields effective heat dissipation, as evidenced by a 28% decrease in device operation temperature.[173] Furthermore, Zhao *et al.* incorporated doped charge transport layers, additional heat spreaders and sinks, and optimized device geometries to address joule heating within infrared PeLEDs.[174] First, by doping both charge transport layers, efficiency roll-off was reduced while achieving EQEs similar to undoped PeLEDs. Specifically, phenyldi(pyren-2-yl)phosphine oxide ($POPy_2$; ETL) was n-doped with (pentamethylcyclopentadienyl)(1,3,5-trimethylbenzene)ruthenium dimer ($[RuCp^*Mes]_2$) while Poly-TPD (HTL) was p-doped with 2,3,5,6-tetrafluoro-7,7,8,8-tetracyanoquinodimethane ($F_4$-TCNQ). Next, integrating graphite (K =

1300 W $m^{-1}$ $K^{-1}$) or polycrystalline diamond heat spreaders (K > 2000 W $m^{-1}$ $K^{-1}$) with copper heat sinks also significantly reduced efficiency roll-off, especially at current densities above 500 mA $cm^{-2}$ where joule heating can induce rapid roll-off in PeLEDs. Lastly, by adjusting the PeLED device area to a line-shape geometry (i.e., 4 μm x 1 mm), heat dissipation was improved, yielding additional reductions to efficiency roll-off. Altogether, addressing joule heating effects within the emissive layer and corresponding PeLED device stack layers can lessen efficiency roll-off and should be carefully considered for V/UV PeLEDs.

One of the largest obstacles limiting the efficiency in V/UV PeLEDs is balanced charge carrier injection. Many of the conventional charge transport layers used within PeLEDs (**Figure 6a**) are not optimized for wide bandgap MHPs capable of V/UV light emission. In particular, the reliance on PEDOT:PSS for hole injection is problematic for V/UV PeLEDs given its misaligned energy levels. While PA modifiers provide a promising solution to circumvent this obstacle, modifications to PEDOT:PSS itself can also improve hole charge carrier transport. For instance, Lu *et al.* fabricated an ultrathin (6.9 nm) PEDOT:PSS HTL using a water stripping method and consequently achieved EQE enhancements in 3D, quasi-3D, and quasi-2D PeLEDs.[175] Notably, this ultrathin PEDOT:PSS HTL achieves a shallower energy level that is better aligned with V/UV MHPs. Introducing additives into PEDOT:PSS can also achieve similar effects. Incorporating additives like sodium polystyrenesulfonate[176] and guanidinium iodide[177] within PEDOT:PSS improved charge transport within PeLEDs and perovskite solar cells, respectively, while also achieving lower energy levels as compared to pure PEDOT:PSS. While modifications to PEDOT:PSS could yield more efficient V/UV PeLEDs, broader exploration of charge transport layers that can support efficient injection into V/UV PeLEDs is needed.

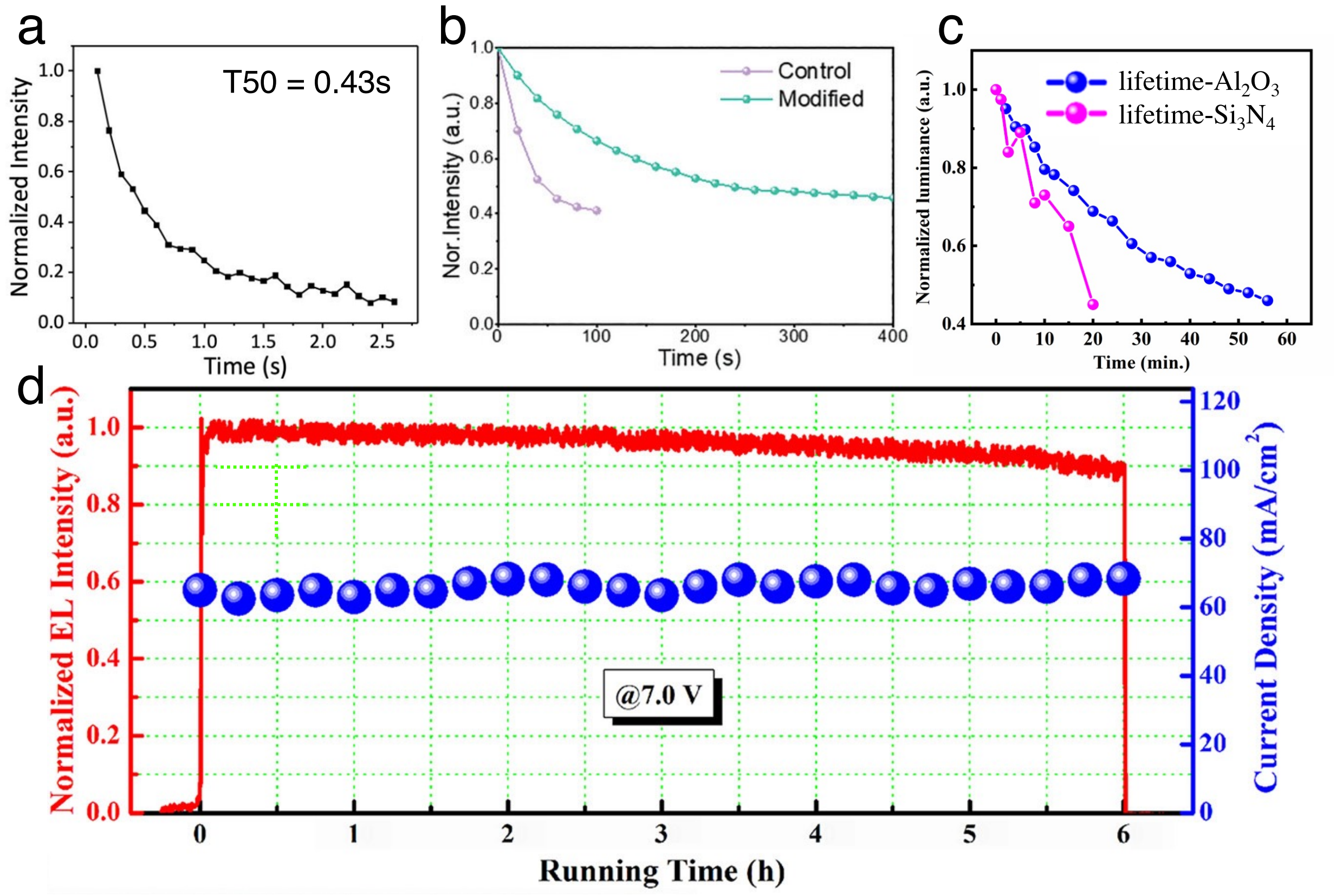

**Figure 11.** Operational stability of V/UV PeLEDs. **(a)** Normalized EL evolution in time of a $PEA_2PbCl_1Br_3$ UV PeLED operated at a constant current bias of 10 mA/cm$^2$. The resulting half-lifetime is 0.43 seconds. Adapted with permission from ref. [121]. Copyright 2024 The Authors. **(b)** Normalized EL evolution in time of both the control and modified (i.e., champion) $Ce^{3+}$-doped $CsPbCl_3$ violet PeLEDs operated at a constant voltage bias of 6 V. The resulting half-lifetime of the modified (i.e., champion) PeLED is 240 seconds. Adapted with permission from ref. [90]. Copyright 2024 John Wiley and Sons. **(c)** Normalized EL evolution in time of both the $Si_3N_4$-based (i.e., unoptimized electron blocking layer) and $Al_2O_3$-based (i.e., optimized electron blocking layer and resulting champion PeLED) $Cs_3CeI_6$ violet PeLEDs operated at a constant voltage bias of 5 V. The resulting half-lifetimes of the $Si_3N_4$-based and $Al_2O_3$-based $Cs_3CeI_6$ violet PeLEDs are ~18 and ~45 minutes, respectively. Adapted with permission from ref. [99]. Copyright

2022 The Authors. **(d)** Normalized EL evolution in time of a $Cs_3Sb_2Br_9$ violet PeLED operated at a constant voltage bias of 7 V. The resulting $T_{90}$ lifetime (i.e., intensity decay of 10%) is 6 hours. Adapted with permission from ref. [107]. Copyright 2019 American Chemical Society.

*Operational Stability.* PeLED operational lifetime is arguably the most important obstacle to overcome in order to ensure their commercialization. Within visible PeLEDs, green PeLEDs have achieved half lifetimes ($T_{50}$) as high as 520 hours[178] with an initial brightness of 1000 cd $m^{-2}$, while deep-blue PeLED stability has demonstrated a $T_{50}$ as high as 59 hours at an initial brightness of 100 cd $m^{-2}$.[152] Unfortunately, the operational stability of V/UV PeLEDs lags behind deep-blue PeLEDs. In **Figures 11a and 11b**, $PEA_2PbCl_1Br_3$ UV PeLEDs achieve a $T_{50}$ of 0.43 seconds while inorganic Ce-doped $CsPbCl_3$ violet PeLEDs demonstrate a $T_{50}$ of 240 seconds. From **Figures 11c and 11d**, lead-free violet PeLEDs based on $Cs_3CeI_6$ and $Cs_3Sb_2Br_9$ emitters achieved a $T_{50}$ and $T_{90}$ of 45 minutes and 6 hours, respectively. While the superior operational stability displayed by the lead-free V/UV MHPs as compared to their lead-based counterparts is promising from a toxicity perspective, significant improvements are still needed to realize this promising technology. The poor operational stability for MHPs, including V/UV MHPs, likely stems from ion migration effects, poor material stability, and lack of defect passivation.

Given the mixed electronic-ionic nature of MHPs and low activation energies of the constituent ionic species, ion migration significantly contributes to the poor operational stability of PeLEDs.[179] During operation, the voltage bias imposed on the PeLED can induce a large electric field within the emissive MHP layer and causes ion to redistribute themselves within the device stack. One of the more common consequences of this effect include the migration of halide anions towards the metal electrodes, since the activation energy of halide anions tends to be lower than other constituent ions within the MHP.[179] Given the higher voltage biases required to operate V/UV PeLEDs, ion migration effects are likely enhanced and thus must be carefully considered in order to improve the operational stability within this energy regime. As mentioned earlier, B-site doping can improve the PLQY of MHPs, but they can also address harmful ion migration effects in PeLEDs. Futscher *et al.* demonstrated how $Mn^{2+}$ doping can decrease ion migration effects within PeLEDs, and consequently improve its operational stability, by increasing the activation energy of mobile halide anions by over 200%.[180] Strategies to limit ion migration, including those related to compositional engineering and defect passivation, should be further explored within V/UV PeLEDs.

Compositional engineering, particularly at the A-site, has been demonstrated to enhance the operational stability of PeLEDs. For example, Cao *et al.* incorporated 5-aminovaleric acid (5AVA), a large cation introduced at the A-site, into red $FAPbI_3$ PeLEDs to prolong their lifetime to a $T_{50}$ of ~20 hours operated at a current density of 100 mA $cm^{-2}$.[181] Additionally, Jiang *et al.* developed a rubidium-cesium alloyed deep blue PeLED and reported an increase in the $T_{50}$ lifetime from 5.2 to 14.5 minutes as a result of rubidium incorporation at the A-site.[182] Moreover, while organic A-site cations, like methylammonium, can yield excellent PeLED EQE performance, their poor thermal stability can also lead to reduced stability as a result of operation-induced degradation mechanisms.[183–185] Thus, moving towards inorganic MHPs could improve the operational stability of V/UV PeLEDs. Lastly, **Figure 11** suggests that lead-free MHPs could lead to better operational stability within V/UV PeLEDs and should be further investigated.

Effective defect passivation can boost the operational stability of PeLEDs. Particularly, the use of zwitterions[186,187] and phosphine oxides[188,189] as additives within MHPs has greatly enhanced PeLED device performance, including stability. Regarding zwitterion additives, Guo *et al.* fabricated near-infrared $FAPbI_3$ PeLEDs and incorporated sulfobetaine 10 (SFB10) additives to yield record stability $T_{50}$ lifetimes of 11,539 and 32,675 hours operated at current densities of ~5.0 mA $cm^{-2}$ and ~3.2 mA $cm^{-2}$, respectively.[187] These authors attribute the excellent device stability to SFB10's ability to bind with the $FA^+$, $Pb^{2+}$, and $I^-$ ions at the perovskite grain boundaries, which in turn provides surface passivation of the MHP and mitigates ion migration of $Pb^{2+}$ and $I^-$ ions. Phosphine oxide additives have also improved PeLED stability, particularly within thermally-evaporated PeLEDs. Given the comparatively higher use of thermal evaporation for depositing V/UV MHPs as compared to visible light-emitting MHPs, this additive class system is particularly well-poised for V/UV PeLEDs. For instance, Li *et al.* showed that co-evaporating triphenylphosphine oxide (TPPO) with CsBr and $PbBr_2$ precursors improved the $T_{50}$ lifetime of green thermally-evaporated PeLEDs from 25 minutes to 78 minutes with an initial brightness of 1600 cd $m^{-2}$.[189] This is largely attributed to the surface passivation of $CsPbBr_3$ via TPPO. Similarly, thermally evaporating tris(trifluoromethyl)phosphine oxide (TFPPO) with quasi-2D MHP precursors yielded a green PeLED with an enhanced $T_{50}$ lifetime of 125 minutes, as compared to 34 minutes for non-TFPPO PeLEDs.[188] However, our previous work[136] and similar investigations[190] have shown that phosphine oxide additives can also lead to operational stability impairments. Thus, further investigation is warranted before universal incorporation of phosphine oxides within PeLED technologies, including those in the V/UV range.

*Toxicity and Scalability.* As V/UV PeLEDs are integrated into applications that interface with humans, such as those related to public health and sanitization, addressing toxicity concerns from these materials is critical. So far, lead-free V/UV PeLEDs have largely been limited to either $Sb^{3+}$- or $Ce^{3+}$-based emitters, while **Table 2** showcases lead-free V/UV MHP emitters employing $Cu^{+}$, $Bi^{3+}$, $Sb^{3+}$, $Ce^{3+}$, $Mn^{2+}$, $Zn^{2+}$, $Pr^{3+}$, $Eu^{2+}$, $Cd^{2+}$, and $In^{3+}$ halide perovskites. This discrepancy between the limited lead-free V/UV PeLEDs demonstrations and the numerous previously studied emitters suggests that more investigations are needed to integrate these materials into functional PeLEDs. Some of the obstacles related to the integration of these materials within PeLEDs include mismatched energy level alignment leading to poor charge injection, nonuniform thin film formation, and low thin film PLQY. Nevertheless, many of the strategies outlined in this section are equipped to address these obstacles and should help lead to an expanded lead-free V/UV PeLED class. Furthermore, to push deeper into the UV spectrum, both $Pr^{3+}$- and $Gd^{3+}$-based halide perovskites deserve additional investigations. Specifically, $Gd^{3+}$ demonstrates strong PL emission at 313 nm due to the spin-allowed ${}^6P_{7/2} \rightarrow {}^8S_{7/2}$ transition.[191,192] Similarly, $Pr^{3+}$ displays three major PL peaks at 265, 276, and 302 nm due to the electronic transitions from the $4f^15d^1$ excited state to $4f^2$ levels.[117] Overall, there are substantial opportunities to study lead-free V/UV PeLEDs and venture towards even shorter wavelength emission.

Additionally, the pursuit of industrially-compatible fabrication methods to promote the scalability of V/UV PeLEDs should be prioritized. While laboratory-scale spin-coating techniques have dominated MHP deposition within PeLEDs, thermally-evaporated PeLEDs are quickly progressing as a promising alternative to increase their scalability. As previously noted, this also extends to V/UV PeLEDs. There are several scalability advantages that thermally-evaporated

PeLEDs have compared to spin-coated PeLEDs.[193] First, since many commercial organic LEDs (OLEDs) utilize thermal evaporation in their production lines, transitioning PeLEDs to thermal evaporation can reduce capital investment. Moreover, insights from thermally-evaporated OLEDs could be directly applied to PeLEDs in industrial settings. Second, the use of quartz crystal microbalances as well as clean and consistent vacuum environments can help ensure the repeatability and reliability of thermally-evaporated PeLEDs - including those with larger device areas as compared to spin-coated PeLEDs. Lastly, since thermal evaporation can process many types of substrates (e.g., glass, silicon, flexible plastics, etc.), there are substantial opportunities for heterogeneous integration of thermally-evaporated PeLEDs. While spin-coated PeLEDs may be largely limited to laboratory-scale demonstrations, other solution-processing deposition techniques should be explored to access the strong potential of solution-processed V/UV MHPs. For example, blade-coating, a mass-fabrication solution-processing technique, has been used to fabricate large-area PeLEDs within the visible range.[194,195] Chu *et al.* blade-coated red $MAPbI_3$ PeLEDs and achieved a large device area of 28 $cm^2$ with uniform electroluminescence emission across the entire device area, showcasing the possibilities of scalable large-area PeLEDs with solution-processing techniques.[194] Altogether, both thermal evaporation and mass-fabrication solution-processing techniques are promising deposition processes that could improve the scalability of V/UV PeLEDs.

## Emerging Trends and Future Directions

*Novel Materials Systems.* As the field moves to address the toxicity of lead and the challenges of accessing deep-UV emission, lead-free double perovskites ($A_2B'B''X_6$) have emerged as a leading solution. Specifically, sodium-lanthanide double perovskite nanocrystals ($Cs_2NaLnCl_6$) represent a versatile platform for wide-bandgap optoelectronics (**Figure 12a**). Unlike lead-halide systems, which rely on band-edge emission, these materials utilize the localized intra-configurational f→f or d→f transitions of trivalent lanthanide ions ($Ln^{3+}$) to achieve precise spectral control.[117] Recent work has demonstrated that by varying the lanthanide dopant, emission can be tuned across the entire UV-Vis-NIR spectrum, including the difficult-to-access UV-C range; for instance, praseodymium-based $Cs_2NaPrCl_6$ nanocrystals have shown sharp emission peaks at 265 nm and 276 nm with PLQYs reaching 14%, while cerium-based $Cs_2NaCeCl_6$ enables efficient UV-A emission at 370 nm. These materials benefit from large bandgaps (>5 eV) that suppress non-radiative recombination and high environmental stability due to their all-inorganic nature.

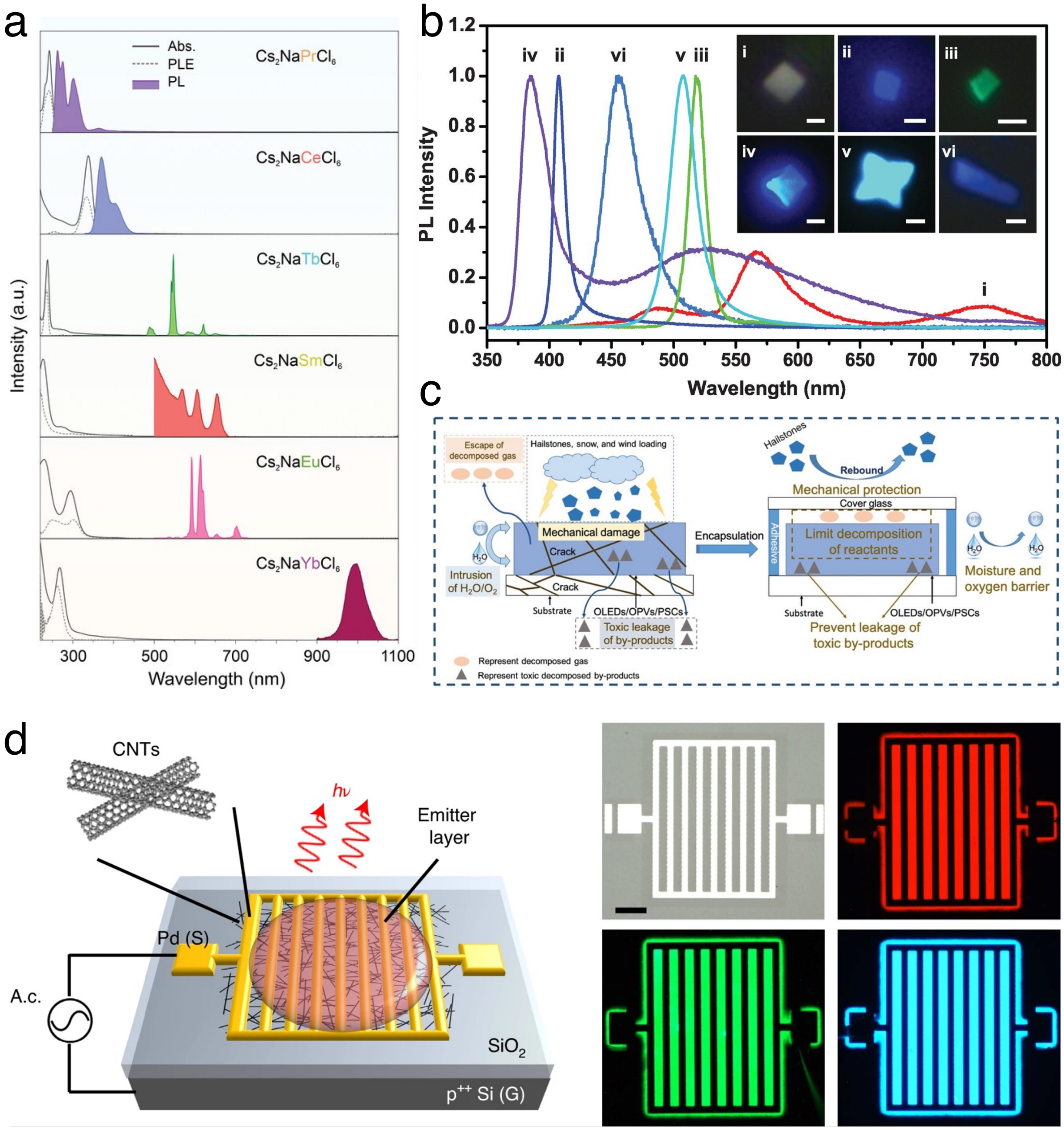

**Figure 12.** Emerging materials and device considerations for V/UV PeLEDs. **(a)** Absorption, PL, and PL excitation spectra of $Cs_2NaLnCl_6$ (Ln = Pr, Ce, Tb, Sm, Eu, and Yb) lanthanide double perovskite nanocrystals. Adapted with permission from ref. [117]. Copyright 2023 John Wiley and Sons. **(b)** PL spectra of various 2D hybrid MHPs. The emissive 2D sheets are as follows: (i) $(C_4H_9NH_3)_2PbCl_4$, (ii) $(C_4H_9NH_3)_2PbBr_4$, (iii) $(C_4H_9NH_3)_2PbI_4$, (iv) $(C_4H_9NH_3)_2PbCl_2Br_2$, (v)

$(C_4H_9NH_3)_2PbBr_2I_2$, and (vi) $(C_4H_9NH_3)_2(CH_3NH_3)Pb_2Br_7$. Inset shows the corresponding optical PL images of the six 2D hybrid MHPs. Scale bars are 2 μm for (i) to (v) and 10 μm for (vi). Adapted with permission from ref. [125]. Copyright 2015 American Association for the Advancement of Science. **(c)** Schematic diagram of encapsulation significance in thin film optoelectronic technologies. Adapted with permission from ref. [196]. Copyright 2021 John Wiley and Sons. **(d)** (left) Device structure with an assembled carbon nanotube (CNT) network on $SiO_2$/Si and patterned source contact. The device operates by applying AC voltage between the metal grid on the CNT network (source) and Si backgate (gate). (right) Uniform EL is achieved with various types of emitters including $Ru(bpy)_3(PF_6)_2$ (red emitter), CdSSe/ZnS alloyed quantum dots (520 nm, green emitter), and poly(9,9-dioctyl-9H-fluorene-2,7-diyl) (i.e., PFO, blue emitter). Devices are measured with a 15 V 100 kHz square wave. Scale bar is 0.1 mm. Adapted with permission from ref. [197]. Copyright 2020 Springer Nature.

Concurrently, the exploration of low-dimensional perovskites (2D and 0D) continues to offer a robust strategy for enhancing UV emission efficiency (**Figure 12b**). By reducing the dimensionality of the perovskite lattice to atomically thin 2D sheets or isolated 0D octahedra, dielectric and quantum confinement effects are dramatically amplified. This results in substantially increased exciton binding energies, which are critical for maintaining high radiative recombination rates at room temperature in wide-bandgap materials.[125] While early work focused on 2D lead-halide sheets like $(C_4H_9NH_3)_2PbBr_4$, current trends are converging towards low-dimensional lead-free systems. The $Cs_2NaLnCl_6$ double perovskites described above are structurally 0D, consisting of isolated $[LnCl_6]^{3-}$ octahedra separated by $Cs^+$ and $Na^+$ cations. This zero-dimensional isolation minimizes charge carrier diffusion to defect sites, providing a "defect-tolerant" lattice that can sustain high PLQY without the need for the rigorous surface passivation required in 3D nanocrystals. Future research lies in optimizing the charge transport layers for these highly confined, often insulating, low-dimensional materials to translate their exceptional photophysical properties into efficient electroluminescence.

*Encapsulation.* For V/UV PeLEDs to transition from the lab to real-world applications, advanced encapsulation is non-negotiable. As depicted in **Figure 12c**, the degradation pathways for these devices are multifaceted, involving not just moisture and oxygen intrusion, but also mechanical damage from environmental stressors and the potential leakage of toxic decomposition products (such as lead or organic volatiles).[196] Traditional encapsulation methods often fail to address all these vectors simultaneously.

Emerging strategies are therefore moving beyond simple epoxy barriers to multi-functional architectures that integrate mechanical robustness with chemical inertness. These next-generation encapsulants employ graded layers or hybrid organic-inorganic structures that provide mechanical rebound protection against physical stress while simultaneously serving as a chemical barrier.

Crucially, for perovskite devices, the encapsulation must also function "internally" to prevent the escape of volatile decomposition gases and toxic lead leakage, ensuring both device longevity and environmental safety. Developing UV-transparent, hydrophobic, and mechanically robust encapsulation materials—capable of withstanding the higher photon energies of V/UV light without degrading—will be critical for the deployment of PeLEDs in harsh environments.

*Alternative Device Architectures.* The development of efficient V/UV PeLEDs is fundamentally hindered by the dual challenge of injecting both electrons and holes into wide-bandgap perovskites. As the bandgap widens to enable UV emission (>3.0 eV), the band edges shift significantly: the valence band maximum (VBM) drops to very deep energy levels, while the conduction band minimum (CBM) moves closer to the vacuum level. This creates severe energy mismatches with standard charge transport materials. Finding stable hole transport layers (HTLs) with sufficiently deep ionization potentials and electron transport layers (ETLs) with appropriate electron affinities to match these V/UV perovskites simultaneously is a critical bottleneck. Consequently, V/UV devices often suffer from unbalanced carrier injection or require complex multi-layer architectures.

To bypass these injection constraints entirely, alternating current electroluminescence (AC-EL) offers a compelling alternative architecture. As demonstrated in recent work by Zhao *et al.* (**Figure 12d**), AC-driven devices operate on a field-induced carrier generation mechanism rather than continuous charge injection from external electrodes. In this architecture, a gate voltage creates transient, steep band bending at the interface between a source contact (e.g., a carbon nanotube network) and the emissive layer.[197]

This field-induced bending thins the injection barrier sufficiently to allow bipolar carrier tunneling directly into the emitter, regardless of the intrinsic band alignment of either the valence or conduction bands. During one half-cycle of the AC drive, electrons tunnel in; during the next, holes tunnel in, facilitating radiative recombination without the need for perfectly aligned ETLs or HTLs.

This mechanism enables electroluminescence across the entire spectrum—from infrared (0.13 eV) to ultraviolet (3.3 eV)—using a single, generic device structure. For V/UV perovskites, this is particularly advantageous as it eliminates the need to identify exotic transport materials. Furthermore, AC operation inherently mitigates the detrimental effects of ion migration—a phenomenon exacerbated by constant DC electric fields in halide perovskites. By periodically reversing the bias, ion accumulation at interfaces is minimized, potentially extending the operational lifetime of V/UV devices significantly while simplifying the device stack.

## Conclusion and Outlook

The rapid ascent of metal halide perovskites has redefined the landscape of solution-processable optoelectronics, yet the extension of this success into the V/UV spectral regimes remains a formidable frontier. As detailed in this review, the challenges governing V/UV PeLEDs are fundamentally distinct from their visible counterparts, necessitating a departure from standard composition and device engineering strategies. To bridge the performance gap between current V/UV PeLEDs and commercial requirements, the field must navigate a strategic evolution in both materials science and device architecture (**Figure 13**).

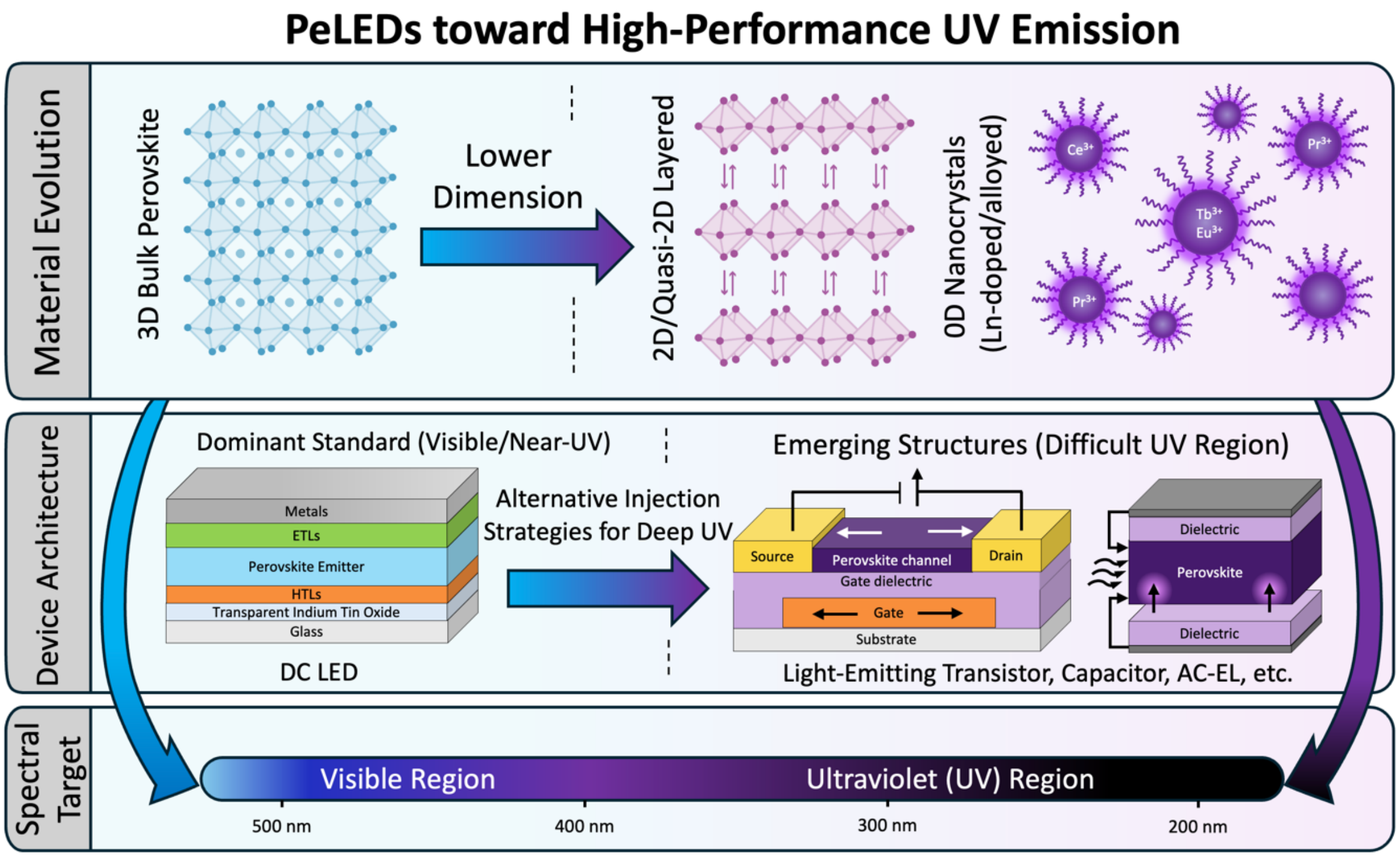

**Figure 13**. Evolutionary pathway of metal halide perovskite materials and corresponding device architectures toward high-performance UV light emission.

*Material Evolution: From Band-Edge to Activator-Based Emission.* While 3D bulk lead-halide perovskites (e.g., $CsPbCl_3$) have provided a foundational understanding of violet emission, their utility diminishes rapidly as the target wavelength shifts from UV-A into the UV-B and UV-C regions. The "bandgap engineering" approach—relying solely on chloride-bromide alloying—faces intrinsic limitations regarding phase segregation and defect intolerance. The evolutionary pathway forward points decisively toward dimensional reduction and rare-earth incorporation. Low-dimensional (2D and quasi-2D) structures utilize quantum confinement to sharpen emission linewidths and enhance exciton binding energies, critical for overcoming thermal quenching. However, to access the deep UV (UV-C), the field is shifting toward lead-free double perovskite nanocrystals (e.g., $Cs_2NaLnCl_6$) doped with lanthanides ($Ce^{3+}$, $Pr^{3+}$, $Gd^{3+}$). By exploiting the localized f-f or d-f transitions of these activators, rather than relying on band-edge recombination, researchers can bypass the defect-sensitivity of the host lattice and achieve precise spectral control down to 265 nm.

*Device Architecture: Overcoming the Injection Bottleneck.* High-efficiency UV electroluminescence is currently limited by charge injection, an inherent property of the standard DC p-i-n architecture. As the perovskite bandgap widens toward the UV-C region, the valence-band maximum deepens significantly. Identifying stable hole transport layers that align energetically with these deep levels remains elusive. Consequently, future engineering must pivot toward architectures that fundamentally circumvent this energetic impedance. Promising alternatives include light-emitting transistors and alternating current electroluminescence configurations. AC-EL structures sandwich the emitter between insulators, utilizing field-induced carrier generation rather than direct charge injection. This effectively eliminates the need for ohmic alignment with transport layers. Similarly, field-emission concepts utilize high electric fields to achieve ballistic electron injection directly into the conduction band. However, while these devices bypass current roadblocks, they introduce new challenges regarding operating voltage, high-field stability, and fabrication complexity. Therefore, the field urgently requires architectures designed specifically for the harsh realities of UV operation. Future research must explore radical departures from the standard visible-LED stack to withstand UV degradation and achieve the paradigm shift required for commercialization.

*Stability and Scalability: The Path to Commercialization.* The transition from lab to industry requires overcoming stability and scalability challenges. UV PeLEDs suffer a unique degradation from high-energy photons, accelerating the decomposition of organic cations and transport layers. This necessitates developing UV-stable, inorganic charge transport layers and robust encapsulation techniques to mitigate ion migration and toxic materials leakage. Furthermore, scalable deposition methods like thermal evaporation or blade-coating are essential for integration into large-area applications such as photolithography and sterilization arrays.

While V/UV PeLEDs are less mature than green and red devices, the fundamental roadblocks are identifiable and surmountable. By synergizing low-dimensional material design, rare-earth doping, and novel injection architectures, metal halide perovskites are positioned to disrupt the UV optoelectronics market. They offer a versatile, low-cost alternative to III-V semiconductors and mercury lamps for critical applications in sensing, nanofabrication, public health, and more.

# Acknowledgements

S.F. acknowledges the support from Stanford University as a Diversifying Academia, Recruiting Excellence (DARE) Fellow, the U.S. Department of Energy (DOE) Building Technologies Office (BTO) as an IBUILD Graduate Research Fellow, Stanford Graduate Fellowship in Science & Engineering (SGF) as a P. Michael Farmwald Fellow, and of the National GEM Consortium as a GEM Fellow. M.H. acknowledges the support of the Department of Electrical Engineering at Stanford University. T.K.C. acknowledges the support from Stanford University as a Knight-Hennessy Scholar and from the National Science Foundation Graduate Research Fellowship Program. D.M. acknowledges the support from Stanford University as an Enhancing Diversity in Graduate Education (EDGE) Fellow. This work was performed under an appointment to the Building Technologies Office (BTO) IBUILD Graduate Research Fellowship administered by the Oak Ridge Institute for Science and Education (ORISE) and managed by Oak Ridge National Laboratory (ORNL) for the U.S. Department of Energy (DOE). ORISE is managed by Oak Ridge Associated Universities (ORAU). All opinions expressed in this paper are the author's and do not necessarily reflect the policies and views of DOE, EERE, BTO, ORISE, ORAU or ORNL.

## Vocabulary

Metal Halide Perovskites (MHPs)

A broad semiconductor class primarily characterized by the formula for 3D $ABX_3$ materials where A is a monovalent cation (e.g., $Cs^+$ and $MA^+$), B is a divalent metal cation (e.g., $Pb^{2+}$ and $Eu^{2+}$), and X is a halide anion (e.g., $I^-$, $Br^-$, and $Cl^-$). However, derivations from this $ABX_3$ crystal structure that employ these ionic octahedra, resulting in 2D, 1D, and 0D crystal structures are classified as metal halide perovskite derivatives.

Violet Light-Emitting Perovskite

A metal halide perovskite that emits light between 400 and 435 nm.

Ultraviolet Light-Emitting Perovskite

A metal halide perovskite that emits light below 400 nm.

Photoluminescence Quantum Yield (PLQY)

The ratio of the number of photons emitted to the number of photons absorbed by a material. A metric to quantify the efficiency of photoluminescence.

Light-Emitting Diode (LED)

A semiconductor device that emits light when electric current flows through the device. Light is produced by electroluminescence via electron-hole recombination.

External Quantum Efficiency (EQE)

The ratio of the number of photons emitted from the LED to the number of electrons injected into the LED. A metric to quantify the efficiency of electroluminescence.

# References

(1) Kim, J. Y.; Lee, J.-W.; Jung, H. S.; Shin, H.; Park, N.-G. High-Efficiency Perovskite Solar Cells. *Chem. Rev.* **2020**, *120* (15), 7867–7918. https://doi.org/10.1021/acs.chemrev.0c00107.

(2) Stranks, S. D.; Snaith, H. J. Metal-Halide Perovskites for Photovoltaic and Light-Emitting Devices. *Nat. Nanotechnol.* **2015**, *10* (5), 391–402. https://doi.org/10.1038/nnano.2015.90.

(3) Jena, A. K.; Kulkarni, A.; Miyasaka, T. Halide Perovskite Photovoltaics: Background, Status, and Future Prospects. *Chem. Rev.* **2019**, *119* (5), 3036–3103. https://doi.org/10.1021/acs.chemrev.8b00539.

(4) Correa-Baena, J.-P.; Saliba, M.; Buonassisi, T.; Grätzel, M.; Abate, A.; Tress, W.; Hagfeldt, A. Promises and Challenges of Perovskite Solar Cells. *Science* **2017**, *358* (6364), 739–744. https://doi.org/10.1126/science.aam6323.

(5) Zhou, J.; Huang, J. Photodetectors Based on Organic–Inorganic Hybrid Lead Halide Perovskites. *Adv. Sci.* **2018**, *5* (1), 1700256. https://doi.org/10.1002/advs.201700256.

(6) Lee, J.; Chai, S.-P.; Tan, L.-L. Perovskite Paradigm Shift: A Green Revolution with Lead-Free Alternatives in Photocatalytic CO2 Reduction. *ACS Energy Lett.* **2024**, *9* (4), 1932–1975. https://doi.org/10.1021/acsenergylett.4c00416.

(7) Zou, C.; Ren, Z.; Hui, K.; Wang, Z.; Fan, Y.; Yang, Y.; Yuan, B.; Zhao, B.; Di, D. Electrically Driven Lasing from a Dual-Cavity Perovskite Device. *Nature* **2025**, *645* (8080), 369–374. https://doi.org/10.1038/s41586-025-09457-2.

(8) Zhang, Q.; Shang, Q.; Su, R.; Do, T. T. H.; Xiong, Q. Halide Perovskite Semiconductor Lasers: Materials, Cavity Design, and Low Threshold. *Nano Lett.* **2021**, *21* (5), 1903–1914. https://doi.org/10.1021/acs.nanolett.0c03593.

(9) Fakharuddin, A.; Gangishetty, M. K.; Abdi-Jalebi, M.; Chin, S.-H.; bin Mohd Yusoff, A. R.; Congreve, D. N.; Tress, W.; Deschler, F.; Vasilopoulou, M.; Bolink, H. J. Perovskite Light-Emitting Diodes. *Nat. Electron.* **2022**, *5* (4), 203–216. https://doi.org/10.1038/s41928-022-00745-7.

(10) Kim, Y.-H.; Cho, H.; Lee, T.-W. Metal Halide Perovskite Light Emitters. *Proc. Natl. Acad. Sci.* **2016**, *113* (42), 11694–11702. https://doi.org/10.1073/pnas.1607471113.

(11) Liu, X.-K.; Xu, W.; Bai, S.; Jin, Y.; Wang, J.; Friend, R. H.; Gao, F. Metal Halide Perovskites for Light-Emitting Diodes. *Nat. Mater.* **2021**, *20* (1), 10–21. https://doi.org/10.1038/s41563-020-0784-7.

(12) Kim, Y.-H.; Kim, S.; Jo, S. H.; Lee, T.-W. Metal Halide Perovskites: From Crystal Formations to Light-Emitting-Diode Applications. *Small Methods* **2018**, *2* (11), 1800093. https://doi.org/10.1002/smtd.201800093.

(13) Quan, L. N.; García de Arquer, F. P.; Sabatini, R. P.; Sargent, E. H. Perovskites for Light Emission. *Adv. Mater.* **2018**, *30* (45), 1801996. https://doi.org/10.1002/adma.201801996.

(14) Liu, Y.; Ma, Z.; Zhang, J.; He, Y.; Dai, J.; Li, X.; Shi, Z.; Manna, L. Light-Emitting Diodes Based on Metal Halide Perovskite and Perovskite Related Nanocrystals. *Adv. Mater.* **2025**, *37* (25), 2415606. https://doi.org/10.1002/adma.202415606.

(15) Han, T.-H.; Jang, K. Y.; Dong, Y.; Friend, R. H.; Sargent, E. H.; Lee, T.-W. A Roadmap for the Commercialization of Perovskite Light Emitters. *Nat. Rev. Mater.* **2022**, *7* (10), 757–777. https://doi.org/10.1038/s41578-022-00459-4.

(16) Feng, S.-C.; Shen, Y.; Hu, X.-M.; Su, Z.-H.; Zhang, K.; Wang, B.-F.; Cao, L.-X.; Xie, F.-M.; Li, H.-Z.; Gao, X.; Tang, J.-X.; Li, Y.-Q. Efficient and Stable Red Perovskite Light-Emitting Diodes via Thermodynamic Crystallization Control. *Adv. Mater.* **2024**, *36* (44), 2410255. https://doi.org/10.1002/adma.202410255.
(17) Sun, S.-Q.; Tai, J.-W.; He, W.; Yu, Y.-J.; Feng, Z.-Q.; Sun, Q.; Tong, K.-N.; Shi, K.; Liu, B.-C.; Zhu, M.; Wei, G.; Fan, J.; Xie, Y.-M.; Liao, L.-S.; Fung, M.-K. Enhancing Light Outcoupling Efficiency via Anisotropic Low Refractive Index Electron Transporting Materials for Efficient Perovskite Light-Emitting Diodes. *Adv. Mater.* **2024**, *36* (24), 2400421. https://doi.org/10.1002/adma.202400421.
(18) Li, M.; Yang, Y.; Kuang, Z.; Hao, C.; Wang, S.; Lu, F.; Liu, Z.; Liu, J.; Zeng, L.; Cai, Y.; Mao, Y.; Guo, J.; Tian, H.; Xing, G.; Cao, Y.; Ma, C.; Wang, N.; Peng, Q.; Zhu, L.; Huang, W.; Wang, J. Acceleration of Radiative Recombination for Efficient Perovskite LEDs. *Nature* **2024**, *630* (8017), 631–635. https://doi.org/10.1038/s41586-024-07460-7.
(19) Gao, Y.; Cai, Q.; He, Y.; Zhang, D.; Cao, Q.; Zhu, M.; Ma, Z.; Zhao, B.; He, H.; Di, D.; Ye, Z.; Dai, X. Highly Efficient Blue Light-Emitting Diodes Based on Mixed-Halide Perovskites with Reduced Chlorine Defects. *Sci. Adv.* **2024**, *10* (29), eado5645. https://doi.org/10.1126/sciadv.ado5645.
(20) Tan, Z.-K.; Moghaddam, R. S.; Lai, M. L.; Docampo, P.; Higler, R.; Deschler, F.; Price, M.; Sadhanala, A.; Pazos, L. M.; Credgington, D.; Hanusch, F.; Bein, T.; Snaith, H. J.; Friend, R. H. Bright Light-Emitting Diodes Based on Organometal Halide Perovskite. *Nat. Nanotechnol.* **2014**, *9* (9), 687–692. https://doi.org/10.1038/nnano.2014.149.
(21) Schmidt, L. C.; Pertegás, A.; González-Carrero, S.; Malinkiewicz, O.; Agouram, S.; Mínguez Espallargas, G.; Bolink, H. J.; Galian, R. E.; Pérez-Prieto, J. Nontemplate Synthesis of $CH_3$ $NH_3$ $PbBr_3$ Perovskite Nanoparticles. *J. Am. Chem. Soc.* **2014**, *136* (3), 850–853. https://doi.org/10.1021/ja4109209.
(22) Hu, M.; Fernández, S.; Zhou, Q.; Narayanan, P.; Saini, B.; Schloemer, T. H.; Lyu, J.; Gallegos, A. O.; Ahmed, G. H.; Congreve, D. N. Water Additives Improve the Efficiency of Violet Perovskite Light-Emitting Diodes. *Matter* **2023**, *6* (7), 2356–2367. https://doi.org/10.1016/j.matt.2023.05.018.
(23) Hinds, L. M.; O'Donnell, C. P.; Akhter, M.; Tiwari, B. K. Principles and Mechanisms of Ultraviolet Light Emitting Diode Technology for Food Industry Applications. *Innov. Food Sci. Emerg. Technol.* **2019**, *56*, 102153. https://doi.org/10.1016/j.ifset.2019.04.006.
(24) Chiappa, F.; Frascella, B.; Vigezzi, G. P.; Moro, M.; Diamanti, L.; Gentile, L.; Lago, P.; Clementi, N.; Signorelli, C.; Mancini, N.; Odone, A. The Efficacy of Ultraviolet Light-Emitting Technology against Coronaviruses: A Systematic Review. *J. Hosp. Infect.* **2021**, *114*, 63–78. https://doi.org/10.1016/j.jhin.2021.05.005.
(25) Ramos, C. C. R.; Roque, J. L. A.; Sarmiento, D. B.; Suarez, L. E. G.; Sunio, J. T. P.; Tabungar, K. I. B.; Tengco, G. S. C.; Rio, P. C.; Hilario, A. L. Use of Ultraviolet-C in Environmental Sterilization in Hospitals: A Systematic Review on Efficacy and Safety. *Int. J. Health Sci.* **2020**, *14* (6), 52–65.
(26) Won, W.-S.; Tran, L. G.; Park, W.-T.; Kim, K.-K.; Shin, C. S.; Kim, N.; Kim, Y.-J.; Yoon, Y.-J. UV-LEDs for the Disinfection and Bio-Sensing Applications. *Int. J. Precis. Eng. Manuf.* **2018**, *19* (12), 1901–1915. https://doi.org/10.1007/s12541-018-0218-5.
(27) Shahrubudin, N.; Lee, T. C.; Ramlan, R. An Overview on 3D Printing Technology: Technological, Materials, and Applications. *Procedia Manuf.* **2019**, *35*, 1286–1296. https://doi.org/10.1016/j.promfg.2019.06.089.
(28) Zheng, L.; Zywietz, U.; Birr, T.; Duderstadt, M.; Overmeyer, L.; Roth, B.; Reinhardt, C. UV-LED Projection Photolithography for High-Resolution Functional Photonic

Components. *Microsyst. Nanoeng.* **2021**, *7* (1), 64. https://doi.org/10.1038/s41378-021-00286-7.
(29) Mo, X.; Liang, Y.; Wang, L.; Qi, Z.; Zhou, J.; Zhang, L.; Liu, P.; Gou, J.; Wang, Y. Solar-Blind Deep Ultraviolet Light Communication Network. *Opt. Express Vol 33 Issue 17 Pp 36280-36292* **2025**. https://doi.org/10.1364/OE.570896.
(30) Amano, H.; Collazo, R.; Santi, C. D.; Einfeldt, S.; Funato, M.; Glaab, J.; Hagedorn, S.; Hirano, A.; Hirayama, H.; Ishii, R.; Kashima, Y.; Kawakami, Y.; Kirste, R.; Kneissl, M.; Martin, R.; Mehnke, F.; Meneghini, M.; Ougazzaden, A.; Parbrook, P. J.; Rajan, S.; Reddy, P.; Römer, F.; Ruschel, J.; Sarkar, B.; Scholz, F.; Schowalter, L. J.; Shields, P.; Sitar, Z.; Sulmoni, L.; Wang, T.; Wernicke, T.; Weyers, M.; Witzigmann, B.; Wu, Y.-R.; Wunderer, T.; Zhang, Y. The 2020 UV Emitter Roadmap. *J. Phys. Appl. Phys.* **2020**, *53* (50), 503001. https://doi.org/10.1088/1361-6463/aba64c.
(31) Xiao, D.; Zhao, S.; Yuan, H.; Yang, X. CE Detector Based on Light-Emitting Diodes. *ELECTROPHORESIS* **2007**, *28* (1–2), 233–242. https://doi.org/10.1002/elps.200600473.
(32) De Oliveira, B. P.; Souza Rastelli, A. N.; Salvador Bagnato, V.; Hugo Panhoca, V. Dental Bleaching Using Violet Light Alone: Clinical Case Report. *Dentistry* **2017**, *7* (11). https://doi.org/10.4172/2161-1122.1000459.
(33) Shur, M. S.; Gaska, R. Deep-Ultraviolet Light-Emitting Diodes. *IEEE Trans. Electron Devices* **2010**, *57* (1), 12–25. https://doi.org/10.1109/TED.2009.2033768.
(34) Khan, A.; Balakrishnan, K.; Katona, T. Ultraviolet Light-Emitting Diodes Based on Group Three Nitrides. *Nat. Photonics* **2008**, *2* (2), 77–84. https://doi.org/10.1038/nphoton.2007.293.
(35) Moe, A. E.; Marx, S.; Banani, N.; Liu, M.; Marquardt, B.; Wilson, D. M. Improvements in LED-Based Fluorescence Analysis Systems. *Sens. Actuators B Chem.* **2005**, *111–112*, 230–241. https://doi.org/10.1016/j.snb.2005.01.057.
(36) Feng, F.; Liu, Y.; Zhang, K.; Yang, H.; Hyun, B.-R.; Xu, K.; Kwok, H.-S.; Liu, Z. High-Power AlGaN Deep-Ultraviolet Micro-Light-Emitting Diode Displays for Maskless Photolithography. *Nat. Photonics* **2025**, *19* (1), 101–108. https://doi.org/10.1038/s41566-024-01551-7.
(37) Bhattarai, T.; Ebong, A.; Raja, M. Y. A. A Review of Light-Emitting Diodes and Ultraviolet Light-Emitting Diodes and Their Applications. *Photonics* **2024**, *11* (6), 491. https://doi.org/10.3390/photonics11060491.
(38) Dou, L. Emerging Two-Dimensional Halide Perovskite Nanomaterials. *J. Mater. Chem. C* **2017**, *5* (43), 11165–11173. https://doi.org/10.1039/C7TC02863F.
(39) Yong, Z.-J.; Guo, S.-Q.; Ma, J.-P.; Zhang, J.-Y.; Li, Z.-Y.; Chen, Y.-M.; Zhang, B.-B.; Zhou, Y.; Shu, J.; Gu, J.-L.; Zheng, L.-R.; Bakr, O. M.; Sun, H.-T. Doping-Enhanced Short-Range Order of Perovskite Nanocrystals for Near-Unity Violet Luminescence Quantum Yield. *J. Am. Chem. Soc.* **2018**, *140* (31), 9942–9951. https://doi.org/10.1021/jacs.8b04763.
(40) Zhang, Y.; Cheng, X.; Tu, D.; Gong, Z.; Li, R.; Yang, Y.; Zheng, W.; Xu, J.; Deng, S.; Chen, X. Engineering the Bandgap and Surface Structure of CsPbCl3 Nanocrystals to Achieve Efficient Ultraviolet Luminescence. *Angew. Chem. Int. Ed.* **2021**, *60* (17), 9693–9698. https://doi.org/10.1002/anie.202017370.
(41) Zhang, J.; Yang, Y.; Deng, H.; Farooq, U.; Yang, X.; Khan, J.; Tang, J.; Song, H. High Quantum Yield Blue Emission from Lead-Free Inorganic Antimony Halide Perovskite

Colloidal Quantum Dots. *ACS Nano* **2017**, *11* (9), 9294–9302. https://doi.org/10.1021/acsnano.7b04683.
(42) Leng, M.; Yang, Y.; Zeng, K.; Chen, Z.; Tan, Z.; Li, S.; Li, J.; Xu, B.; Li, D.; Hautzinger, M. P.; Fu, Y.; Zhai, T.; Xu, L.; Niu, G.; Jin, S.; Tang, J. All-Inorganic Bismuth-Based Perovskite Quantum Dots with Bright Blue Photoluminescence and Excellent Stability. *Adv. Funct. Mater.* **2018**, *28* (1), 1704446. https://doi.org/10.1002/adfm.201704446.
(43) Wang, L.; Guo, Q.; Duan, J.; Xie, W.; Ji, G.; Li, S.; Chen, C.; Li, J.; Yang, L.; Tan, Z.; Xu, L.; Xiao, Z.; Luo, J.; Tang, J. Exploration of Nontoxic Cs3CeBr6 for Violet Light-Emitting Diodes. *ACS Energy Lett.* **2021**, *6* (12), 4245–4254. https://doi.org/10.1021/acsenergylett.1c02022.
(44) Creason, T. D.; Yangui, A.; Roccanova, R.; Strom, A.; Du, M.-H.; Saparov, B. Rb2CuX3 (X = Cl, Br): 1D All-Inorganic Copper Halides with Ultrabright Blue Emission and Up-Conversion Photoluminescence. *Adv. Opt. Mater.* **2020**, *8* (2), 1901338. https://doi.org/10.1002/adom.201901338.
(45) Lian, L.; Zheng, M.; Zhang, P.; Zheng, Z.; Du, K.; Lei, W.; Gao, J.; Niu, G.; Zhang, D.; Zhai, T.; Jin, S.; Tang, J.; Zhang, X.; Zhang, J. Photophysics in Cs3Cu2X5 (X = Cl, Br, or I): Highly Luminescent Self-Trapped Excitons from Local Structure Symmetrization. *Chem. Mater.* **2020**, *32* (8), 3462–3468. https://doi.org/10.1021/acs.chemmater.9b05321.
(46) Creason, T. D.; McWhorter, T. M.; Bell, Z.; Du, M.-H.; Saparov, B. $K_2$ $CuX_3$ (X = Cl, Br): All-Inorganic Lead-Free Blue Emitters with Near-Unity Photoluminescence Quantum Yield. *Chem. Mater.* **2020**, *32* (14), 6197–6205. https://doi.org/10.1021/acs.chemmater.0c02098.
(47) Shi, E.; Gao, Y.; Finkenauer, B. P.; Akriti; Coffey, A. H.; Dou, L. Two-Dimensional Halide Perovskite Nanomaterials and Heterostructures. *Chem. Soc. Rev.* **2018**, *47* (16), 6046–6072. https://doi.org/10.1039/C7CS00886D.
(48) Cresp, M.; Liu, M.; Rager, M.-N.; Zheng, D.; Pauporté, T. 2D Ruddlesden-Popper versus 2D Dion-Jacobson Perovskites: Of the Importance of Determining the "True" Average n-Value of Annealed Layers. *Adv. Funct. Mater.* **2025**, *35* (3), 2413671. https://doi.org/10.1002/adfm.202413671.
(49) Mauck, C. M.; Tisdale, W. A. Excitons in 2D Organic–Inorganic Halide Perovskites. *Trends Chem.* **2019**, *1* (4), 380–393. https://doi.org/10.1016/j.trechm.2019.04.003.
(50) Zheng, H.; Loh, K. P. Ferroics in Hybrid Organic–Inorganic Perovskites: Fundamentals, Design Strategies, and Implementation. *Adv. Mater.* **2024**, *36* (14), 2308051. https://doi.org/10.1002/adma.202308051.
(51) Deng, W.; Jin, X.; Lv, Y.; Zhang, X.; Zhang, X.; Jie, J. 2D Ruddlesden–Popper Perovskite Nanoplate Based Deep-Blue Light-Emitting Diodes for Light Communication. *Adv. Funct. Mater.* **2019**, *29* (40), 1903861. https://doi.org/10.1002/adfm.201903861.
(52) Protesescu, L.; Yakunin, S.; Bodnarchuk, M. I.; Krieg, F.; Caputo, R.; Hendon, C. H.; Yang, R. X.; Walsh, A.; Kovalenko, M. V. Nanocrystals of Cesium Lead Halide Perovskites (CsPbX3, X = Cl, Br, and I): Novel Optoelectronic Materials Showing Bright Emission with Wide Color Gamut. *Nano Lett.* **2015**, *15* (6), 3692–3696. https://doi.org/10.1021/nl5048779.
(53) Kambhampati, P. Nanoparticles, Nanocrystals, and Quantum Dots: What Are the Implications of Size in Colloidal Nanoscale Materials? *J. Phys. Chem. Lett.* **2021**, *12* (20), 4769–4779. https://doi.org/10.1021/acs.jpclett.1c00754.

(54) Liang, D.; Peng, Y.; Fu, Y.; Shearer, M. J.; Zhang, J.; Zhai, J.; Zhang, Y.; Hamers, R. J.; Andrew, T. L.; Jin, S. Color-Pure Violet-Light-Emitting Diodes Based on Layered Lead Halide Perovskite Nanoplates. *ACS Nano* **2016**, *10* (7), 6897–6904. https://doi.org/10.1021/acsnano.6b02683.
(55) Leng, M.; Chen, Z.; Yang, Y.; Li, Z.; Zeng, K.; Li, K.; Niu, G.; He, Y.; Zhou, Q.; Tang, J. Lead-Free, Blue Emitting Bismuth Halide Perovskite Quantum Dots. *Angew. Chem. Int. Ed.* **2016**, *55* (48), 15012–15016. https://doi.org/10.1002/anie.201608160.
(56) Momma, K.; Izumi, F. VESTA 3 for Three-Dimensional Visualization of Crystal, Volumetric and Morphology Data. *J. Appl. Crystallogr.* **2011**, *44* (6), 1272–1276. https://doi.org/10.1107/S0021889811038970.
(57) Jin, H.; Debroye, E.; Keshavarz, M.; G. Scheblykin, I.; J. Roeffaers, M. B.; Hofkens, J.; A. Steele, J. It's a Trap! On the Nature of Localised States and Charge Trapping in Lead Halide Perovskites. *Mater. Horiz.* **2020**, *7* (2), 397–410. https://doi.org/10.1039/C9MH00500E.
(58) Smith, M. D.; Crace, E. J.; Jaffe, A.; Karunadasa, H. I. The Diversity of Layered Halide Perovskites. *Annu. Rev. Mater. Res.* **2018**, *48* (Volume 48, 2018), 111–136. https://doi.org/10.1146/annurev-matsci-070317-124406.
(59) Matheu, R.; Vigil, J. A.; Crace, E. J.; Karunadasa, H. I. The Halogen Chemistry of Halide Perovskites. *Trends Chem.* **2022**, *4* (3), 206–219. https://doi.org/10.1016/j.trechm.2021.12.002.
(60) Song, J.; Li, J.; Li, X.; Xu, L.; Dong, Y.; Zeng, H. Quantum Dot Light-Emitting Diodes Based on Inorganic Perovskite Cesium Lead Halides (CsPbX3). *Adv. Mater.* **2015**, *27* (44), 7162–7167. https://doi.org/10.1002/adma.201502567.
(61) Nedelcu, G.; Protesescu, L.; Yakunin, S.; Bodnarchuk, M. I.; Grotevent, M. J.; Kovalenko, M. V. Fast Anion-Exchange in Highly Luminescent Nanocrystals of Cesium Lead Halide Perovskites (CsPbX3, X = Cl, Br, I). *Nano Lett.* **2015**, *15* (8), 5635–5640. https://doi.org/10.1021/acs.nanolett.5b02404.
(62) Akkerman, Q. A.; D'Innocenzo, V.; Accornero, S.; Scarpellini, A.; Petrozza, A.; Prato, M.; Manna, L. Tuning the Optical Properties of Cesium Lead Halide Perovskite Nanocrystals by Anion Exchange Reactions. *J. Am. Chem. Soc.* **2015**, *137* (32), 10276–10281. https://doi.org/10.1021/jacs.5b05602.
(63) Tao, S.; Schmidt, I.; Brocks, G.; Jiang, J.; Tranca, I.; Meerholz, K.; Olthof, S. Absolute Energy Level Positions in Tin- and Lead-Based Halide Perovskites. *Nat. Commun.* **2019**, *10* (1), 2560. https://doi.org/10.1038/s41467-019-10468-7.
(64) Amat, A.; Mosconi, E.; Ronca, E.; Quarti, C.; Umari, P.; Nazeeruddin, Md. K.; Grätzel, M.; De Angelis, F. Cation-Induced Band-Gap Tuning in Organohalide Perovskites: Interplay of Spin–Orbit Coupling and Octahedra Tilting. *Nano Lett.* **2014**, *14* (6), 3608–3616. https://doi.org/10.1021/nl5012992.
(65) Wang, H.; Tal, A.; Bischoff, T.; Gono, P.; Pasquarello, A. Accurate and Efficient Band-Gap Predictions for Metal Halide Perovskites at Finite Temperature. *Npj Comput. Mater.* **2022**, *8* (1), 237. https://doi.org/10.1038/s41524-022-00869-6.
(66) Ko, P. K.; Ge, J.; Ding, P.; Chen, D.; Tsang, H. L. T.; Kumar, N.; Halpert, J. E. The Deepest Blue: Major Advances and Challenges in Deep Blue Emitting Quasi-2D and Nanocrystalline Perovskite LEDs. *Adv. Mater.* **2025**, *37* (23), 2407764. https://doi.org/10.1002/adma.202407764.

(67) Li, X.; Wu, Y.; Zhang, S.; Cai, B.; Gu, Y.; Song, J.; Zeng, H. CsPbX3 Quantum Dots for Lighting and Displays: Room-Temperature Synthesis, Photoluminescence Superiorities, Underlying Origins and White Light-Emitting Diodes. *Adv. Funct. Mater.* **2016**, *26* (15), 2435–2445. https://doi.org/10.1002/adfm.201600109.
(68) Pan, G.; Bai, X.; Yang, D.; Chen, X.; Jing, P.; Qu, S.; Zhang, L.; Zhou, D.; Zhu, J.; Xu, W.; Dong, B.; Song, H. Doping Lanthanide into Perovskite Nanocrystals: Highly Improved and Expanded Optical Properties. *Nano Lett.* **2017**, *17* (12), 8005–8011. https://doi.org/10.1021/acs.nanolett.7b04575.
(69) Imran, M.; Caligiuri, V.; Wang, M.; Goldoni, L.; Prato, M.; Krahne, R.; De Trizio, L.; Manna, L. Benzoyl Halides as Alternative Precursors for the Colloidal Synthesis of Lead-Based Halide Perovskite Nanocrystals. *J. Am. Chem. Soc.* **2018**, *140* (7), 2656–2664. https://doi.org/10.1021/jacs.7b13477.
(70) Mondal, N.; De, A.; Samanta, A. Achieving Near-Unity Photoluminescence Efficiency for Blue-Violet-Emitting Perovskite Nanocrystals. *ACS Energy Lett.* **2019**, *4* (1), 32–39. https://doi.org/10.1021/acsenergylett.8b01909.
(71) Ahmed, T.; Seth, S.; Samanta, A. Boosting the Photoluminescence of CsPbX3 (X = Cl, Br, I) Perovskite Nanocrystals Covering a Wide Wavelength Range by Postsynthetic Treatment with Tetrafluoroborate Salts. *Chem. Mater.* **2018**, *30* (11), 3633–3637. https://doi.org/10.1021/acs.chemmater.8b01235.
(72) Ahmed, G. H.; El-Demellawi, J. K.; Yin, J.; Pan, J.; Velusamy, D. B.; Hedhili, M. N.; Alarousu, E.; Bakr, O. M.; Alshareef, H. N.; Mohammed, O. F. Giant Photoluminescence Enhancement in CsPbCl3 Perovskite Nanocrystals by Simultaneous Dual-Surface Passivation. *ACS Energy Lett.* **2018**, *3* (10), 2301–2307. https://doi.org/10.1021/acsenergylett.8b01441.
(73) Zhai, Y.; Bai, X.; Pan, G.; Zhu, J.; Shao, H.; Dong, B.; Xu, L.; Song, H. Effective Blue-Violet Photoluminescence through Lanthanum and Fluorine Ions Co-Doping for CsPbCl 3 Perovskite Quantum Dots. *Nanoscale* **2019**, *11* (5), 2484–2491. https://doi.org/10.1039/C8NR09794A.
(74) Chen, J.-K.; Ma, J.-P.; Guo, S.-Q.; Chen, Y.-M.; Zhao, Q.; Zhang, B.-B.; Li, Z.-Y.; Zhou, Y.; Hou, J.; Kuroiwa, Y.; Moriyoshi, C.; Bakr, O. M.; Zhang, J.; Sun, H.-T. High-Efficiency Violet-Emitting All-Inorganic Perovskite Nanocrystals Enabled by Alkaline-Earth Metal Passivation. *Chem. Mater.* **2019**, *31* (11), 3974–3983. https://doi.org/10.1021/acs.chemmater.9b00442.
(75) De, A.; Das, S.; Mondal, N.; Samanta, A. Highly Luminescent Violet- and Blue-Emitting Stable Perovskite Nanocrystals. *ACS Mater. Lett.* **2019**, *1* (1), 116–122. https://doi.org/10.1021/acsmaterialslett.9b00101.
(76) Dutta, A.; Behera, R. K.; Pal, P.; Baitalik, S.; Pradhan, N. Near-Unity Photoluminescence Quantum Efficiency for All CsPbX3 (X=Cl, Br, and I) Perovskite Nanocrystals: A Generic Synthesis Approach. *Angew. Chem. Int. Ed.* **2019**, *58* (17), 5552–5556. https://doi.org/10.1002/anie.201900374.
(77) Lai, R.; Wu, K. Picosecond Electron Trapping Limits the Emissivity of CsPbCl3 Perovskite Nanocrystals. *J. Chem. Phys.* **2019**, *151* (19), 194701. https://doi.org/10.1063/1.5127887.
(78) Chen, Y.-C.; Chou, H.-L.; Lin, J.-C.; Lee, Y.-C.; Pao, C.-W.; Chen, J.-L.; Chang, C.-C.; Chi, R.-Y.; Kuo, T.-R.; Lu, C.-W.; Wang, D.-Y. Enhanced Luminescence and Stability of Cesium Lead Halide Perovskite CsPbX3 Nanocrystals by Cu2+-Assisted Anion Exchange

Reactions. *J. Phys. Chem. C* **2019**, *123* (4), 2353–2360. https://doi.org/10.1021/acs.jpcc.8b11535.
(79) Das, S.; De, A.; Samanta, A. Ambient Condition Mg2+ Doping Producing Highly Luminescent Green- and Violet-Emitting Perovskite Nanocrystals with Reduced Toxicity and Enhanced Stability. *J. Phys. Chem. Lett.* **2020**, *11* (3), 1178–1188. https://doi.org/10.1021/acs.jpclett.9b03831.
(80) Bai, L.; Wang, S.; Zhang, Y.; Zhang, K.; Li, H.; Ou, K.; Yi, L. Investigation on Violet/Blue All-Inorganic Light-Emitting Diodes Based on CsPbCl3 Films. *J. Lumin.* **2020**, *226*, 117422. https://doi.org/10.1016/j.jlumin.2020.117422.
(81) Naresh, V.; Lee, N. Zn(II)-Doped Cesium Lead Halide Perovskite Nanocrystals with High Quantum Yield and Wide Color Tunability for Color-Conversion Light-Emitting Displays. *ACS Appl. Nano Mater.* **2020**, *3* (8), 7621–7632. https://doi.org/10.1021/acsanm.0c01254.
(82) Zhang, C.; Wan, Q.; Ono, L. K.; Liu, Y.; Zheng, W.; Zhang, Q.; Liu, M.; Kong, L.; Li, L.; Qi, Y. Narrow-Band Violet-Light-Emitting Diodes Based on Stable Cesium Lead Chloride Perovskite Nanocrystals. *ACS Energy Lett.* **2021**, *6* (10), 3545–3554. https://doi.org/10.1021/acsenergylett.1c01380.
(83) Paul, S.; Ahmed, T.; Das, S.; Samanta, A. Effect of Lead:Halide Precursor Ratio on the Photoluminescence and Carrier Dynamics of Violet- and Blue-Emitting Lead Halide Perovskite Nanocrystals. *J. Phys. Chem. C* **2021**, *125* (42), 23539–23547. https://doi.org/10.1021/acs.jpcc.1c07740.
(84) Hu, Q.; Guo, J.; Lu, M.; Lu, P.; Zhang, Y.; Yu, W. W.; Bai, X. Efficient and Stable Mg2+-Doped CsPbCl3 Nanocrystals for Violet LEDs. *J. Phys. Chem. Lett.* **2021**, *12* (34), 8203–8211. https://doi.org/10.1021/acs.jpclett.1c02416.
(85) Das, S.; Samanta, A. On Direct Synthesis of High Quality APbX3 (A = Cs+, MA+ and FA+; X = Cl−, Br− and I−) Nanocrystals Following a Generic Approach. *Nanoscale* **2022**, *14* (26), 9349–9358. https://doi.org/10.1039/D2NR01305C.
(86) Ji, Y.; Wang, M.; Yang, Z.; Wang, H.; Amin Padhiar, M.; Qiu, H.; Dang, J.; Miao, Y.; Zhou, Y.; Saleem Bhatti, A. Strong Violet Emission from Ultra-Stable Strontium-Doped CsPbCl 3 Superlattices. *Nanoscale* **2022**, *14* (6), 2359–2366. https://doi.org/10.1039/D1NR07848H.
(87) Lin, M.; Zhang, X.; Dong, Y.; Xu, S.; Li, X.; Yu, H.; Sun, J.; Cheng, L.; Chen, B. Controlled Growth and Spectroscopy Characterization of Blue Violet Perovskite Quantum Dots in Borate Glasses. *Ceram. Int.* **2023**, *49* (12), 20281–20289. https://doi.org/10.1016/j.ceramint.2023.03.151.
(88) Das, S.; Hossain, M.; Samanta, A. Stable and Intense Violet-Emitting CsPbCl3 Nanocrystals for Light-Emitting Diodes: Directly Obtained by L-Type Surface Passivation. *ACS Appl. Nano Mater.* **2023**, *6* (6), 4812–4820. https://doi.org/10.1021/acsanm.3c00371.
(89) Xie, C.; Zhang, X.; Chen, H. S.; Yang, P. Synthesis-Kinetics of Violet- and Blue-Emitting Perovskite Nanocrystals with High Brightness and Superior Stability toward Flexible Conversion Layer. *Small* **2024**, *20* (19), 2308896. https://doi.org/10.1002/smll.202308896.
(90) Wang, Y.; Zhou, D.; Sun, R.; Wang, Y.; Wang, T.; Li, W.; Song, R.; Bai, X.; Xu, W.; Song, H. Folic Acid Passivated Efficient Cerium Doped Perovskite Nanocrystals for High-Performance Light-Emitting Diodes. *Adv. Opt. Mater.* **2024**, *12* (20), 2400338. https://doi.org/10.1002/adom.202400338.

(91) Zhang, W.; Wang, S.; Wen, Z.; Zhao, Q.; Yi, L. Effect of Mg0.71Zn0.29O Electron Transport Layer on Violet/Blue Emission of CsPbCl3 LED. *J. Photochem. Photobiol. Chem.* **2025**, *463*, 116309. https://doi.org/10.1016/j.jphotochem.2025.116309.
(92) Sadhanala, A.; Ahmad, S.; Zhao, B.; Giesbrecht, N.; Pearce, P. M.; Deschler, F.; Hoye, R. L. Z.; Gödel, K. C.; Bein, T.; Docampo, P.; Dutton, S. E.; De Volder, M. F. L.; Friend, R. H. Blue-Green Color Tunable Solution Processable Organolead Chloride–Bromide Mixed Halide Perovskites for Optoelectronic Applications. *Nano Lett.* **2015**, *15* (9), 6095–6101. https://doi.org/10.1021/acs.nanolett.5b02369.
(93) Sun, K. W.; Shellaiah, M. Deep-Blue Hybrid Perovskite Light Emitting Diode with High Color Purity Based on CH3NH3PbCl3. *Org. Electron.* **2023**, *119*, 106829. https://doi.org/10.1016/j.orgel.2023.106829.
(94) Hu, M.; Murrietta, N.; Lyu, J.; Congreve, D. N. Two-Dimensional Perovskite Materials towards Violet and Ultraviolet Light-Emitting Diodes. In *UV and Higher Energy Photonics: From Materials to Applications 2023*; SPIE, 2023; Vol. 12652, pp 29–35. https://doi.org/10.1117/12.2676613.
(95) Leng, M.; Yang, Y.; Chen, Z.; Gao, W.; Zhang, J.; Niu, G.; Li, D.; Song, H.; Zhang, J.; Jin, S.; Tang, J. Surface Passivation of Bismuth-Based Perovskite Variant Quantum Dots To Achieve Efficient Blue Emission. *Nano Lett.* **2018**, *18* (9), 6076–6083. https://doi.org/10.1021/acs.nanolett.8b03090.
(96) Wu, L.; Pan, Y.; Zhang, Y.; Geng, Y.; Cao, J.; Su, X.; Xu, J.; Xie, H.; Gao, D. Violet Perovskite Quantum Dots of MA3Bi2Br9 and MA3Bi2Br6Cl3 Synthesized by the Cb-LARP Method with Tunable Emission Wavelengths in Range of 379–400 Nm. *Adv. Funct. Mater.* **2025**, *35* (7), 2415315. https://doi.org/10.1002/adfm.202415315.
(97) Shen, Y.; Yin, J.; Cai, B.; Wang, Z.; Dong, Y.; Xu, X.; Zeng, H. Lead-Free, Stable, High-Efficiency (52%) Blue Luminescent FA3Bi2Br9 Perovskite Quantum Dots. *Nanoscale Horiz.* **2020**, *5* (3), 580–585. https://doi.org/10.1039/C9NH00685K.
(98) Zhuravleva, M.; Yang, K.; Rothfuss, H.; Melcher, C. L. Crystal Growth and Scintillation Properties of Cs3CeX6 and CsCe2X7 (X = Cl, Br). In *IEEE Nuclear Science Symposium & Medical Imaging Conference*; 2010; pp 1296–1299. https://doi.org/10.1109/NSSMIC.2010.5873977.
(99) Guo, Q.; Wang, L.; Yang, L.; Duan, J.; Du, H.; Ji, G.; Liu, N.; Zhao, X.; Chen, C.; Xu, L.; Gao, L.; Luo, J.; Tang, J. Spectra Stable Deep-Blue Light-Emitting Diodes Based on Cryolite-like Cerium(III) Halides with Nanosecond d-f Emission. *Sci. Adv.* **2022**, *8* (50), eabq2148. https://doi.org/10.1126/sciadv.abq2148.
(100) Wang, Q.; Bai, T.; Ji, S.; Zhao, H.; Meng, X.; Zhang, R.; Jiang, J.; Liu, F. Ultraviolet Emission from Cerium-Based Organic-Inorganic Hybrid Halides and Their Abnormal Anti-Thermal Quenching Behavior. *Adv. Funct. Mater.* **2023**, *33* (34), 2303399. https://doi.org/10.1002/adfm.202303399.
(101) Yang, B.; Yin, L.; Niu, G.; Yuan, J.-H.; Xue, K.-H.; Tan, Z.; Miao, X.-S.; Niu, M.; Du, X.; Song, H.; Lifshitz, E.; Tang, J. Lead-Free Halide Rb2CuBr3 as Sensitive X-Ray Scintillator. *Adv. Mater.* **2019**, *31* (44), 1904711. https://doi.org/10.1002/adma.201904711.
(102) Zhao, X.; Niu, G.; Zhu, J.; Yang, B.; Yuan, J.-H.; Li, S.; Gao, W.; Hu, Q.; Yin, L.; Xue, K.-H.; Lifshitz, E.; Miao, X.; Tang, J. All-Inorganic Copper Halide as a Stable and Self-Absorption-Free X-Ray Scintillator. *J. Phys. Chem. Lett.* **2020**, *11* (5), 1873–1880. https://doi.org/10.1021/acs.jpclett.0c00161.

(103) Gao, W.; Niu, G.; Yin, L.; Yang, B.; Yuan, J.-H.; Zhang, D.; Xue, K.-H.; Miao, X.; Hu, Q.; Du, X.; Tang, J. One-Dimensional All-Inorganic $K_2$ $CuBr_3$ with Violet Emission as Efficient X-Ray Scintillators. *ACS Appl. Electron. Mater.* **2020**, *2* (7), 2242–2249. https://doi.org/10.1021/acsaelm.0c00414.
(104) Gao, W.; Yin, L.; Yuan, J.-H.; Xue, K.-H.; Niu, G.; Yang, B.; Hu, Q.; Liu, X.; Tang, J. Lead-Free Violet-Emitting K2CuCl3 Single Crystal with High Photoluminescence Quantum Yield. *Org. Electron.* **2020**, *86*, 105903. https://doi.org/10.1016/j.orgel.2020.105903.
(105) Shi, Y.; Liang, D.; Mo, Q.; Lu, S.; Sun, Z.; Xiao, H.; Qian, Q.; Zang, Z. Highly Efficient Copper-Based Halide Single Crystals with Violet Emission for Visible Light Communication. *Chem. Commun.* **2023**, *59* (5), 583–586. https://doi.org/10.1039/D2CC05965G.
(106) Liu, M.; Matta, S. K.; Ali-Löytty, H.; Matuhina, A.; Grandhi, G. K.; Lahtonen, K.; Russo, S. P.; Vivo, P. Moisture-Assisted near-UV Emission Enhancement of Lead-Free Cs4CuIn2Cl12 Double Perovskite Nanocrystals. *Nano Lett.* **2022**, *22* (1), 311–318. https://doi.org/10.1021/acs.nanolett.1c03822.
(107) Ma, Z.; Shi, Z.; Yang, D.; Zhang, F.; Li, S.; Wang, L.; Wu, D.; Zhang, Y.; Na, G.; Zhang, L.; Li, X.; Zhang, Y.; Shan, C. Electrically-Driven Violet Light-Emitting Devices Based on Highly Stable Lead-Free Perovskite $Cs_3$ $Sb_2$ $Br_9$ Quantum Dots. *ACS Energy Lett.* **2020**, *5* (2), 385–394. https://doi.org/10.1021/acsenergylett.9b02096.
(108) Yang, L.; Du, H.; Li, J.; Luo, Y.; Lin, X.; Pang, J.; Liu, Y.; Gao, L.; He, S.; Kang, J.-W.; Liang, W.; Song, H.; Luo, J.; Tang, J. Efficient Deep-Blue Electroluminescence from Ce-Based Metal Halide. *Nat. Commun.* **2024**, *15* (1), 6240. https://doi.org/10.1038/s41467-024-50508-5.
(109) Samanta, T.; Han, J. H.; Lee, H. U.; Cha, B. K.; Park, Y. M.; Viswanath, N. S. M.; Cho, H. B.; Kim, H. W.; Cho, S. B.; Im, W. B. Large-Scale Mechanochemical Synthesis of Cesium Lanthanide Chloride for Radioluminescence. *Inorg. Chem.* **2024**, *63* (35), 16483–16490. https://doi.org/10.1021/acs.inorgchem.4c02766.
(110) Dutta, S.; Yoo, J. H.; Kwon, S. B.; Dastgeer, G.; Yoon, D. H. Harnessing Dual Violet Emission in Cerium-Based Perovskite Derivatives for Solution-Processed Next-Generation Lighting. *ACS Appl. Opt. Mater.* **2025**, *3* (5), 1070–1077. https://doi.org/10.1021/acsaom.5c00027.
(111) Zhu, D.; Qin, W.; Chen, X.; Wang, M.; Zhu, J.; Jia, M.; Han, Y.; Lian, L.; Ma, Z.; Zhang, J.; Liu, Y.; Li, X.; Xu, W.; Shan, C.; Shi, Z. Environmentally Friendly Synthesis of Near-Unity UV/Violet-Emitting Cerium-Based Metal Halides with Reversible Structural Switching for Smart Anticounterfeiting. *Nano Lett.* **2025**. https://doi.org/10.1021/acs.nanolett.5c04689.
(112) Sun, H.; Yang, X.; Li, P.; Bai, Y.; Meng, Q.; Zhao, H.; Wang, Q.; Wen, Z.; Huang, L.; Huang, D.; Yu, W. W.; Chen, H.; Liu, F. Solution Synthesis and Light-Emitting Applications of One-Dimensional Lead-Free Cerium(III) Metal Halides. *Nano Lett.* **2024**, *24* (33), 10355–10361. https://doi.org/10.1021/acs.nanolett.4c03019.
(113) Du, H.; Yang, L.; Pang, J.; Shen, Z.; Li, J.; Dong, X.; Luo, Y.; Luo, J.; Tang, J. Vacuum-Deposited $Rb_3CeI_6$ for Deep-Blue-Light-Emitting Diodes. *Opt. Lett. Vol 48 Issue 11 Pp 2777-2780* **2023**. https://doi.org/10.1364/OL.486168.
(114) Dutta, S.; Yoo, J. H.; Kwon, S. B.; Panchanan, S.; Yoo, H. C.; Yoon, D. H. Green Synthesis of Ce Doped Cs3MnBr5 for Highly Stable Violet Light Emitting Diodes.

*Electron. Mater. Lett.* **2023**, *19* (6), 518–526. https://doi.org/10.1007/s13391-023-00411-w.
(115) Wen, Z.; Bai, Y.; Meng, Q.; Zhao, H.; Wang, Q.; Sun, H.; Huang, L.; Huang, D.; Yu, W. W.; Zhu, J.; Liu, F. Optimizing Energy Transfer: Suppressing Cs2ZnCl4 Self-Trapped States and Boosting Ce3+ Ion Luminescence Efficiency. *Laser Photonics Rev.* **2024**, *18* (12), 2400525. https://doi.org/10.1002/lpor.202400525.
(116) Zhu, D.; Chen, X.; Wang, M.; Li, G.; Pan, G.; Jia, M.; Han, Y.; Liu, Y.; Yang, D.; Xu, W.; Li, X.; Shan, C.; Shi, Z. Efficient UV Emission with Anti-Thermal Quenching from Ce3+-Activated 0D All-Inorganic Zinc Halides. *Adv. Opt. Mater.* **2025**, *13* (17), 2500227. https://doi.org/10.1002/adom.202500227.
(117) Saghy, P.; Brown, A. M.; Chu, C.; Dube, L. C.; Zheng, W.; Robinson, J. R.; Chen, O. Lanthanide Double Perovskite Nanocrystals with Emissions Covering the UV-C to NIR Spectral Range. *Adv. Opt. Mater.* **2023**, *11* (12), 2300277. https://doi.org/10.1002/adom.202300277.
(118) Kitazawa, N. Excitons in Two-Dimensional Layered Perovskite Compounds: (C6H5C2H4NH3)2Pb(Br,I)4 and (C6H5C2H4NH3)2Pb(Cl,Br)4. *Mater. Sci. Eng. B* **1997**, *49* (3), 233–238. https://doi.org/10.1016/S0921-5107(97)00132-3.
(119) Kaewurai, P.; Ponchai, J.; Amratisha, K.; Naikaew, A.; Swe, K. Z.; Pinsuwan, K.; Boonthum, C.; Sahasithiwat, S.; Kanjanaboos, P. Enhancing Violet Photoluminescence of 2D Perovskite Thin Films via Swift Cation Doping and Grain Size Reduction. *Appl. Phys. Express* **2018**, *12* (1), 015506. https://doi.org/10.7567/1882-0786/aaf286.
(120) Jang, G.; Han, H.; Ma, S.; Lee, J.; Lee, C. U.; Jeong, W.; Son, J.; Cho, D.; Kim, J.-H.; Park, C.; Moon, J. Rapid Crystallization-Driven High-Efficiency Phase-Pure Deep-Blue Ruddlesden–Popper Perovskite Light-Emitting Diodes. *Adv. Photonics* **2023**, *5* (1), 016001. https://doi.org/10.1117/1.AP.5.1.016001.
(121) Hu, M.; Lyu, J.; Murrietta, N.; Fernández, S.; Michaels, W.; Zhou, Q.; Narayanan, P.; Congreve, D. N. 2D Mixed Halide Perovskites for Ultraviolet Light-Emitting Diodes. *Device* **2024**, *2* (11). https://doi.org/10.1016/j.device.2024.100511.
(122) Ni, J.-S.; Vasudevan, T.; Chen, L.-C. Enhanced Short-Wavelength Perovskite UVLEDs: Fabrication and Optimization of (PEA)2PbBr3Cl Devices with Poly-TPD Hole Transport Layers. *ACS Appl. Electron. Mater.* **2025**. https://doi.org/10.1021/acsaelm.5c01267.
(123) Fernández, S.; Mbachu, D.; Hu, M.; Cui, H.; Michaels, W.; Narayanan, P.; Colenbrander, T. K.; Zhou, Q.; Lin, D.; Lecina, O. S.; Hong, G.; Congreve, D. N. Lead-Free Europium Halide Perovskite Nanoplatelets. arXiv November 14, 2025. https://doi.org/10.48550/arXiv.2511.10873.
(124) Kumar, S.; Jagielski, J.; Yakunin, S.; Rice, P.; Chiu, Y.-C.; Wang, M.; Nedelcu, G.; Kim, Y.; Lin, S.; Santos, E. J. G.; Kovalenko, M. V.; Shih, C.-J. Efficient Blue Electroluminescence Using Quantum-Confined Two-Dimensional Perovskites. *ACS Nano* **2016**, *10* (10), 9720–9729. https://doi.org/10.1021/acsnano.6b05775.
(125) Dou, L.; Wong, A. B.; Yu, Y.; Lai, M.; Kornienko, N.; Eaton, S. W.; Fu, A.; Bischak, C. G.; Ma, J.; Ding, T.; Ginsberg, N. S.; Wang, L.-W.; Alivisatos, A. P.; Yang, P. Atomically Thin Two-Dimensional Organic-Inorganic Hybrid Perovskites. *Science* **2015**, *349* (6255), 1518–1521. https://doi.org/10.1126/science.aac7660.
(126) Sun, K. W. Centimeter-Scale Violet Light Emitting Diode with Two-Dimensional BA2PbBr4 Perovskite Emitter. *J. Electrochem. Soc.* **2023**, *170* (6), 065501. https://doi.org/10.1149/1945-7111/acd811.

(127) Kucheriv, O. I.; Sirenko, V. Y.; Shova, S.; Gural'skiy, I. A. 2D Hybrid Organic-Inorganic Perovskite Displaying Narrow-Band Violet-Blue Photoluminescence. *J. Lumin.* **2024**, *275*, 120753. https://doi.org/10.1016/j.jlumin.2024.120753.
(128) Cheng, T.; Xie, Y.; Lin, Y.; Dong, Y.; Lan, Y.; Chen, R.; Li, J.; Cui, B.-B. Narrow-Band Violet Light-Emitting Diodes Based on One-Dimensional Lead Bromides. *J. Lumin.* **2023**, *260*, 119872. https://doi.org/10.1016/j.jlumin.2023.119872.
(129) Sun, C.; Jiang, K.; Han, M.-F.; Liu, M.-J.; Lian, X.-K.; Jiang, Y.-X.; Shi, H.-S.; Yue, C.-Y.; Lei, X.-W. A Zero-Dimensional Hybrid Lead Perovskite with Highly Efficient Blue-Violet Light Emission. *J. Mater. Chem. C* **2020**, *8* (34), 11890–11895. https://doi.org/10.1039/D0TC02351E.
(130) Roccanova, R.; Ming, W.; Whiteside, V. R.; McGuire, M. A.; Sellers, I. R.; Du, M.-H.; Saparov, B. Synthesis, Crystal and Electronic Structures, and Optical Properties of (CH3NH3)2CdX4 (X = Cl, Br, I). *Inorg. Chem.* **2017**, *56* (22), 13878–13888. https://doi.org/10.1021/acs.inorgchem.7b01986.
(131) Zhou, B.; Yan, D. Simultaneous Long-Persistent Blue Luminescence and High Quantum Yield within 2D Organic–Metal Halide Perovskite Micro/Nanosheets. *Angew. Chem. Int. Ed.* **2019**, *58* (42), 15128–15135. https://doi.org/10.1002/anie.201909760.
(132) Zhang, C.; Feng, X.; Song, Q.; Zhou, C.; Peng, L.; Chen, J.; Liu, X.; Chen, H.; Lin, J.; Chen, X. Blue-Violet Emission with Near-Unity Photoluminescence Quantum Yield from Cu(I)-Doped Rb3InCl6 Single Crystals. *J. Phys. Chem. Lett.* **2021**, *12* (33), 7928–7934. https://doi.org/10.1021/acs.jpclett.1c01751.
(133) Yuan, Z.; Shu, Y.; Tian, Y.; Xin, Y.; Ma, B. A Facile One-Pot Synthesis of Deep Blue Luminescent Lead Bromide Perovskite Microdisks. *Chem. Commun.* **2015**, *51* (91), 16385–16388. https://doi.org/10.1039/C5CC06750B.
(134) Yang, B.; Mao, X.; Hong, F.; Meng, W.; Tang, Y.; Xia, X.; Yang, S.; Deng, W.; Han, K. Lead-Free Direct Band Gap Double-Perovskite Nanocrystals with Bright Dual-Color Emission. *J. Am. Chem. Soc.* **2018**, *140* (49), 17001–17006. https://doi.org/10.1021/jacs.8b07424.
(135) Yamanoi, K.; Nishi, R.; Takeda, K.; Shinzato, Y.; Tsuboi, M.; Luong, M. V.; Nakazato, T.; Shimizu, T.; Sarukura, N.; Cadatal-Raduban, M.; Pham, M. H.; Nguyen, H. D.; Kurosawa, S.; Yokota, Y.; Yoshikawa, A.; Togashi, T.; Nagasono, M.; Ishikawa, T. Perovskite Fluoride Crystals as Light Emitting Materials in Vacuum Ultraviolet Region. *Opt. Mater.* **2014**, *36* (4), 769–772. https://doi.org/10.1016/j.optmat.2013.11.023.
(136) Fernández, S.; Michaels, W.; Hu, M.; Narayanan, P.; Murrietta, N.; Gallegos, A. O.; Ahmed, G. H.; Lyu, J.; Gangishetty, M. K.; Congreve, D. N. Trade-off between Efficiency and Stability in Mn2+-Doped Perovskite Light-Emitting Diodes. *Device* **2023**, *1* (2). https://doi.org/10.1016/j.device.2023.100017.
(137) Ahmed, G. H.; Liu, Y.; Bravić, I.; Ng, X.; Heckelmann, I.; Narayanan, P.; Fernández, M. S.; Monserrat, B.; Congreve, D. N.; Feldmann, S. Luminescence Enhancement Due to Symmetry Breaking in Doped Halide Perovskite Nanocrystals. *J. Am. Chem. Soc.* **2022**, *144* (34), 15862–15870. https://doi.org/10.1021/jacs.2c07111.
(138) Cui, X.; Li, Y.; Chen, Z.; Zou, Y.; Liu, Y.; Sun, B.; Bo, Z. Metal Oxide Charge Transport Layer Targeting Efficient and Stable Perovskite Light-Emitting Diodes. *J. Alloys Compd.* **2023**, *960*, 170823. https://doi.org/10.1016/j.jallcom.2023.170823.
(139) Shi, Y.; Wu, W.; Dong, H.; Li, G.; Xi, K.; Divitini, G.; Ran, C.; Yuan, F.; Zhang, M.; Jiao, B.; Hou, X.; Wu, Z. A Strategy for Architecture Design of Crystalline Perovskite Light-Emitting Diodes with High Performance. *Adv. Mater.* **2018**, *30* (25), 1800251. https://doi.org/10.1002/adma.201800251.
(140) Lin, K.; Xing, J.; Quan, L. N.; de Arquer, F. P. G.; Gong, X.; Lu, J.; Xie, L.; Zhao, W.; Zhang, D.; Yan, C.; Li, W.; Liu, X.; Lu, Y.; Kirman, J.; Sargent, E. H.; Xiong, Q.; Wei, Z. Perovskite Light-Emitting Diodes with External Quantum Efficiency Exceeding 20 per Cent. *Nature* **2018**, *562* (7726), 245–248. https://doi.org/10.1038/s41586-018-0575-3.
(141) Sun, Y.; Chen, S.; Huang, J.-Y.; Wu, Y.-R.; Greenham, N. C. Device Physics of Perovskite Light-Emitting Diodes. *Appl. Phys. Rev.* **2024**, *11* (4), 041418. https://doi.org/10.1063/5.0228117.
(142) Zhao, B.; Bai, S.; Kim, V.; Lamboll, R.; Shivanna, R.; Auras, F.; Richter, J. M.; Yang, L.; Dai, L.; Alsari, M.; She, X.-J.; Liang, L.; Zhang, J.; Lilliu, S.; Gao, P.; Snaith, H. J.; Wang, J.; Greenham, N. C.; Friend, R. H.; Di, D. High-Efficiency Perovskite–Polymer Bulk Heterostructure Light-Emitting Diodes. *Nat. Photonics* **2018**, *12* (12), 783–789. https://doi.org/10.1038/s41566-018-0283-4.
(143) Lee, S.; Kim, D. B.; Hamilton, I.; Daboczi, M.; Nam, Y. S.; Lee, B. R.; Zhao, B.; Jang, C. H.; Friend, R. H.; Kim, J.-S.; Song, M. H. Control of Interface Defects for Efficient and Stable Quasi-2D Perovskite Light-Emitting Diodes Using Nickel Oxide Hole Injection Layer. *Adv. Sci.* **2018**, *5* (11), 1801350. https://doi.org/10.1002/advs.201801350.
(144) Huang, C.-Y.; Chang, S.-P.; Ansay, A. G.; Wang, Z.-H.; Yang, C.-C. Ambient-Processed, Additive-Assisted CsPbBr3 Perovskite Light-Emitting Diodes with Colloidal NiOx Nanoparticles for Efficient Hole Transporting. *Coatings* **2020**, *10* (4), 336. https://doi.org/10.3390/coatings10040336.
(145) Chen, R.; Chen, H.; Yu, Y.; Luo, C.; Xu, C.; Zhou, X.; Ren, Z.; Chen, Y. Sufficient Hole Injection for High-Performance Blue Perovskite Light-Emitting Diodes. *ACS Photonics* **2024**, *11* (12), 5380–5387. https://doi.org/10.1021/acsphotonics.4c01767.
(146) Zhao, L.; Roh, K.; Kacmoli, S.; Kurdi, K. A.; Jhulki, S.; Barlow, S.; Marder, S. R.; Gmachl, C.; Rand, B. P. Thermal Management Enables Bright and Stable Perovskite Light-Emitting Diodes. *Adv. Mater.* **2020**, *32* (25), 2000752. https://doi.org/10.1002/adma.202000752.
(147) Zhao, L.; Lee, K. M.; Roh, K.; Khan, S. U. Z.; Rand, B. P. Improved Outcoupling Efficiency and Stability of Perovskite Light-Emitting Diodes Using Thin Emitting Layers. *Adv. Mater.* **2019**, *31* (2), 1805836. https://doi.org/10.1002/adma.201805836.
(148) Subramanian, A.; Pan, Z.; Zhang, Z.; Ahmad, I.; Chen, J.; Liu, M.; Cheng, S.; Xu, Y.; Wu, J.; Lei, W.; Khan, Q.; Zhang, Y. Interfacial Energy-Level Alignment for High-Performance All-Inorganic Perovskite CsPbBr3 Quantum Dot-Based Inverted Light-Emitting Diodes. *ACS Appl. Mater. Interfaces* **2018**, *10* (15), 13236–13243. https://doi.org/10.1021/acsami.8b01684.
(149) Hsu, C.; Tian, S.; Lian, Y.; Zhang, G.; Zhou, Q.; Cao, X.; Zhao, B.; Di, D. Efficient Mini/Micro-Perovskite Light-Emitting Diodes. *Cell Rep. Phys. Sci.* **2021**, *2* (9). https://doi.org/10.1016/j.xcrp.2021.100582.
(150) Hoye, R. L. Z.; Lai, M.-L.; Anaya, M.; Tong, Y.; Gałkowski, K.; Doherty, T.; Li, W.; Huq, T. N.; Mackowski, S.; Polavarapu, L.; Feldmann, J.; MacManus-Driscoll, J. L.; Friend, R. H.; Urban, A. S.; Stranks, S. D. Identifying and Reducing Interfacial Losses to Enhance Color-Pure Electroluminescence in Blue-Emitting Perovskite Nanoplatelet Light-Emitting Diodes. *ACS Energy Lett.* **2019**, *4* (5), 1181–1188. https://doi.org/10.1021/acsenergylett.9b00571.

(151) Chen, Z.; Li, Z.; Hopper, T. R.; Bakulin, A. A.; Yip, H.-L. Materials, Photophysics and Device Engineering of Perovskite Light-Emitting Diodes. *Rep. Prog. Phys.* **2021**, *84* (4), 046401. https://doi.org/10.1088/1361-6633/abefba.
(152) Yao, Z.; Bi, C.; Liu, A.; Zhang, M.; Tian, J. High Brightness and Stability Pure-Blue Perovskite Light-Emitting Diodes Based on a Novel Structural Quantum-Dot Film. *Nano Energy* **2022**, *95*, 106974. https://doi.org/10.1016/j.nanoen.2022.106974.
(153) Wang, H.; Zhang, B.; Wang, B.; Bai, S.; Cheng, L.; Hu, Y.; Lu, S. Efficient Quasi-2D Perovskite Based Blue Light-Emitting Diodes with Carbon Dots Modified Hole Transport Layer. *Nano Lett.* **2024**, *24* (28), 8702–8708. https://doi.org/10.1021/acs.nanolett.4c02110.
(154) Zhang, L.; Yuan, F.; Xi, J.; Jiao, B.; Dong, H.; Li, J.; Wu, Z. Suppressing Ion Migration Enables Stable Perovskite Light-Emitting Diodes with All-Inorganic Strategy. *Adv. Funct. Mater.* **2020**, *30* (40), 2001834. https://doi.org/10.1002/adfm.202001834.
(155) Yang, L.; Luo, J.; Gao, L.; Song, B.; Tang, J. Inorganic Lanthanide Compounds with f–d Transition: From Materials to Electroluminescence Devices. *J. Phys. Chem. Lett.* **2022**, *13* (19), 4365–4373. https://doi.org/10.1021/acs.jpclett.2c00927.
(156) Luo, J.; Yang, L.; Tan, Z.; Xie, W.; Sun, Q.; Li, J.; Du, P.; Xiao, Q.; Wang, L.; Zhao, X.; Niu, G.; Gao, L.; Jin, S.; Tang, J. Efficient Blue Light Emitting Diodes Based On Europium Halide Perovskites. *Adv. Mater.* **2021**, *33* (38), 2101903. https://doi.org/10.1002/adma.202101903.
(157) Miura, N.; Kawanishi, M.; Matsumoto, H.; Nakano, R. High-Luminance Blue-Emitting BaAl2S4:Eu Thin-Film Electroluminescent Devices. *Jpn. J. Appl. Phys.* **1999**, *38* (11B), L1291. https://doi.org/10.1143/JJAP.38.L1291.
(158) Leising, Gün.; Tasch, S.; Brandstatter, C.; Meghdadi, F.; Froyer, G.; Athouel, L. Red–Green–Blue Light Emission from a Thin Film Electroluminescence Device Based on Parahexaphenyl. *Adv. Mater.* **1997**, *9* (1), 33–36. https://doi.org/10.1002/adma.19970090105.
(159) Ando, M.; Ono, Y. A. Role of Eu2+ Luminescent Centers in the Electro-optical Characteristics of Red-emitting CaS:Eu Thin-film Electroluminescent Devices with Memory. *J. Appl. Phys.* **1990**, *68* (7), 3578–3583. https://doi.org/10.1063/1.346317.
(160) Haecke, J. E. V.; Smet, P. F.; Poelman, D. The Formation of Eu2 + Clusters in Saturated Red Ca0.5Sr0.5S : Eu Electroluminescent Devices. *J. Electrochem. Soc.* **2005**, *152* (12), H225. https://doi.org/10.1149/1.2118207.
(161) Zhang, W.; Wang, S.; Yi, L. Optimization of CsPbCl3 Violet/Blue All-Inorganic Light-Emitting Diodes Devices. *Mater. Sci. Semicond. Process.* **2025**, *192*, 109403. https://doi.org/10.1016/j.mssp.2025.109403.
(162) Shah, K. S.; Glodo, J.; Higgins, W.; van Loef, E. V. D.; Moses, W. W.; Derenzo, S. E.; Weber, M. J. CeBr/Sub 3/ Scintillators for Gamma-Ray Spectroscopy. In *IEEE Symposium Conference Record Nuclear Science 2004.*; 2004; Vol. 7, pp 4278–4281. https://doi.org/10.1109/NSSMIC.2004.1466835.
(163) Xing, G.; Wu, B.; Wu, X.; Li, M.; Du, B.; Wei, Q.; Guo, J.; Yeow, E. K. L.; Sum, T. C.; Huang, W. Transcending the Slow Bimolecular Recombination in Lead-Halide Perovskites for Electroluminescence. *Nat. Commun.* **2017**, *8* (1), 14558. https://doi.org/10.1038/ncomms14558.
(164) Wang, L.; Zhao, Z.; Zhan, G.; Fang, H.; Yang, H.; Huang, T.; Zhang, Y.; Jiang, N.; Duan, L.; Liu, Z.; Bian, Z.; Lu, Z.; Huang, C. Deep-Blue Organic Light-Emitting Diodes Based on a Doublet d–f Transition Cerium(III) Complex with 100% Exciton Utilization Efficiency. *Light Sci. Appl.* **2020**, *9* (1), 157. https://doi.org/10.1038/s41377-020-00395-4.
(165) Zou, W.; Li, R.; Zhang, S.; Liu, Y.; Wang, N.; Cao, Y.; Miao, Y.; Xu, M.; Guo, Q.; Di, D.; Zhang, L.; Yi, C.; Gao, F.; Friend, R. H.; Wang, J.; Huang, W. Minimising Efficiency Roll-off in High-Brightness Perovskite Light-Emitting Diodes. *Nat. Commun.* **2018**, *9* (1), 608. https://doi.org/10.1038/s41467-018-03049-7.
(166) Shen, G.; Zhang, Y.; Juarez, J.; Contreras, H.; Sindt, C.; Xu, Y.; Kline, J.; Barlow, S.; Reichmanis, E.; Marder, S. R.; Ginger, D. S. Increased Brightness and Reduced Efficiency Droop in Perovskite Quantum Dot Light-Emitting Diodes Using Carbazole-Based Phosphonic Acid Interface Modifiers. *ACS Nano* **2025**, *19* (1), 1116–1127. https://doi.org/10.1021/acsnano.4c13036.
(167) Fakharuddin, A.; Qiu, W.; Croes, G.; Devižis, A.; Gegevičius, R.; Vakhnin, A.; Rolin, C.; Genoe, J.; Gehlhaar, R.; Kadashchuk, A.; Gulbinas, V.; Heremans, P. Reduced Efficiency Roll-Off and Improved Stability of Mixed 2D/3D Perovskite Light Emitting Diodes by Balancing Charge Injection. *Adv. Funct. Mater.* **2019**, *29* (37), 1904101. https://doi.org/10.1002/adfm.201904101.
(168) Wang, Y.; Teng, Y.; Lu, P.; Shen, X.; Jia, P.; Lu, M.; Shi, Z.; Dong, B.; Yu, W. W.; Zhang, Y. Low Roll-Off Perovskite Quantum Dot Light-Emitting Diodes Achieved by Augmenting Hole Mobility. *Adv. Funct. Mater.* **2020**, *30* (19), 1910140. https://doi.org/10.1002/adfm.201910140.
(169) Kim, H.; Zhao, L.; Price, J. S.; Grede, A. J.; Roh, K.; Brigeman, A. N.; Lopez, M.; Rand, B. P.; Giebink, N. C. Hybrid Perovskite Light Emitting Diodes under Intense Electrical Excitation. *Nat. Commun.* **2018**, *9* (1), 4893. https://doi.org/10.1038/s41467-018-07383-8.
(170) Wang, Q.; Bian, W.; Si, J.; Ding, F.; Zhao, Z.; Hu, B.; Wang, K.; Tan, J.; Chen, D.; Xu, S.; Sun, T.; Cai, M.; Liu, Z. Efficient and High-Conductivity Perovskite LEDs with Low Operating Voltage. *ACS Nano* **2025**, *19* (44), 38340–38349. https://doi.org/10.1021/acsnano.5c10266.
(171) Lin, Y.; Zhang, Y.; Zhang, J.; Marcinskas, M.; Malinauskas, T.; Magomedov, A.; Nugraha, M. I.; Kaltsas, D.; Naphade, D. R.; Harrison, G. T.; El-Labban, A.; Barlow, S.; De Wolf, S.; Wang, E.; McCulloch, I.; Tsetseris, L.; Getautis, V.; Marder, S. R.; Anthopoulos, T. D. 18.9% Efficient Organic Solar Cells Based on n-Doped Bulk-Heterojunction and Halogen-Substituted Self-Assembled Monolayers as Hole Extracting Interlayers. *Adv. Energy Mater.* **2022**, *12* (45), 2202503. https://doi.org/10.1002/aenm.202202503.
(172) Gedda, M.; Gkeka, D.; Nugraha, M. I.; Scaccabarozzi, A. D.; Yengel, E.; Khan, J. I.; Hamilton, I.; Lin, Y.; Deconinck, M.; Vaynzof, Y.; Laquai, F.; Bradley, D. D. C.; Anthopoulos, T. D. High-Efficiency Perovskite–Organic Blend Light-Emitting Diodes Featuring Self-Assembled Monolayers as Hole-Injecting Interlayers. *Adv. Energy Mater.* **2023**, *13* (33), 2201396. https://doi.org/10.1002/aenm.202201396.
(173) Li, H.; Zhu, X.; Zhang, D.; Gao, Y.; Feng, Y.; Ma, Z.; Huang, J.; He, H.; Ye, Z.; Dai, X. Thermal Management towards Ultra-Bright and Stable Perovskite Nanocrystal-Based Pure Red Light-Emitting Diodes. *Nat. Commun.* **2024**, *15* (1), 6561. https://doi.org/10.1038/s41467-024-50634-0.
(174) Zhao, L.; Roh, K.; Kacmoli, S.; Al Kurdi, K.; Jhulki, S.; Barlow, S.; Marder, S. R.; Gmachl, C.; Rand, B. P. Thermal Management Enables Bright and Stable Perovskite

Light-Emitting Diodes. *Adv. Mater.* **2020**, *32* (25), 2000752. https://doi.org/10.1002/adma.202000752.
(175) Lu, J.; Feng, W.; Mei, G.; Sun, J.; Yan, C.; Zhang, D.; Lin, K.; Wu, D.; Wang, K.; Wei, Z. Ultrathin PEDOT:PSS Enables Colorful and Efficient Perovskite Light-Emitting Diodes. *Adv. Sci.* **2020**, *7* (11), 2000689. https://doi.org/10.1002/advs.202000689.
(176) Mukherjee, S.; Panigrahi, A.; Perumal, A. Altering the Optoelectronic Properties of the PEDOT:PSS Hole Transport Layer with Sodium Polystyrenesulfonate to Enhance the Efficiency and Stability of Perovskite Light-Emitting Diode. *ACS Omega* **2025**, *10* (22), 23348–23358. https://doi.org/10.1021/acsomega.5c01768.
(177) Chen, X.; Zheng, F.; Hou, Y.; Yang, B.; Zhao, R.; Ma, G.; Wu, J.; Shafique, S.; Ghiggino, K. P.; Hu, Z. Guanidinium Iodide-Modified PEDOT:PSS Hole Transport Layer for Improving the Performance of 2D Perovskite Solar Cells. *Langmuir* **2025**, *41* (16), 10715–10725. https://doi.org/10.1021/acs.langmuir.5c01113.
(178) Kim, J. S.; Heo, J.-M.; Park, G.-S.; Woo, S.-J.; Cho, C.; Yun, H. J.; Kim, D.-H.; Park, J.; Lee, S.-C.; Park, S.-H.; Yoon, E.; Greenham, N. C.; Lee, T.-W. Ultra-Bright, Efficient and Stable Perovskite Light-Emitting Diodes. *Nature* **2022**, *611* (7937), 688–694. https://doi.org/10.1038/s41586-022-05304-w.
(179) Li, N.; Jia, Y.; Guo, Y.; Zhao, N. Ion Migration in Perovskite Light-Emitting Diodes: Mechanism, Characterizations, and Material and Device Engineering. *Adv. Mater.* **2022**, *34* (19), 2108102. https://doi.org/10.1002/adma.202108102.
(180) Futscher, M. H.; Gangishetty, M. K.; Congreve, D. N.; Ehrler, B. Manganese Doping Stabilizes Perovskite Light-Emitting Diodes by Reducing Ion Migration. *ACS Appl. Electron. Mater.* **2020**, *2* (6), 1522–1528. https://doi.org/10.1021/acsaelm.0c00125.
(181) Cao, Y.; Wang, N.; Tian, H.; Guo, J.; Wei, Y.; Chen, H.; Miao, Y.; Zou, W.; Pan, K.; He, Y.; Cao, H.; Ke, Y.; Xu, M.; Wang, Y.; Yang, M.; Du, K.; Fu, Z.; Kong, D.; Dai, D.; Jin, Y.; Li, G.; Li, H.; Peng, Q.; Wang, J.; Huang, W. Perovskite Light-Emitting Diodes Based on Spontaneously Formed Submicrometre-Scale Structures. *Nature* **2018**, *562* (7726), 249–253. https://doi.org/10.1038/s41586-018-0576-2.
(182) Jiang, Y.; Qin, C.; Cui, M.; He, T.; Liu, K.; Huang, Y.; Luo, M.; Zhang, L.; Xu, H.; Li, S.; Wei, J.; Liu, Z.; Wang, H.; Kim, G.-H.; Yuan, M.; Chen, J. Spectra Stable Blue Perovskite Light-Emitting Diodes. *Nat. Commun.* **2019**, *10* (1), 1868. https://doi.org/10.1038/s41467-019-09794-7.
(183) Conings, B.; Drijkoningen, J.; Gauquelin, N.; Babayigit, A.; D'Haen, J.; D'Olieslaeger, L.; Ethirajan, A.; Verbeeck, J.; Manca, J.; Mosconi, E.; Angelis, F. D.; Boyen, H.-G. Intrinsic Thermal Instability of Methylammonium Lead Trihalide Perovskite. *Adv. Energy Mater.* **2015**, *5* (15), 1500477. https://doi.org/10.1002/aenm.201500477.
(184) Yuan, B.; Zhong, H.; Shan, Q.; Hu, T.; Wu, X.; Cai, X.; Huang, M.; Zeng, H. Recent Advances in Stability Enhancement of Metal Halide Perovskite Light-Emitting Diodes. *Chem. Commun.* **2025**. https://doi.org/10.1039/D5CC05017K.
(185) Zhang, L.; Yang, X.; Jiang, Q.; Wang, P.; Yin, Z.; Zhang, X.; Tan, H.; Yang, Y. (Michael); Wei, M.; Sutherland, B. R.; Sargent, E. H.; You, J. Ultra-Bright and Highly Efficient Inorganic Based Perovskite Light-Emitting Diodes. *Nat. Commun.* **2017**, *8* (1), 15640. https://doi.org/10.1038/ncomms15640.
(186) Zhang, J.; Xing, X.; Qian, D.; Wang, A.; Gu, L.; Kuang, Z.; Wang, J.; Zhang, H.; Wen, K.; Xu, W.; Niu, M.; Du, X.; Yuan, L.; Cao, C.; Cao, Y.; Zhu, L.; Wang, N.; Yi, C.; Huang, W.; Wang, J. Sulfonic Zwitterion for Passivating Deep and Shallow Level Defects in Perovskite Light-Emitting Diodes. *Adv. Funct. Mater.* **2022**, *32* (22), 2111578. https://doi.org/10.1002/adfm.202111578.
(187) Guo, B.; Lai, R.; Jiang, S.; Zhou, L.; Ren, Z.; Lian, Y.; Li, P.; Cao, X.; Xing, S.; Wang, Y.; Li, W.; Zou, C.; Chen, M.; Hong, Z.; Li, C.; Zhao, B.; Di, D. Ultrastable Near-Infrared Perovskite Light-Emitting Diodes. *Nat. Photonics* **2022**, *16* (9), 637–643. https://doi.org/10.1038/s41566-022-01046-3.
(188) Shen, Z.; Xie, H.; Li, J.; Yan, S.; Ou, J.; Zheng, G.; Li, L.; Song, B.; Luo, J.; Tang, J. Phase Distribution Regulation of Thermally Evaporated Quasi-2D Perovskite Light-Emitting Diodes. *ACS Energy Lett.* **2025**, *10* (10), 4941–4947. https://doi.org/10.1021/acsenergylett.5c02099.
(189) Li, J.; Du, P.; Guo, Q.; Sun, L.; Shen, Z.; Zhu, J.; Dong, C.; Wang, L.; Zhang, X.; Li, L.; Yang, C.; Pan, J.; Liu, Z.; Xia, B.; Xiao, Z.; Du, J.; Song, B.; Luo, J.; Tang, J. Efficient All-Thermally Evaporated Perovskite Light-Emitting Diodes for Active-Matrix Displays. *Nat. Photonics* **2023**, *17* (5), 435–441. https://doi.org/10.1038/s41566-023-01177-1.
(190) Qin, X.; He, Y.; Tao, C. K.; Sergeev, A.; Wong, K. S.; Ren, Z.; Liang, Q.; Leung, T. L.; Li, G.; Popović, J.; Ng, A. M. C.; Djurišić, A. B. Effect of Passivating Molecules and Antisolvents on Lifetime of Green Dion–Jacobson Perovskite Light-Emitting Diodes. *ACS Appl. Mater. Interfaces* **2023**, *15* (25), 30083–30092. https://doi.org/10.1021/acsami.3c02170.
(191) Shah, K.; Murthy, K. V. R.; Chakrabarty, B. S. Investigation of UV Emission and Energy Transfer Process in Ce3+, Gd3+, Pr3+ and Their Combination Doped Nano Crystallite La2O3 Phosphors. *Results Opt.* **2023**, *11*, 100413. https://doi.org/10.1016/j.rio.2023.100413.
(192) Jobe, S.; Siddig, L. A.; Najar, A.; Saleh, N.; Khaleel, A.; Hassan, F. M. 0D Cesium Lanthanide Chlorides: Spectroscopic Insights and Optical Properties for Emerging Optoelectronic Applications. *Adv. Opt. Mater.* **2025**, *n/a* (n/a), e02940. https://doi.org/10.1002/adom.202502940.
(193) Luo, J.; Li, J.; Grater, L.; Guo, R.; Mohd Yusoff, A. R. bin; Sargent, E.; Tang, J. Vapour-Deposited Perovskite Light-Emitting Diodes. *Nat. Rev. Mater.* **2024**, *9* (4), 282–294. https://doi.org/10.1038/s41578-024-00651-8.
(194) Chu, S.; Chen, W.; Fang, Z.; Xiao, X.; Liu, Y.; Chen, J.; Huang, J.; Xiao, Z. Large-Area and Efficient Perovskite Light-Emitting Diodes via Low-Temperature Blade-Coating. *Nat. Commun.* **2021**, *12* (1), 147. https://doi.org/10.1038/s41467-020-20433-4.
(195) Shi, G.; Huang, Z.; Qiao, R.; Chen, W.; Li, Z.; Li, Y.; Mu, K.; Si, T.; Xiao, Z. Manipulating Solvent Fluidic Dynamics for Large-Area Perovskite Film-Formation and White Light-Emitting Diodes. *Nat. Commun.* **2024**, *15* (1), 1066. https://doi.org/10.1038/s41467-024-45488-5.
(196) Lu, Q.; Yang, Z.; Meng, X.; Yue, Y.; Ahmad, M. A.; Zhang, W.; Zhang, S.; Zhang, Y.; Liu, Z.; Chen, W. A Review on Encapsulation Technology from Organic Light Emitting Diodes to Organic and Perovskite Solar Cells. *Adv. Funct. Mater.* **2021**, *31* (23), 2100151. https://doi.org/10.1002/adfm.202100151.
(197) Zhao, Y.; Wang, V.; Lien, D.-H.; Javey, A. A Generic Electroluminescent Device for Emission from Infrared to Ultraviolet Wavelengths. *Nat. Electron.* **2020**, *3* (10), 612–621. https://doi.org/10.1038/s41928-020-0459-z.